\documentclass[prx,twocolumn,superscriptaddress,longbibliography, aps, 10pt, showkeys]{revtex4-2}

\usepackage[]{fontenc}

\usepackage{graphicx}
\usepackage{subcaption}
\usepackage{ragged2e} 
\usepackage{bm}
\usepackage{mathtools} 
\usepackage{hyperref}
\usepackage{chngcntr} 
\usepackage{multirow}
\usepackage{subeqnarray}

\usepackage[british]{babel}
\usepackage{amsmath}
\usepackage{amssymb}
\usepackage{framed}
\usepackage{soul}
\usepackage{color}
\usepackage{alltt}
\usepackage{pstricks,pst-node,pst-tree,pst-grad,pst-text,graphics}
\usepackage{dcolumn}
\usepackage{url}
\usepackage{cases}
\usepackage{natbib}

\usepackage{ifthen}
\usepackage{afterpage}
\usepackage{comment}
\usepackage{supertabular}
\usepackage{xspace}
\usepackage{setspace}
\usepackage{textcomp}
\usepackage{tikz}
\usetikzlibrary{decorations.pathmorphing}
\usetikzlibrary{decorations.markings}
\usetikzlibrary{arrows,shapes,decorations,automata,backgrounds,petri}

\newcommand{\Xgpcomment}[1]{\PackageError{commentCommand}{Don't use comments in production LaTeX}{}}

\newcommand{\latticeVector}{\avec}
\newcommand{\latticeVectorCompo}{a}

\newcommand{\BMatrix}{\gpmatrix{B}}
\newcommand{\CMatrix}{\gpmatrix{C}}
\newcommand{\SMatrixFourier}[2]{\gpmatrix{S}_{#1 #2}}
\newcommand{\SMatrixEval}{s}

\newcommand{\plaind}{\mathrm{d}}
\newcommand{\dint}[1]{\mathchoice{\!\plaind#1\,}{\!\plaind#1\,}{\!\plaind#1\,}{\!\plaind#1\,}}
\newcommand{\ddint}[1]{\ddintx{#1}{d}}
\newcommand{\dTWOint}[1]{\ddintx{#1}{2}}
\newcommand{\ddintx}[2]{\mathchoice{\!\plaind^{#2}#1\,}{\!\plaind^{#2}#1\,}{\!\plaind^{#2}#1\,}{\!\plaind^{#2}#1\,}}

\newcommand{\dbar}{\text{\dj}}

\DeclareFontFamily{U}{wncy}{}
\DeclareFontShape{U}{wncy}{m}{n}{<->wncyr10}{}
\DeclareSymbolFont{mcy}{U}{wncy}{m}{n}
\DeclareMathSymbol{\sha}{\mathord}{mcy}{"58}

\usepackage{dsfont}

\newcommand{\gpset}[1]{\mathds{#1}}

\newcommand{\canetset}[1]{{\mathchoice {\hbox{$\sf\textstyle #1\kern-0.4em #1$}}
{\hbox{$\sf\textstyle #1\kern-0.4em #1$}}
{\hbox{$\sf\scriptstyle #1\kern-0.3em #1$}}
{\hbox{$\sf\scriptscriptstyle #1\kern-0.2em #1$}}}}

\newcommand{\Pset}{\gpset{P}}
\newcommand{\Zset}{\gpset{Z}}

\newcommand{\Rset}{\gpset{R}}

\newcommand{\Sset}{\gpset{S}}

\def\nbZ{{\mathchoice {\hbox{$\sf\textstyle Z\kern-0.4em Z$}}
{\hbox{$\sf\textstyle Z\kern-0.4em Z$}}
{\hbox{$\sf\scriptstyle Z\kern-0.3em Z$}}
{\hbox{$\sf\scriptscriptstyle Z\kern-0.2em Z$}}}}

\newcommand{\gpvec}[1]{\mathbf{#1}}

\newcommand{\zerovec}{\gpvec{0}}
\newcommand{\nullvec}{\zerovec}
\newcommand{\avec}{\gpvec{a}}

\newcommand{\evec}{\gpvec{e}}
\newcommand{\fvec}{\gpvec{f}}

\newcommand{\kvec}{\gpvec{k}}

\newcommand{\mvec}{\gpvec{m}}
\newcommand{\nvec}{\gpvec{n}}
\newcommand{\pvec}{\gpvec{p}}
\newcommand{\qvec}{\gpvec{q}}
\newcommand{\rvec}{\gpvec{r}}

\newcommand{\uvec}{\gpvec{u}}
\newcommand{\vvec}{\gpvec{v}}
\newcommand{\wvec}{\gpvec{w}}
\newcommand{\xvec}{\gpvec{x}}

\newcommand{\zvec}{\gpvec{z}}

\newcommand{\Fvec}{\gpvec{F}}

\newcommand{\Nvec}{\gpvec{N}}

\newcommand{\Rvec}{\gpvec{R}}

\newcommand{\phivec}{\bm{\phi}}

\newcommand{\xivec}{\bm{\xi}}

\newcommand{\transpose}{\mathsf{T}}
\renewcommand{\det}[1]{\mathchoice
{\operatorname{det}\left(#1\right)}
{\operatorname{det}(#1)}
{\operatorname{det}(#1)}
{\operatorname{det}(#1)}
}
\newcommand{\ident}{\mathds{1}}

\newcommand{\Trace}{\operatorname{Tr}}

\newcommand{\CC}{\mathcal{C}}

\newcommand{\EC}{\mathcal{E}}

\newcommand{\NC}{\mathcal{N}}
\newcommand{\OC}{\mathcal{O}}

\newcommand{\GammaTilde}{\tilde{\Gamma}}

\newcommand{\half}{\mathchoice{\frac{1}{2}}{(1/2)}{\frac{1}{2}}{(1/2)}}

\newcommand{\fourth}{\mathchoice{\frac{1}{4}}{(1/4)}{\frac{1}{4}}{(1/4)}}

\newcommand{\quarter}{\fourth}

\renewcommand{\exp}[1]{\mathchoice%
{\mathrm{e}^{#1}}%
{\operatorname{exp}(#1)}
{\operatorname{exp}\left(#1\right)}%
{\operatorname{exp}\left(#1\right)}}

\newcommand{\elabel}[1]{\label{eq:#1}}
\newcommand{\eref}[1]{(\ref{eq:#1})}
\newcommand{\Eref}[1]{\mbox{Eq.~(\ref{eq:#1})}}
\newcommand{\Erefs}[1]{\mbox{Eqs.~(\ref{eq:#1})}}

\newcommand{\seclabel}[1]{\label{sec:#1}}

\newcommand{\Sref}[1]{Section~\ref{sec:#1}}

\newcommand{\flabel}[1]{\label{fig:#1}}
\newcommand{\fref}[1]{\ref{fig:#1}}
\newcommand{\Frefs}[1]{Figs.~\ref{fig:#1}}
\newcommand{\Fref}[1]{Fig.~\ref{fig:#1}}

\newcommand{\Tref}[1]{Table~\ref{tab:#1}}

\newcommand{\latin}[1]{{\it #1}}
\newcommand{\ie}{i.e.\@\xspace}
\newcommand{\eg}{e.g.\@\xspace}
\newcommand{\cf}{cf.\@\xspace}
\newcommand{\etc}{etc.\@\xspace}

\newlength \standardfigwidth
\DeclareMathAlphabet{\matheub}{U}{eur}{m}{n}

\usepackage{tikz}
\usetikzlibrary{decorations.pathmorphing}
\usetikzlibrary{decorations.markings}
\usetikzlibrary{arrows,shapes,decorations,automata,backgrounds,petri}
\usetikzlibrary{calc}
\usetikzlibrary{patterns}
\usetikzlibrary{decorations.text}

\newcommand{\ave}[2][]{\mathchoice%
{\left\langle #2 \right\rangle_{#1}}%
{\langle #2\rangle_{#1}}
{\langle #2\rangle_{#1}}%
{\langle #2\rangle_{#1}}}

\makeatletter
\newcommand{\creatX}[2][]{a^{\@ifempty{#1}{\dagger}{\dagger\,#1}}\@ifempty{#2}{}{(#2)}}
\makeatother
\newcommand{\creatDS}{\tilde{a}}
\makeatletter
\newcommand{\creatDSX}[2][]{\@ifempty{#1}{\creatDS}{\creatDS^{#1}}\@ifempty{#2}{}{(#2)}}
\makeatother

\makeatletter
\newcommand{\annihX}[2][]{a\@ifempty{#1}{}{^{#1}}\@ifempty{#2}{}{(#2)}}
\newcommand{\kernel}[1]{K\@ifempty{#1}{}{\,\!\!\left(#1\right)}}
\newcommand{\kernelDeri}[1]{K'\@ifempty{#1}{}{\left(#1\right)}}

\newcommand{\tableofappendixcontents}{\@starttoc{toa}}
\newcommand{\l@toact}[2]{\@dottedtocline{1}{1.5em}{5.5em}{#1}{#2}}
\newcommand{\l@toactspace}[2]{\ \hfil}
\makeatother

\newcommand{\APref}[1]{\mbox{App.~\ref{sec:#1}}}

\newcommand{\imag}{\mathring{\imath}}

\renewcommand{\exp}[1]{\mathchoice%
{e^{#1}}%
{\operatorname{exp}(#1)}%
{\operatorname{exp}(#1)}%
{\operatorname{exp}(#1)}}
\newcommand{\corresponding}{\hat{=}}

\usepackage{xspace}

\renewcommand{\st}[1]{}

\newcommand{\entropyProduction}{\dot{S}_{\text{int}}}
\newcommand{\entropyProductionDensity}{\dot{\sigma}}
\newcommand{\entropyProductionDensityFourier}{\dot{\hat{\sigma}}}

\newcommand{\density}{\rho}

\newcommand{\densityFourier}{\hat{\rho}}

\makeatletter
\newcommand{\HMdensity}[2]{\rho_{#2}^{\@ifempty{#1}{}{(#1)}}}
\newcommand{\HMdensityBar}[2]{\overline{\rho}_{#2}^{\@ifempty{#1}{}{(#1)}}}
\newcommand{\HMdensityDot}[2]{\dot{\rho}_{#2}^{\@ifempty{#1}{}{(#1)}}}
\makeatother

\usepackage{textcomp}

\newcommand{\pairPot}{U}
\newcommand{\pairPotCoeff}{u}

\newcommand{\RLVec}{\qvec}
\newcommand{\BrillouinVec}{\kvec}
\newcommand{\BriVec}[1]{\BrillouinVec_{#1}}

\newcommand{\gpmatrix}[1]{\bm{\mathsf{#1}}}
\newcommand{\Wmatrix}{\gpmatrix{W}}
\makeatletter

\newcommand{\displacement}[3][]{u^{#1}_{#2}\@ifempty{#3}{}{(#3)}}
\newcommand{\displacementFourier}[3][]{\hat{u}^{#1}_{#2}\@ifempty{#3}{}{(#3)}}
\newcommand{\displacementVec}[3][]{\uvec^{#1}_{#2}\@ifempty{#3}{}{(#3)}}
\newcommand{\displacementVecDot}[2]{\dot{\uvec}_{#1}\@ifempty{#2}{}{(#2)}}
\newcommand{\displacementDot}[3][]{\dot{u}^{#1}_{#2}\@ifempty{#3}{}{(#3)}}
\newcommand{\displacementDotFourier}[3][]{\dot{\hat{u}}^{#1}_{#2}\@ifempty{#3}{}{(#3)}}
\newcommand{\displacementRel}[3][]
{v^{#1}_{#2}\@ifempty{#3}{}{(#3)}}

\newcommand{\displacementVecFourier}[2]{\hat{\uvec}_{#1}\@ifempty{#2}{}{(#2)}}
\newcommand{\displacementVecDotFourier}[2]{\dot{\hat{\uvec}}_{#1}\@ifempty{#2}{}{(#2)}}
\newcommand{\displacementBar}[1]{\overline{u}\@ifempty{#1}{}{(#1)}}

\newcommand{\commaseparated}[2]{#1\@ifempty{#2}{}{,#2}}
\newcommand{\dynMatrixArgs}[2]{\commaseparated{#1}{#2}}

\newcommand{\dynamicalMatrixFourier}[2]{\hat{\gpmatrix{D}}_{\dynMatrixArgs{#1}{#2}}}
\newcommand{\dynamicalScalar}[2]{D_{\dynMatrixArgs{#1}{#2}}}
\newcommand{\dynamicalScalarFourier}[2]{\hat{D}_{\dynMatrixArgs{#1}{#2}}}
\newcommand{\curlyDynamicalMatrix}{\gpmatrix{P}}
\newcommand{\oneMcos}[2]{\kappa_{#1\@ifempty{#2}{}{,#2}}}
\newcommand{\oneMcosFunc}{f_\kappa}
\newcommand{\oneMcosCont}[1]{\kappa_{#1}}

\newcommand{\thermalNoiseFourier}{\hat{\xi}}
\newcommand{\thermalNoiseVecFourier}{\hat{\xivec}}
\newcommand{\thermalNoiseFourierPM}[1][]{\hat{\xi}^{#1}}

\newcommand{\activeNoiseFourier}{\hat{w}}
\newcommand{\activeNoiseFourierPM}[1][]{\hat{w}^{#1}}
\newcommand{\activeNoiseVecFourier}{\hat{\wvec}}

\newcommand{\wavenumber}[1]{k_{#1}}

\newcommand{\displacementVecBar}[2][]{\overline{\uvec^{#1}}\@ifempty{#2}{}{(#2)}}

\newcommand{\displacementVecRel}[3][]{\vvec^{#1}_{#2}\@ifempty{#3}{}{(#3)}}
\newcommand{\displacementVecRelDot}[2]{\dot{\vvec}_{#1}\@ifempty{#2}{}{(#2)}}
\newcommand{\displacementRelDot}[2]{\dot{v}_{#1}\@ifempty{#2}{}{(#2)}}

\newcommand{\Fuvec}{\hat{\uvec}}

\newcommand{\displacementFVec}[2]{\Fuvec_{#1}\@ifempty{#2}{}{(#2)}}
\newcommand{\displacementFVecDot}[2]{\dot{\Fuvec}_{#1}\@ifempty{#2}{}{(#2)}}
\newcommand{\displacementFDot}[2]{\dot{\Fu}_{#1}\@ifempty{#2}{}{(#2)}}
\newcommand{\displacementFRel}[2]{\Fv_{#1}\@ifempty{#2}{}{(#2)}}

\newcommand{\displacementFVecBar}[1]{\overline{\Fuvec}\@ifempty{#1}{}{(#1)}}

\newcommand{\displacementFVecRel}[2]{\Fvvec_{#1}\@ifempty{#2}{}{(#2)}}
\newcommand{\displacementFVecRelDot}[2]{\dot{\Fvvec}_{#1}\@ifempty{#2}{}{(#2)}}
\newcommand{\displacementFRelDot}[2]{\dot{\Fv}_{#1}\@ifempty{#2}{}{(#2)}}

\newcommand{\dynamicalMatrix}[2]{\gpmatrix{D}_{#1\@ifempty{#2}{}{,#2}}}

\makeatother

\newcommand{\internalEnergy}{\EC}
\newcommand{\internalEnergyHarmonic}{\EC_\text{harm.}}

\newcommand{\siteVec}{\Rvec}
\newcommand{\latticeMatrix}{\gpmatrix{A}}
\newcommand{\latticeConstant}{\ell}
\newcommand{\latticeLength}{L}
\newcommand{\eval}{\lambda}
\newcommand{\NOne}{N^1}
\newcommand{\NTwo}{N^2}
\newcommand{\NOneTwo}{N^{1,2}}
\newcommand{\latticeLengthOneTwo}{L^{1,2}}

\newcommand{\Ematrix}{\gpmatrix{E}}

\newcommand{\thermalNoise}{\xi}
\newcommand{\thermalNoiseVec}{\xivec}
\newcommand{\activeNoise}{w}
\newcommand{\activeNoiseVec}{\wvec}
\newcommand{\activityCorrInt}{\mathcal{W}}
\newcommand{\thermalCorrInt}{\Xi}

\newcommand{\siteSet}{\Sset}
\newcommand{\activity}{\nu} 
\newcommand{\noiseRate}{\mu}
\newcommand{\springConstant}{\kappa}
\newcommand{\relaxedLength}{z_0}

\newcommand{\circmark}{\odot}
\newcommand{\sdots}{\ifmmode\mathinner{\ldotp\kern-0.2em\ldotp\kern-0.2em\ldotp}\else.\kern-0.13em.\kern-0.13em.\fi}

\usepackage{color}
\definecolor{darkgreen}{rgb}{0,0.6,0}
\definecolor{darkblue}{rgb}{0,0,0.6}
\definecolor{darkred}{rgb}{0.6,0,0}
\definecolor{darkpurple}{rgb}{0.5,0,0.5}
\hypersetup{
bookmarksopen=true,
pdftitle="",
pdfauthor="", 
pdftoolbar=false, 
pdfstartview={FitH},		
pdfmenubar=true,			
pdfhighlight=/O,			
colorlinks=true,			
urlcolor=darkblue,
citecolor=darkblue,		
linkcolor=darkblue}

\makeatletter
\newcommand{\customlabel}[2]{%
   \protected@write \@auxout {}{\string \newlabel {#1}{{#2}{\thepage}{#2}{#1}{}} }%
   \hypertarget{#1}{\hspace{0pt}}
}
\makeatother

\begin{document}

\newcommand{\titleText}{Exact results and instabilities in the harmonic approximation of active crystals}
\title{\titleText}

\author{Connor Roberts}
\affiliation{%
Department of Mathematics
and Centre of Complexity Science, 
Imperial College London, London SW7 2AZ, United Kingdom}%

\author{Gunnar Pruessner}
\email{g.pruessner@imperial.ac.uk}
\affiliation{%
Department of Mathematics
and Centre of Complexity Science, 
Imperial College London, London SW7 2AZ, United Kingdom}%

\date{29 November 2025}

\begin{abstract}
Condensates of active particles such as cells form almost-crystalline lattices which play a central role in many biological systems. Typically, their properties have been determined merely by analogy to the rather trivial one-dimensional case, leaving a gap between experimentally accessible observables and suitable theoretical models. Within a harmonic approximation, we characterise analytically a two-dimensional triangular lattice of active particles that interact with their nearest neighbours through a general pair potential, obtaining exact expressions for the correlators. We study this ``active crystal’’ as a means of characterising active matter in the dense phase.
Our treatment correctly approximates arbitrary pair potentials, rather than demanding an unphysical non-singular bilinear form. We retain ``off-diagonal’’ terms that are routinely neglected despite quantifying the anisotropy of the particles' local potential. From the exact expressions for the correlation matrices, we derive exact results that shed light on the presence (or absence) of crystalline order. We further calculate the mean-squared particle separation, energy, entropy production rate and the onset of a pressure-induced instability resulting in the breakdown of the harmonic approximation. The entropy production rate is found to have a general form that is valid for generic active particles and lattice geometries, while resembling that of non-interacting ``active modes’’. 
\end{abstract}

                              
\maketitle

\section{Introduction}
Active matter encompasses systems whose constituents perpetually dissipate energy to perform directed motion. This frequently produces phenomena with no equilibrium counterpart, such as boundary aggregation of active particles \cite{angelani2017confined, li2009accumulation, razin2020entropy, malakar2018steady, roberts2022exact} and autonomous ratcheting \cite{di2010bacterial,angelani2011active,reichhardt2017ratchet,roberts2023run}. Perhaps the most striking phenomenon in active matter is motility-induced phase separation, \ie the formation of a condensed phase from active particles that interact only through repulsive forces \cite{cates2015motility}. When these condensates are densely packed in two dimensions, they form an almost-perfect triangular lattice or ``active crystal" \cite{caprini2019comparative, caprini2023entropons, caprini2023entropy, caprini2023inhomogeneous}. In a mean-field spirit \cite{caprini2020spontaneous, caprini2020active, caprini2021spatial}, much of the understanding of the condensed phase in the literature is based on the much simpler analysis of the one-dimensional counterpart \cite{winkler2020physics, vandebroek2015dynamics, singh2021crossover, samanta2016chain, put2019non, osmanovic2018properties, osmanovic2017dynamics, goswami2022reconfiguration, ghosh2014dynamics, dutta2023entropy, chaki2019enhanced, caprini2020time}. As a result, there is relatively little quantitative understanding of these crystals in higher dimensions as well as apparent or supposed clashes with established concepts such as the Mermin-Wagner theorem \cite{shi2023extreme, galliano2023two, mermin1968crystalline, mermin1966absence}. Hence, to address these issues, the present work characterises such active crystals in two dimensions.

Starting from a setup of dense, non-linearly repulsively interacting active particles, we show in \APref{triangular_active_crystal} how a linear Langevin description of an active crystal is obtained in the \emph{harmonic approximation}. 
In this description, the equations of motion are exactly solvable, so that correlation functions can be determined in exact form. These can then be used to quantify observables that are physically more relevant, such as the entropy production \cite{ro2022model, cocconi2020entropy, razin2020entropy, garcia2021run} and fluctuations \cite{zhang2022field}. Our results relate the macroscopic properties of the active crystal directly to the underlying microscopic interactions. Crucially, we avoid entirely two commonly used approximations, namely 1) to assume an unphysical, bilinear and diagonal, form of the pair potential \cite{caprini2020active,caprini2021spatial,caprini2023inhomogeneous}, whereby forces act along the mutual relative displacement, rather than the connecting vector, and 2) to adopt a mean-field-like approach of casting the dynamics as an effective one-dimensional equation \cite{caprini2023entropons, caprini2023entropy, caprini2023inhomogeneous} (originally derived in \cite{caprini2020spontaneous, caprini2020active, caprini2021spatial}), characterised by a single effective coupling.
Instead, we keep the full lattice anisotropy and show in which sense crystalline order is or is not sustained in two dimensions as well as when and how it breaks down. Such a quantitative characterisation of active crystals can also inform our understanding of more complex biological processes, such as extrusion events in cell monolayers \cite{monfared2023mechanical}.

The remainder of this work is structured as follows: in \Sref{model}, we describe the model in detail and highlight the features that have been ignored or unnecessarily approximated in the literature. Much of the technical details of the derivation are relegated to the appendices.
In \Sref{Results}, we state the main results of the derivations, 
such as in which sense crystalline order exists in the harmonic approximation, 
\Eref{densityFourier_result}, 
the general displacement correlators, \Eref{displacement_corr}, 
and the mean squared particle separation (i.e.\ the variance of the lattice spacing), \Eref{2D_crystalline_integrity}. We find the triangular lattice sustains external pressure, but eventually undergoes a buckling transition, 
beyond which the harmonic approximation breaks down.
We further demonstrate that activity spoils
equipartition, \Eref{expected_energy_main}, and finally calculate the entropy production, \Eref{EPR_derivation_step5_main}. To our knowledge, the present work is the first to calculate in closed form the entropy production of many interacting degrees of freedom.
The contribution to the entropy production from each normal mode is found to have the same form as that of non-interacting active particles moving in a harmonic potential \cite{garcia2021run, frydel2023entropy}. In fact the derived expression is generally applicable to any system of active particles subject to linear forces.
Again, the details of the derivation can be found in the extensive appendix, in particular \APref{Entropy}. We finally conclude and discuss our results in the light of previous work in \Sref{Discussion}.

\section{Model}\seclabel{model}
In the following, we study the bulk properties of a triangular lattice composed of active particles, \Fref{ActiveCrystalSchematic}. Such an arrangement occurs, for example, when self-propelled particles undergo motility-induced phase separation \cite{RednerHaganBaskaran:2013, cates2015motility}, or when they are confined to a finite volume by suitable boundaries. The lattice may be subject to very significant pressure fluctuations \cite{RednerHaganBaskaran:2013} --- either from the bath or the boundaries --- so that their rest positions are determined by the balancing of all forces exerted by neighbouring particles, rather than those forces all vanishing. 
Hence, the pair potentials $\pairPot$ between any neighbouring particles residing at their rest positions are generally \emph{not} at a minimum, \Fref{ActiveCrystalSchematic}a, but rather the forces sum to zero. In other words, the particles are ``hemmed in''.

\begin{figure}
    \includegraphics[width=\columnwidth, trim = 0cm 8.1cm 0cm 0cm, clip]{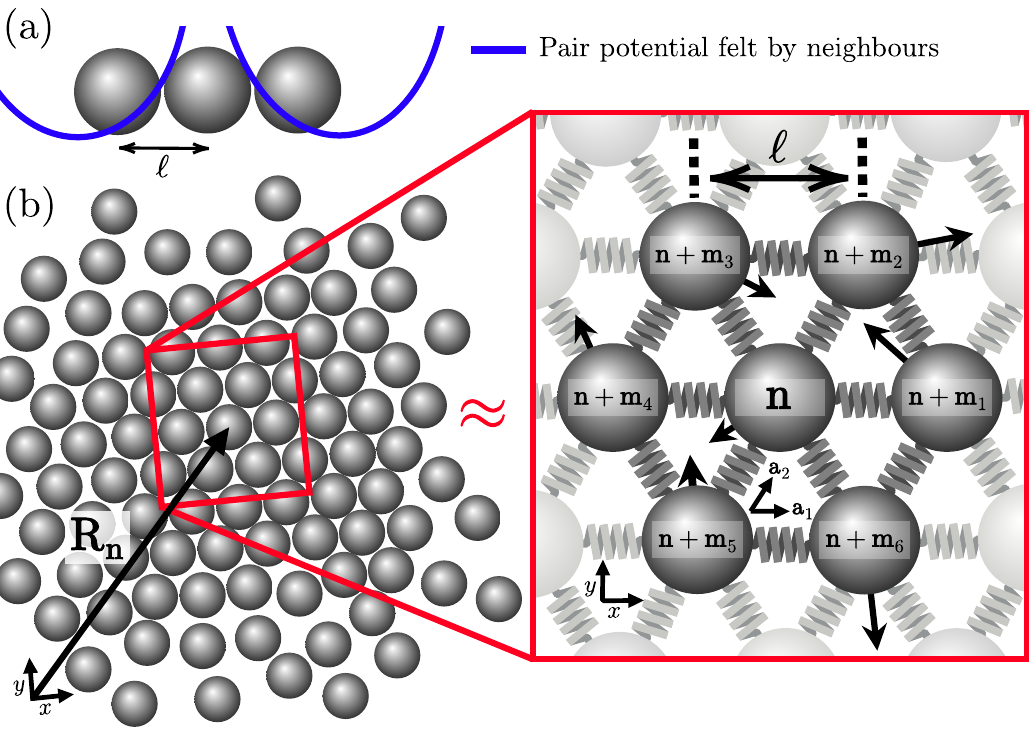}
    \caption{\justifying Schematic representation of how the bulk of a condensate of densely packed active particles, interacting through repulsive pair potentials, can be approximated by a harmonic active crystal. 
    (a)~Illustration of the repulsive pair potential felt by nearest neighbours of the central particle. (b) In the bulk of the condensate, particles form an ordered lattice. Particle interactions are illustrated by springs, which are typically under pressure. Arrows indicate the direction and magnitude of each particle's self-propulsion $\activeNoiseVec_{\nvec'}(t)$.
    }
    \flabel{ActiveCrystalSchematic}
\end{figure}

Following a notation similar to that of Ashcroft and Mermin \cite{ashcroft1976solid}, the active particles are 
indexed by $\nvec=(n^1,n^2)\in\{1,2,\ldots,\NOne\}\otimes\{1,2,\ldots,\NTwo\} = \siteSet$
and are described in terms of their displacement $\displacementVec{\nvec}{t}$ from lattice positions $\siteVec_\nvec=(\latticeVector_1 n^1 + \latticeVector_2 n^2)\latticeConstant$, where the primitive lattice vectors are dimensionless, i.e.\ $|\latticeVector_{1,2}|=1$, and $\latticeConstant$ represents the dimensionful lattice spacing. The setup is illustrated in \Fref{ActiveCrystalSchematic}b. 
Superscripts will generally denote components of two-dimensional vectors, while subscripts will typically represent (lattice) indices.

We assume there is exactly one particle associated with each lattice site and therefore $V=\NOne\NTwo$ particles in total. For finite $\NOne$ and $\NTwo$, we assume periodic Born-von Karman boundary conditions \cite{ashcroft1976solid,vonKarmanBorn:1912,BornvonKarman:1913}, such that $(n^1,n^2)$ refers to the same particle (and rest position) as $(n^1+\NOne,n^2)$ and $(n^1,n^2+\NTwo)$. 
As a consequence, the lattice cannot freely rotate.
We further assume every particle indexed by $\nvec$ interacts with $Q=6$ nearest neighbouring particles indexed by $\nvec'=\nvec+\mvec_q$ for $q=1,2,\ldots,6$, where $\mvec_q$ represents the allowed index ``displacements'', as shown in \Fref{ActiveCrystalSchematic}b and discussed in \APref{triangular_active_crystal}. They can be used to define the physical displacements of nearest neighbours via $\latticeVector_q=\latticeVector_1m_q^1+\latticeVector_2m_q^2$, so that $\siteVec_{\nvec+\mvec_q}=\siteVec_\nvec+\latticeConstant\latticeVector_q$.

In order to keep displacements well-defined, we assume particles are initially located at $\siteVec_\nvec$, so that $\displacementVec{\nvec}{t=0}=\nullvec$. Crucially, and in contrast to much of the literature, the pair potential $\pairPot(|\displacementVec{\nvec}{t}-\displacementVec{\nvec+\mvec_q}{t}-\latticeConstant \latticeVector_q|)$ acting between nearest-neighbouring particles is a function of the \emph{distance} between particles. Regardless of whether the pair potential is quadratic, and in this sense harmonic, the singular nature of the distance renders the interaction anharmonic. Expanding the potential, and the distance featuring in its argument, to second order results in the \emph{harmonic approximation}, \APref{triangular_active_crystal}, that we pursue in the following. As a matter of mechanical stability, we demand $\pairPot''(\latticeConstant)>0$, while the sign of $\pairPot'(\latticeConstant)$ depends on whether the lattice is under tension, $\pairPot'(\latticeConstant)>0$, or under pressure, $\pairPot'(\latticeConstant)<0$.

Given the periodic boundary conditions, there is a rigid global translation 
\begin{equation}
	\displacementVecBar{t} = \frac{1}{V} \sum_{\nvec} \displacementVec{\nvec}{t} 
\end{equation}
of all particles that does not result in a restoring force, so that the  $\displacementVec{\nvec}{t}$ become much larger than the lattice spacing $\latticeConstant$. Most of the relevant physics is thus captured by the \emph{relative} displacement, 
\begin{equation}
	\displacementVecRel{\nvec}{t} = \displacementVec{\nvec}{t} - \displacementVecBar{t}\ .
\end{equation}
The stochastic equation of motion in the harmonic approximation, \APref{triangular_active_crystal}, for the triangular lattice is
\begin{equation}\elabel{2d_SDE_main}
    \displacementVecDot{\nvec}{t} = - \sum_{\nvec'} \dynamicalMatrix{\nvec}{\nvec'} \displacementVec{\nvec'}{t}
    + \thermalNoiseVec_\nvec(t) + \activeNoiseVec_\nvec(t)\ ,
\end{equation}
with dynamical $2\times2$ matrix $\dynamicalMatrix{\nvec}{\nvec'}$ \cite{ashcroft1976solid} as defined in \Eref{def_dynamicalMatrix_HarmonicAppendix} in terms of the derivatives of the pair potential $\pairPot$ and the lattice vectors $\latticeVector_q$. We do not assume that $\dynamicalMatrix{\nvec}{\nvec'}$ is diagonal. What makes \Eref{2d_SDE_main} a \emph{harmonic approximation} is its linearity in $\displacementVec{\nvec'}{}$, which cannot fully account for the dependence of the pair potential on the \emph{singular} distance $|\displacementVec{\nvec}{}-\displacementVec{\nvec+\mvec_q}{}-\latticeConstant\latticeVector_q|$ between particles, even when the interaction between particles is Hookian, say $\pairPot=\springConstant(|\displacementVec{\nvec}{}-\displacementVec{\nvec+\mvec_q}{}-\latticeConstant\latticeVector_q|-\relaxedLength)^2/2$. However, the harmonic approximation systematically approximates the pair force acting along the connecting line between particles, rather than assuming a particularly simple, yet unphysical, form of the potential, say $\pairPot=\springConstant|\displacementVec{\nvec}{}-\displacementVec{\nvec+\mvec_q}{}|^2/2$, independent of the connecting line.

The vectorial thermal noise $\thermalNoiseVec_\nvec(t)\in\Rset^2$ in \Eref{2d_SDE_main} is characterised by the correlator matrix
\begin{equation}\elabel{def_thermalNoise_main}
    \ave{\thermalNoiseVec_\nvec(t) \thermalNoiseVec^\transpose_{\nvec'}(t')}
    = \Gamma^2 \ident_2 \delta_{\nvec,\nvec'} \delta(t-t')\ ,
\end{equation}
where $\thermalNoiseVec_\nvec(t) \thermalNoiseVec^\transpose_{\nvec'}(t')$ is an outer product, $\Gamma^2$ is the square of the noise amplitude $\Gamma$ and $\ident_2$ is the $2\times2$ identity matrix. 
Meanwhile, the vectorial active noise $\activeNoiseVec_\nvec(t)\in\Rset^2$ has correlator matrix
\begin{equation}\elabel{activeNoise_correlator_main}
\ave{\activeNoiseVec_\nvec(t) \activeNoiseVec^\transpose_{\nvec'}(t')}
    = \activity^2 \ident_2 \delta_{\nvec,\nvec'} \exp{-\noiseRate|t-t'|}\ ,
\end{equation}
implementing the self-propulsion of speed $\activity$, so that $\activity^2$ is a square, and correlation time $1/\noiseRate$. 
Equation \eref{activeNoise_correlator_main} represents a general active-noise correlator satisfied by the three canonical active particle models: active Ornstein-Uhlenbeck \cite{FodorETAL:2016}, run-and-tumble \cite{TailleurCates:2008}, and active Brownian \cite{Howse:2007}. More complicated active noises can be considered and implemented by adjusting, in particular, \Eref{def_activityCorrInt} in \APref{correlators}. The identity matrix $\ident_2$ in \Eref{activeNoise_correlator_main} means activity in the $x$-direction does not affect activity in the $y$-direction, and vice versa, which arises from the mirror and rotational symmetry of the statistics of $\activeNoiseVec(t)$.

As we assume nearest-neighbour interactions only, the dynamical matrix $\dynamicalMatrix{\nvec}{\nvec'}$ in \Eref{2d_SDE_main} vanishes for all $\nvec'$ except for $\nvec'=\nvec$ and for the six nearest neighbours $\nvec'=\nvec+\mvec_q$, with $q=1,\ldots,6$. This renders \Eref{2d_SDE_main} amenable to a significant simplification by Fourier transform. However, the matrix $\dynamicalMatrix{\nvec}{\nvec'}$ is generally not diagonal for any of the six $\nvec' = \nvec+\mvec_q$. This is a complication sometimes omitted 
in the literature by effectivevely replacing $\dynamicalMatrix{\nvec}{\nvec'}$ with 
$\sum_q\dynamicalMatrix{\nvec}{\nvec+\mvec_q}/6 = -\dynamicalMatrix{\nvec}{\nvec}/6\propto\ident_2$ 
\cite{caprini2020hidden, caprini2021spatial,caprini2023entropons, caprini2023entropy, caprini2023inhomogeneous},
which amounts to a mean-field like approach by averaging the coupling matrix over the nearest neighbours. As a result, the equation of motion under this simplification is readily diagonal, similar to two non-interacting one-dimensional chains. In the present work, we instead calculate the full displacement correlator, \Eref{displacement_corr}, without resorting to this simplification. We then use the resulting framework to derive some key properties of active crystals in closed form.

\section{Results}\seclabel{Results}
The relative displacements $\displacementVecRel{\nvec}{t}$ can be described in terms of wave vectors located in the first Brillouin zone of indices, 
here equivalently
$\BriVec{\pvec}=(2\pi p^1/\NOne,2\pi p^2/\NTwo)$, where $\pvec=(p^1,p^2)\in\{0,1,\ldots,\NOne-1\}\otimes\{0,1,\ldots,\NTwo-1\}=\Pset$. The wave vectors can be seen as degrees of freedom representing collective excitations, resulting in wave-like $\displacementVecRel{\nvec}{t}$ as they vary between lattice sites $\nvec$.
The wave vectors $\BriVec{\pvec}$, which become arbitrarily dense in the thermodynamic limit, are the Fourier modes of functions $\displacementVecRel{\nvec}{}$ of $\nvec$. In physical space, where $\nvec$ corresponds to $\siteVec_\nvec$, the largest component of such a  Fourier mode has magnitude $2\pi/\latticeConstant$, the smallest, non-vanishing one has magnitude $2\pi/\latticeLengthOneTwo$, with $\latticeLengthOneTwo=\NOneTwo\latticeConstant$ the physical extent of the lattice. 

\subsection{Crystalline long-ranged order}
In the following, we want to make the link to the traditional notion of crystalline long-ranged order, as introduced by Mermin \cite{mermin1968crystalline,Strandburg:1988}, and to established results for active crystals \cite{galliano2023two,MairePlati:2024}. Crystalline order is found if the particle density $\density(\xvec;t)$ displays periodicity.

Although we derive our results in terms of the Fourier modes $\BriVec{\pvec}$, in the present section we need to draw on the (much larger) reciprocal lattice vectors $\RLVec\in(2\pi\Zset/\latticeConstant)^2$. These $\RLVec$ have the convenient property that $\RLVec\cdot\siteVec_\nvec$ is an integer multiple of $2\pi$ for every $\nvec\in\siteSet$. The smallest, non-vanishing reciprocal lattice vector has magnitude $2\pi/\latticeConstant$, while there is no largest, reflecting that the $\displacementVecRel{\nvec}{t}$ are continuous, so that $\density(\xvec;t)$ may be evaluated for any $\xvec\in\Rset^2$ with a period of one unit cell.

Mermin focuses on the expectation of the instantaneous density $\sum_{\nvec}\delta(\siteVec_\nvec+\displacementVec{\nvec}{t}-\xvec)$ in terms of its Fourier transform. However, to suppress the effect of the global rigid displacement $\displacementVecBar{t}$, which trivially ``smears out'' periodicity, we may instead define the density as $\density(\xvec;t)=\sum_{\nvec}\delta(\siteVec_\nvec+\displacementVecRel{\nvec}{t}-\xvec)$. As the $\displacementVecRel{\nvec}{t}$ follow the same statistics for different $\nvec$, the density $\ave{\density(\xvec;t)}$ is invariant under changing the fixed $\nvec'$ in
\begin{equation}
    \ave{\density(\xvec;t)} = \sum_{\nvec}\ave{\delta(\siteVec_\nvec+\displacementVecRel{\nvec'}{t}-\xvec)},
\end{equation}
as the sum solely implements the translational invariance $\ave{\density(\xvec;t)}=\ave{\density(\xvec+\siteVec_\nvec;t)}$ for \emph{any} lattice vector $\siteVec_\nvec$. With this periodicity, the Fourier transform
\begin{equation}\elabel{Mermin_density}
    \ave{\densityFourier(\RLVec;t)} = \int\dTWOint{x}\exp{-\imag\xvec\cdot\RLVec} \ave{\density(\xvec;t)}= V \ave{\exp{-\imag \RLVec\cdot\displacementVecRel{\nvec'}{t}}},
\end{equation}
indicates crystallinity according to Mermin if there is any non-zero reciprocal lattice vector $\RLVec$ for which $\ave{\densityFourier(\RLVec;t)}$ does not vanish \cite{mermin1968crystalline}. 

Equation~\eref{Mermin_density} renders $\ave{\densityFourier(\RLVec;t)}$ proportional to the characteristic function of the local displacements $\displacementVecRel{\nvec}{t}$. In the harmonic approximation used here, the characteristic function is Gaussian provided the noises are also Gaussian, whence we have that
\begin{equation}\elabel{densityFourier_result}
\frac{1}{V}\ave{\densityFourier(\RLVec;t)} =
\ave{\exp{-\imag \RLVec\cdot\displacementVecRel{\nvec'}{t}}} = 
\exp{-\half \RLVec \cdot 
\ave{\displacementVecRel{\nvec'}{t}\displacementVecRel[\transpose]{\nvec'}{t}} 
     \RLVec}\ ,
\end{equation}
which vanishes in the thermodynamic limit $V \to \infty$ for all non-vanishing $\RLVec$ if all eigenvalues of the correlation matrix 
$\ave{\displacementVecRel{\nvec'}{t}\displacementVecRel[\transpose]{\nvec'}{t}}$ diverge in the thermodynamic limit, so that there is no (eigen-) direction that allows $\RLVec\cdot\ave{\displacementVecRel{}{}\displacementVecRel[\transpose]{}{}}\RLVec$ to remain finite for any $\RLVec\ne\nullvec$.

In the more general case of non-Gaussian (active) noise, a divergent $\ave{\displacementVecRel{}{}\cdot\displacementVecRel{}{}}$ indicates that particle positions fluctuate without bound about their regular lattice positions even when allowing for a global rigid displacement. To proceed, we determine $\ave{\displacementVecRel{}{}\displacementVecRel[\transpose]{}{}}$ for the two-dimensional triangular active crystal, \Erefs{2d_SDE_main}--\eref{activeNoise_correlator_main}.

\newcommand\starredsum{\mathop{{\sum\nolimits^{\mathrlap{\star}}}}}

\subsection{Correlators}
Based on the more detailed calculation presented in \APref{correlators}, if a steady state exists, the correlation matrix behaves like
\Eref{displacementRel_corr}
repeated here,
\begin{widetext}
\begin{equation}\elabel{displacementRel_corr_main}
\lim_{t\to\infty}
\ave{\displacementVecRel{\nvec}{t}\displacementVecRel[\transpose]{\nvec}{t}}
=
\frac{1}{V}\starredsum_{\pvec}
\Bigg(
\Ematrix_\pvec^-
\Big[
\frac{\activity^2}{\eval_\pvec^- (\eval_\pvec^-+\noiseRate)}
+
\frac{\Gamma^2}{2\eval_\pvec^-}
\Big]
+
\Ematrix_\pvec^+
\Big[
\frac{\activity^2}{\eval_\pvec^+ (\eval_\pvec^++\noiseRate)}
+
\frac{\Gamma^2}{2\eval_\pvec^+}
\Big]
\Bigg)\ .
\end{equation}
\end{widetext}
where the starred sum $\starredsum_{\pvec}$ runs over $\pvec\in\Pset\setminus\{(0,0)\}$, \ie it omits $\pvec=(0,0)$, and $\Ematrix_\pvec^\pm$ are matrices formed from the eigenvectors of the dynamical matrix, \Erefs{def_evecs} and \eref{def_Ematrix}, which have corresponding eigenvalues $\eval_\pvec^\pm$ that depend on $\pvec$, $\pairPot'(\latticeConstant)/\latticeConstant$, and $\pairPot''(\latticeConstant)$, but not $\noiseRate$ or $\activity$. The existence of a steady state is equivalent to demanding non-negative eigenvalues, $\eval_\pvec^\pm\ge0$ for all $\pvec$, to be further discussed below. Assuming non-negative eigenvalues for the time being, in large $V$ the right-hand side of \Eref{displacementRel_corr_main} is logarithmically divergent. 
With or without active noise, a harmonic crystal in two dimensions therefore does not display crystalline order according to Mermin's criterion above. This notion of crystalline order focuses on the lattice being regular in the sense of the expected density $\ave{\density(\xvec;t)}$ of the particle position displaying long-ranged translational periodic order \cite{MairePlati:2024}. 

A similar result is obtained for a one-dimensional active crystal, with pair potential $\pairPot\big(|\displacement{n}{t}-\displacement{n+m_q}{t}-\latticeConstant \latticeVectorCompo_q|\big)$ acting between neighbouring particles, $q\in\{1,2\}$ and
$m_q=\latticeVectorCompo_q=(-1)^{q-1}$, \APref{one_dimension}. The steady-state variance of the relative displacement is found to be \Eref{VarianceDisplacementRel1D},
\begin{equation}\elabel{VarianceDisplacementRel1D_main}
\lim_{t\to\infty}
\ave{\displacementRel{n}{t}\displacementRel{n}{t}}
=
\frac{1}{N}\sum_{p=1}^{N-1}
\left(
\frac{\activity^2}{\dynamicalScalarFourier{p}{} (\dynamicalScalarFourier{p}{}+\noiseRate)}
+
\frac{\Gamma^2}{2\dynamicalScalarFourier{p}{}}
\right),
\end{equation}
where $\hat{D}_p=2\pairPot''(\latticeConstant)(1-\cos(2\pi p/N))$ are the Fourier modes of the equivalent dynamical interaction in one dimension. However, the variance of the displacement diverges linearly in the total particle number $N$ for the one-dimensional crystal, 
\begin{equation}\elabel{VarianceDisplacementRel1D_LargeN_Final_main}
\lim_{t\to\infty}
\ave{\displacementRel{n}{t}\displacementRel{n}{t}}
=
\frac{N}{12\pairPot''(\latticeConstant)}\left(
\frac{\activity^2}{\noiseRate}
+
\frac{\Gamma^2}{2}
\right) + \OC(N^0)\ .
\end{equation}

\bigskip

\subsection{Crystalline integrity}
A weaker notion of crystallinity probes the integrity of the lattice via the particle separation $\displacementVec{\nvec}{} - \displacementVec{\nvec+\mvec_q}{}-\latticeConstant\latticeVector_q$. A finite particle separation is required to maintain, firstly, the harmonic approximation and, secondly, the local neighbourhood of any given lattice site. In other words, if the particle separation becomes large compared to $\latticeConstant$ so that particles start slipping past each other \cite{SanoriaChelakkotNandi:2021}, then the notion of a lattice of active particles breaks down. For the variance of the particle separation in the steady state of the two-dimensional triangular lattice, we find from \APref{correlators} \Eref{distance_variance_2D}, 
\begin{widetext}\begin{equation}\elabel{2D_crystalline_integrity}
  \lim_{t\to\infty}\ave{|\displacementVec{\nvec}{t} - \displacementVec{\nvec+\mvec_q}{t}|^2} 
  =
\frac{1}{V}\starredsum_{\pvec}
2\big(1- \cos(\BriVec{\pvec} \cdot \mvec_q) \big)
\left(
\left[
\frac{\activity^2}{\eval_\pvec^- (\eval_\pvec^-+\noiseRate)}
+
\frac{\Gamma^2}{2\eval_\pvec^-}
\right]
+
\left[
\frac{\activity^2}{\eval_\pvec^+ (\eval_\pvec^++\noiseRate)}
+
\frac{\Gamma^2}{2\eval_\pvec^+}
\right]
\right).
\end{equation}
Equation \eref{2D_crystalline_integrity} remains finite in the thermodynamic limit $V\to\infty$ provided the eigenvalues $\eval_\pvec^\pm$ are non-negative, \APref{convergence}. The contrast to the behaviour of \Eref{displacementRel_corr_main} can be rationalised by comparing the behaviour in large $L$ of the integral $\int_{2\pi/\latticeLength}^{2\pi/\latticeConstant}\ddint{k}k^{-2}$, corresponding to the displacement variance, $\Eref{displacementRel_corr_main}$, to that of $\int_{2\pi/\latticeLength}^{2\pi/\latticeConstant}\ddint{k}k^{0}$, corresponding to the variance of the gradient of the displacement, \Eref{2D_crystalline_integrity}.

In comparison, the steady-state variance of the particle separation for a one-dimensional active crystal is, \APref{one_dimension},
\begin{equation}\elabel{MeanSquaredSeparation1D_defn_main}
  \lim_{t\to\infty}\ave{|\displacement{n}{t} - \displacement{n+m_q}{t}|^2}
  =
\frac{1}{N}\sum_{p=1}^{N-1}
2\big(1- \cos(\wavenumber{p}) \big)
\left(
\frac{\activity^2}{\dynamicalScalarFourier{p}{} (\dynamicalScalarFourier{p}{}+\noiseRate)}
+
\frac{\Gamma^2}{2\dynamicalScalarFourier{p}{}}
\right),
\end{equation}
independent of $q\in{1,2}$ on the left-hand side, which in large $N$ behaves like
\begin{equation}\elabel{distance_variance_1D_main}
\lim_{N\to\infty}\lim_{t\to\infty}\ave{|\displacement{n}{t} - \displacement{n+m_q}{t}|^2}
  =
\frac{1}{\pairPot''(\latticeConstant)}
\left(
\frac{\activity^2}{\noiseRate\sqrt{1+4\frac{\pairPot''(\latticeConstant)}{\mu}}}
+
\frac{\Gamma^2}{2}
\right),
\end{equation}
clearly indicating the characteristic particle separation remains finite in the thermodynamic limit. For large persistence times $\noiseRate^{-1}$, the low modes in \Eref{MeanSquaredSeparation1D_defn_main} dominate, eventually producing a divergence, as the bracketed term on the right-hand side of the equality in \Eref{distance_variance_1D_main} becomes just $\activity^2/\noiseRate$.
\end{widetext}

The finiteness of the expected squared separation, \Erefs{2D_crystalline_integrity} and \eref{MeanSquaredSeparation1D_defn_main}, challenges the motivation to spatially average the local forces \cite{caprini2020hidden,caprini2021spatial,caprini2023entropy, caprini2023inhomogeneous}. It guarantees crystalline integrity in the sense that the local neighbourhood of any given particle is maintained, even when the squared relative 
displacement, \Erefs{displacementRel_corr_main} and \eref{VarianceDisplacementRel1D_LargeN_Final_main}, of individual sites diverges. In other words, if a steady state exists in the harmonic approximation, a triangular lattice, active or not, does not display regularity in the sense of the long-ranged translational order demanded by Mermin's criterion \cite{mermin1968crystalline,MairePlati:2024}, but the lattice itself stays intact \cite{shi2023extreme}.

\subsection{Buckling instability}
What sets the active crystal apart from its passive counterpart is that an active crystal is typically subject to external pressure, in particular that exerted by the dilute phase on a condensate droplet \cite{RednerHaganBaskaran:2013,pruessner2025field}. This leads to the lattice constant $\latticeConstant$ being less than the relaxed length $\relaxedLength$, where $\pairPot'(\relaxedLength)=0$, and thus to repulsive inter-particle forces $\pairPot'(\latticeConstant) < 0$ even when $\displacementVec{\nvec}{}=\nullvec$ for all $\nvec \in \siteSet$. In the present setup, where the lattice is subject to periodic boundary conditions, $\pairPot'(\latticeConstant)/\latticeConstant<0$ amounts to the lattice being squeezed into a space smaller than $\NOne\relaxedLength\times\NTwo\relaxedLength$. Contrary to the two-dimensional case, the dependence on $\pairPot'(\latticeConstant)$ cancels in one dimension.

It turns out that the sign of the eigenvalues $\eval_\pvec^\pm$, and thus the stability of the harmonic approximation, is determined by $\pairPot'(\latticeConstant)/\latticeConstant$ compared to $\pairPot''(\latticeConstant)$. There is no physical reason for $\pairPot'(\latticeConstant)/\latticeConstant$ to be bounded from above or below and so it can be arbitrarily negative in the presence of external pressure.  We find in \APref{buckling_transition} that, for 
\begin{equation}\elabel{stability_criterion}
    \frac{\pairPot'(\latticeConstant)}{\latticeConstant}<-\frac{1}{3}\pairPot''(\latticeConstant) \ ,
\end{equation}
the sign of $\eval_\pvec^\pm$ is negative for some $\pvec\ne\nullvec$, so that the harmonic approximation breaks down as it no longer admits a steady state consistent with small relative displacements. If the pair potential is Hookian, $\pairPot(z)=\springConstant(z-\relaxedLength)^2/2$, \Eref{harmonic_pot}, the harmonic approximation breaks down if $\latticeConstant<(3/4)\relaxedLength$, \Eref{stability_criterion}. 

The underlying physics differs from that of the square lattice, which immediately buckles under any small external pressure, $\pairPot'(\latticeConstant)<0$, \APref{buckling_transition}. Figure \fref{TriangularLatticeBucklingMechanics} shows three sites of the triangular lattice and the effect of the surrounding external pressure. Given the periodic boundary conditions of our system, we consider the pressure to act equally on each particle. As the particle in the lower row moves by $x$ to the side, and with it the entire row in unison, the blue spring relaxes, while the red spring compresses, \ie spring lengths become $\ell_\pm=\latticeConstant\pm x/2+\OC(x^2)$ resulting in corresponding forces $\Fvec_\pm=-\pairPot'-(\ell_\pm-\latticeConstant)\pairPot''$. In principle, this leads to a restoring force. However, as the geometry changes, the force along the direction of movement increases for the blue spring and decreases for the red spring, \ie $\fvec_\pm=\Fvec_\pm (\latticeConstant/2\pm x)/\ell_\pm$. The net force $\fvec_+-\fvec_-=-x(\pairPot''+3\pairPot'/\latticeConstant)/2$ depends on the pre-tension $\pairPot'$ of the springs and thus on the external pressure. When $\pairPot''+3\pairPot'/\latticeConstant$ is negative (\ie when \Eref{stability_criterion} is satisfied) the force is no longer restoring and, to this order, the row of particles shifts indefinitely.

Beyond the harmonic approximation, higher orders in the expansion of the potential, \APref{triangular_active_crystal} and \APref{higher_order_terms}, take over for sufficiently large displacements --- even in the Hookian case, where the potential is still a function of the distance and thus not fully captured by an expression bilinear in $\displacementVec{\nvec}{}$. For the use of the harmonic approximation in describing experimental phenomena such as motility-induced phase separation 
\cite{RednerHaganBaskaran:2013,cates2015motility},
it is therefore central to determine whether the pressure that builds in the dense phase exceeds the threshold in \Eref{stability_criterion}. If it does, then the clusters are not described by the harmonic approximation.

\begin{figure}
\includegraphics[width=\columnwidth, trim = 1.5cm 14cm 3.2cm 0.6cm, clip]{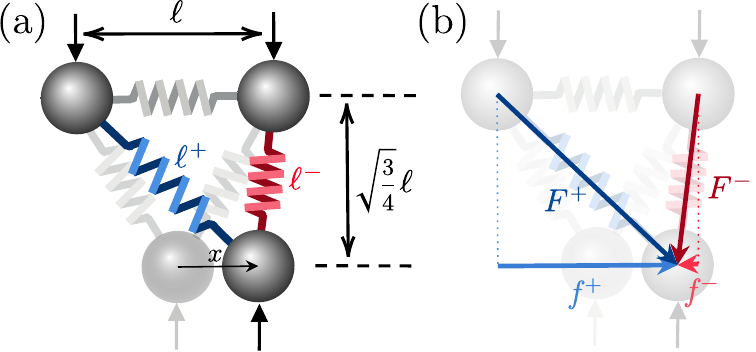}
\caption{
\justifying \flabel{TriangularLatticeBucklingMechanics}
(a) Three particles from a triangular lattice and the response of the lower one to external pressure (thick arrows). The grey springs are initially all equally compressed. As a result of the displacement, the blue spring relaxes and the red spring gets more compressed. (b) At the same time, the projection $\fvec_\pm$ of the forces $\Fvec_\pm$ along the axis of motion changes. 
}
\end{figure}

\subsection{Internal energy}
We now turn our attention to more global features of the crystal, namely internal energy and entropy production in the steady state, provided a steady state exists. Within the harmonic approximation, the energy
\begin{equation}
    \internalEnergy=\half\sum_\nvec\sum_{q=1}^Q\pairPot\big(|\displacementVec{\nvec}{t}-\displacementVec{\nvec+\mvec_q}{t}-\latticeConstant \latticeVector_q|\big)
\end{equation}
is readily expressed in terms of the dynamical matrix, \APref{energy}, whence one obtains 
\begin{multline}\elabel{expected_energy_main}
\lim_{t\to\infty}\ave{\internalEnergyHarmonic(t)} = 
3V\pairPot(\latticeConstant)+
\quarter \Gamma^2 d (V-1)\\
+
\half \activity^2 \sum_\pvec^*
\left(
\frac{1}{\eval_\pvec^+ + \noiseRate}
+
\frac{1}{\eval_\pvec^- + \noiseRate}
\right).
\end{multline}
The first term on the right-hand side is a trivial offset by $V(Q/2)\pairPot(\latticeConstant)$. The second term is solely due to the thermal noise, with $\Gamma^2$, \Eref{def_thermalNoise_main}, playing the role of (twice) a diffusion constant or (twice) the thermal energy $k_b T$. Hence, $\Gamma^2 d (V-1)/4$ apportions an energy of $k_b T / 2$ to each of the $d(V-1)$ degrees of freedom in $d$ dimensions, with the total number of modes $V$ being reduced by one due to the divergent zero-mode not contributing. The term involving a summation over modes $\pvec$ is solely due to the activity $\activity$, with the sum generally containing $d=2$ similar terms involving the $d$ eigenvalues of the $d\times d$ dynamical matrix. If the interaction potential is very soft, so that both $\pairPot''$ and $\pairPot'/\latticeConstant$ are small compared to $\noiseRate$, then the summands in \Eref{expected_energy_main} depend only weakly on $\pvec$ via $\eval_\pvec$, \Eref{eval_formula}. In this case, the sum can be approximated by $\activity^2 d(V-1)/(2\noiseRate)$.

If, on the other hand, the pair potential is stiff compared to the characteristic frequency $\noiseRate$, then the contributions to the overall energy of the active crystal from different modes become more visibly dependent on $\pvec$, signifying the \emph{breakdown of equipartition}, with both very small $\pvec$ and very large $\pvec$ making the most significant contributions to the energy as these correspond to small $\eval_\pvec$.

\newcommand{\specificEnergy}{\epsilon}
\newcommand{\lambdaTilde}{\tilde{\eval}}
From \Eref{expected_energy_main}, the energy per particle in the thermodynamic limit can be expressed in terms of a Riemann sum,
\begin{equation}\elabel{specific_energy}
    \begin{split}
    \ave{\specificEnergy} &=
    \lim_{V\to\infty}\lim_{t\to\infty}\frac{\ave{\internalEnergyHarmonic(t)}}{V} \\
    &= 3 \pairPot(\latticeConstant) + 
    \quarter \Gamma^2 d\\
    &~+\frac{\activity^2}{2\noiseRate} \int_{-\pi}^\pi\frac{\dTWOint{p}}{(2\pi)^2} 
    \left(
    \frac{1}{1+\noiseRate^{-1}\lambdaTilde^+(\kvec)}
+
    \frac{1}{1+\noiseRate^{-1}\lambdaTilde^-(\kvec)}
    \right) 
    \end{split}
\end{equation}
with $\lambdaTilde^\pm(\kvec)=\eval^\pm_\pvec$, \Eref{eval_formula}, neatly expressible in terms of $\kvec$, as $p^{1,2}=2\pi\NOneTwo k^{1,2}$.
Equation~\eref{specific_energy} immediately generalises to higher dimensions and different lattice geometries, as the first term is proportional to the number of nearest neighbours, $3\pairPot=(Q/2)\pairPot$, the second term is due to equilibrium equipartition, and the last term is the integral over $\kvec=[-\pi,\pi]^d$ of the sum over $1/(1+\eval/\noiseRate)$ using the $d$ eigenvalues $\eval$ of the dynamical matrix. The value of the integral thus depends on the lattice, as well as the two dimensionless quantities $\pairPot''(\latticeConstant)/\noiseRate>0$ and $\pairPot'(\latticeConstant)/(\latticeConstant\pairPot''(\latticeConstant))$. Figure~\fref{energy_integral} shows the value of the integral as a function of these two parameters. Even when some eigenvalues become negative, so that there is no steady state, the energy per particle in the harmonic approximation is finite.

\begin{figure}
    \centering
    \begin{tikzpicture}
            \node at (0,0) {\includegraphics[width=1.1\linewidth]{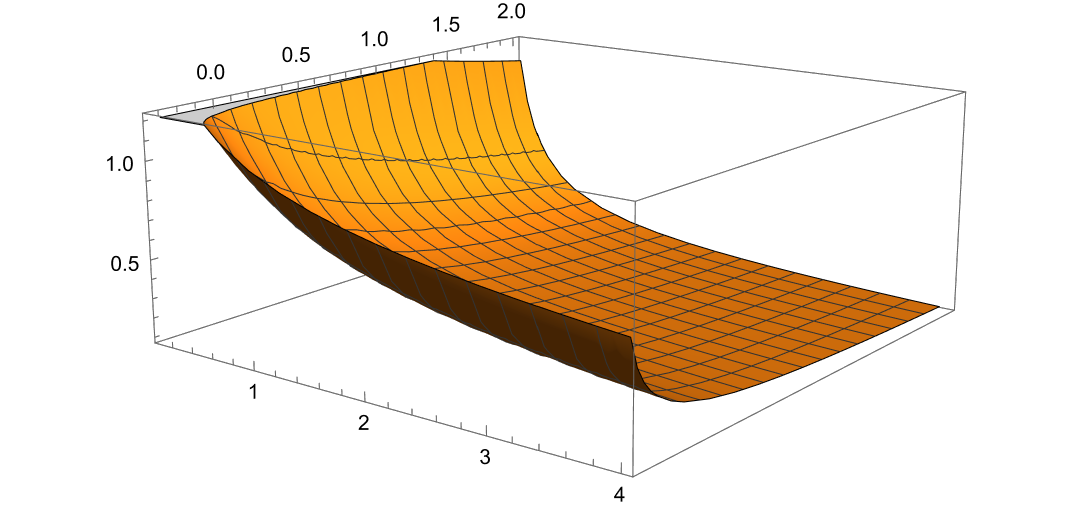}};
            \node[rotate=-10] at (-1.4,-1.9) {$\pairPot''(\latticeConstant)/\noiseRate$};
            \node[rotate=10] at (-1.8,2.3) {$\pairPot'(\latticeConstant)/(\latticeConstant\pairPot''(\latticeConstant))$};
    \end{tikzpicture}
    \caption{\justifying
    Contribution 
    $(2\pi)^{-2}\int_{-\pi}^\pi \dTWOint{p} [(1+\noiseRate^{-1}\lambdaTilde^+(\kvec))^{-1}+(1+\noiseRate^{-1}\lambdaTilde^-(\kvec))^{-1}]$ as a function of $\pairPot''(\latticeConstant)/\noiseRate\in[0.1,4]$ and $\pairPot'(\latticeConstant)/(\latticeConstant\pairPot''(\latticeConstant)\in[-0.333,2]$    
    to the energy per particle, \Eref{specific_energy}. 
    }
    \flabel{energy_integral}
\end{figure}

Unlike the two-dimensional case above, the energy of the one-dimensional crystal can be readily expressed in terms of the mean squared separation, \Eref{MeanSquaredSeparation1D_defn_main}, explicitly
\begin{multline}\elabel{EnergyHarmonic1D_main}
\ave{\internalEnergyHarmonic(t)} = N\pairPot(\latticeConstant)\\
+
\half \sum_{n=1}^{N}\sum_{q=1}^{2}
\half \pairPot''(\latticeConstant)\ave{|\displacement{n}{t} - \displacement{n+m_q}{t}|^2} \ .
\end{multline}
Using \Eref{distance_variance_1D_main}, the steady-state energy per particle in one dimension in the thermodynamic limit is
\begin{equation}\elabel{EnergyHarmonic1D_LargeN_main}
    \begin{split}
\ave{\specificEnergy} &=
\lim_{N\to\infty}\lim_{t\to\infty}\frac{\ave{\internalEnergyHarmonic(t)}}{N}\\
&= \pairPot(\latticeConstant)
+
\half \left(
\frac{\activity^2}{\noiseRate\sqrt{1+4\frac{\pairPot''(\latticeConstant)}{\mu}}}
+
\frac{\Gamma^2}{2}
\right)\ ,
    \end{split}
\end{equation}
in line with \Eref{specific_energy}.
Ignoring the trivial offset $\propto\pairPot(\latticeConstant)$, \Erefs{expected_energy_main} and \eref{EnergyHarmonic1D_LargeN_main} show the energy of a crystal depends on the details of the pair interaction, \ie $\pairPot''(\latticeConstant)$ and $\pairPot'(\latticeConstant)/\latticeConstant$, only if the constituents are active, namely via the eigenvalues $\eval_\pvec^\pm$. Without activity, the energy per particle is assymptotically $(Q/2)\pairPot(\latticeConstant)+d\Gamma^2/4$.

\subsection{Entropy production}
We finally determine the entropy production of the crystal using a framework based on field theory \cite{pruessner2025field}, 
in particular \Eref{EPR_instantaneous} in \APref{Entropy}, which has two contributions: one from the effective velocity squared over the thermal diffusion constant, and one from the second derivative of the pair potential, which is readily expressed in terms of the trace of the dynamical matrix. After rewriting in terms of Fourier modes, we arrive at \Eref{entropyProduction_final} and thus at
\begin{equation}\elabel{EPR_derivation_step5_main}
\ave{\entropyProductionDensity_\nvec} = 
\frac{2\activity^2}{V\Gamma^2}\sum_{\pvec\in\Pset}
\left(
\frac{\noiseRate}{\eval_\pvec^+ + \noiseRate}
+
\frac{\noiseRate}{\eval_\pvec^-+ \noiseRate}
\right)
\end{equation}
for the entropy production per particle. 
Similarly, the entropy production per particle for the one-dimensional crystal is 
\begin{equation}\elabel{Entropy1Dfinal_main}
\ave{\entropyProductionDensity_n} = 
\frac{2\activity^2}{N\Gamma^2}\sum_{p=0}^{N-1}
\frac{\noiseRate}{\dynamicalScalarFourier{p}{}+\noiseRate}
\end{equation}
which follows the same pattern as \Eref{EPR_derivation_step5_main}, which can be generalised to different lattices in different dimensions just like the harmonic approximation of the internal energy, \Eref{expected_energy_main}. In fact, in the harmonic approximation and the thermodynamic limit, generically
\begin{equation}\elabel{generic_eps_sigma}
    \ave{\entropyProductionDensity_\nvec} = 
\frac{4\noiseRate}{\Gamma^2}\Big(\ave{\specificEnergy}-\frac{Q}{2}\pairPot(\latticeConstant)
- \quarter \Gamma^2 d
\Big)
\ ,
\end{equation}
where the final term $4\noiseRate\Gamma^2d/(4\Gamma^2)=d\noiseRate$ deducts the purely thermal contribution to the internal energy for every degree of freedom.

Equations~\eref{EPR_derivation_step5_main} and \eref{Entropy1Dfinal_main} incorporate the entropy production (per particle) due to the rigid translation $\displacementVecBar{t}$ of the lattice as a whole, given by $2d\activity^2/(V\Gamma^2)$, through the $\pvec=\nullvec$ term in the sum, \Erefs{def_ubar}, since $\eval_{\pvec=\nullvec}^\pm=\dynamicalScalarFourier{p=0}{}=0$. Indeed, the lattice as a whole moves like an active Brownian particle, $\ave{\displacementVecBar{t}\cdot\displacementVecBar{t}}= 2d(\Gamma^2/2+\activity^2/\noiseRate)t/V$ for $t\gg1/\noiseRate$, \Eref{var_rigid_mode}, with diffusion constant 
$\Gamma^2/(2V)$ 
and squared self-propulsion speed 
$d\activity^2/V$ \cite{zhang2024field}.

According to \Eref{EPR_derivation_step5_main}, the entropy production vanishes if $\mu$ vanishes, \ie the particles are so persistent that they eventually all get stuck. If $\mu$ diverges, then the sum in \Eref{EPR_derivation_step5_main} converges to $Vd$ and the entropy production per particle becomes $\ave{\entropyProductionDensity_\nvec} = 2d\activity^2/\Gamma^2$, as the particles never get stuck, because the (known) self-propulsion direction constantly changes. In that case, each of the $d$ degrees of freedom generate the entropy of a free self-propelled particle \cite{cocconi2020entropy} with diffusion constant $\Gamma^2/2$.

The entropy production rate, \Eref{EPR_derivation_step5_main}, can also be read as the average over $2V$ independent \emph{active modes}, each behaving like a particle in a harmonic potential \cite{garcia2021run,frydel2023entropy} with stiffness $\eval^\pm_\pvec$, \APref{Entropy}. Given these collective modes are non-interacting, just like the collective modes in a passive harmonic crystal, they appear to offer a meaningful alternative perspective on the active crystal. If the adjacency itself is changing \cite{Strandburg:1988}, such as in the Vicsek Model \cite{VicsekETAL:1995}, such modes may be thought as being themselves subject to change, as if the stiffness of the potential changes as a function of time.

\section{Discussion and conclusion}\seclabel{Discussion}
We have studied two-dimensional, and for reasons of comparison, one-dimensional active crystals in the harmonic approximation, which goes beyond assuming pair potentials bilinear in every component of $\displacementVec{\nvec}{}$ \cite{caprini2020active,caprini2021spatial,caprini2023inhomogeneous}. Such assumptions are unphysical, as they ignore that the potentials are functions of (singular) distances.

In two dimensions, particles are thought to be arranged into a regular triangular lattice subject to thermal and active noise. Applying periodic boundary conditions, the lattice can be stretched or squeezed into a space so small that particles exert repulsive forces on each other. 

We find the harmonic approximation breaks down for a triangular lattice when the particles are pushed together too hard, \Eref{stability_criterion}, \ie a triangular lattice sustains some external pressure in the harmonic approximation, but only up to a threshold when a buckling transition occurs. Beyond this threshold, rows of particles start slipping past each other \cite{SanoriaChelakkotNandi:2021,shi2023extreme}, eventually breaking the assumption of the fixed adjacency reflecting the fixed set of nearest neighbours \cite{Strandburg:1988}. A steady state with fixed adjacency exists only at low-enough repulsive forces, \ie before the threshold. A similar buckling occurs for any finite amount of pressure in a square lattice, but does not occur in one dimension.

In the steady state, the one- and two-dimensional active crystals in the thermodynamic limit display as little long-range order, according to Mermin's criterion \cite{mermin1968crystalline}, as their passive counterparts, \Erefs{densityFourier_result} and \eref{displacementRel_corr}. 
However, provided the external pressure is low enough to allow for a steady state, the harmonic approximation is meaningful for the two-dimensional triangular lattice, as the variance of inter-particle distances do not grow without bound in the thermodynamic limit, \Eref{2D_crystalline_integrity}. In other words, particle density does not display crystalline periodicity on the large scale, \Eref{displacementRel_corr}, as the lattice wobbles and sways \cite{SanoriaChelakkotNandi:2021,shi2023extreme}, but the local adjacency and thus the overall structure is maintained, \Eref{2D_crystalline_integrity}.

The internal energy in the steady state reveals that equipartition breaks down in the presence of activity. While the thermal noise distributes the same amount of energy to each excitable mode, the energy ``stored'' in each mode due to the activity is dependent on the wave vector, \Eref{expected_energy_main}. 
Although this expression is unwieldy, in the thermodynamic limit it can be written as an integral that is readily evaluated numerically, \Fref{energy_integral}. 

The entropy production of the active crystal turns out to be generically that of $dV$ non-interacting ``active modes'', behaving like particles in a harmonic potential, \Erefs{EPR_derivation_step5_main} and \eref{Entropy1Dfinal_main}. In the harmonic approximation, the key contribution to the entropy production is identical to that of the internal energy, so that the two are generically linked, \Eref{generic_eps_sigma}. More concretely, rewriting \Eref{generic_eps_sigma} as 
\begin{equation}
\ave{\specificEnergy} = \frac{Q}{2}\pairPot(\latticeConstant) + \frac{\Gamma^2}{4}\left(d+\frac{\ave{\entropyProductionDensity_\nvec}}{\noiseRate}\right)
\end{equation}
suggests the relative entropy production rate $\ave{\entropyProductionDensity_\nvec}/\noiseRate$ plays the same role as any of the $d$ degrees of freedom of $\displacementVec{\nvec}{}$.

Within the harmonic approximation, the results above are exact. Evaluating the sums over functions of eigenvalues is a matter of numerics. However, the present work also reveals a significant limitation of the harmonic approximation, which breaks down as soon as \Eref{stability_criterion} holds. Whether and how this happens is to a large extent beyond the scope of the present work. The instability is not a consequence of, or is aggravated by, the activity. In other words, the instability is not triggered from within since \Eref{2D_crystalline_integrity} indicates finite particle separation, even in the presence of activity, and the condition \Eref{stability_criterion} is not a consequence of activity. 
However, condition \Eref{stability_criterion} might be triggered by the pressure of active forces, \eg when a condensate of particles is hemmed in by the dense liquid-like layer of particles, which arrive at the cluster with their self-propulsion pointing inwards, towards the bulk.

Our approach to the harmonic approximation can be contrasted with the literature. Firstly, there is some confusion about what amounts to a harmonic approximation. Assuming a harmonic pair potential $\pairPot=\springConstant|\displacementVec{\nvec}{}-\displacementVec{\nvec'}{}|^2/2$ \cite{caprini2020active,caprini2021spatial,caprini2023inhomogeneous} for nearest neighbours $\nvec,\nvec'$ results in unphysical forces minimising $\displacementVec{\nvec}{}-\displacementVec{\nvec'}{}$, rather than acting along the connecting line between particles. Such an approach does not distinguish displacements along the connecting vector between particles from those orthogonal to it. In contrast, in the present work, we harmonically approximate a potential that is a function of the particle separation, say the Hookian $\pairPot=\springConstant(|\displacementVec{\nvec}{}-\displacementVec{\nvec'}{}-\latticeConstant\latticeVector_q|-\relaxedLength)^2/2$. Such a potential allows the relaxed length to deviate from the lattice spacing and results in forces along the connecting line between particles. Even when $\latticeConstant=\relaxedLength$ it does not reduce to the aforementioned unphysical potential. 

Secondly, it seems that in a mean-field-inspired approach $\sum_{\nvec'}
\dynamicalMatrix{\nvec}{\nvec'} \displacementVec{\nvec'}{}$ is sometimes replaced by an average over $Q$ nearest neighbours, 
$Q^{-1} \sum_q
\dynamicalMatrix{\nvec}{\nvec+\mvec_q} \sum_{\nvec'} \displacementVec{\nvec'}{}$ \cite{caprini2020spontaneous,caprini2020active,caprini2021spatial,MairePlati:2024}. Both approximations render the (effective) dynamical matrix $\dynamicalMatrix{\nvec}{\nvec'}$ diagonal, substantially simplifying the calculations, but also obscuring the conditions for stability and the exact results, in particular for the internal energy and entropy production.

When the harmonic approximation breaks down, rather than the catastrophe suggested by the divergence of the integrals in the correlation functions, \Eref{def_WXi}, higher orders in the expansion, as well as higher orders in the potential, stabilise the lattice. If particles do not change their adjacency, the effect of those higher orders corresponds to that of the quartic terms in $\varphi^4$-theory, here a $(\nabla\phi)^4$ term in the field theory of the crumpling membrane \cite{PaczuskiKardarNelson:1988}, \APref{higher_order_modulus}.

Clusters as they appear in motility-induced phase separation, however, rearrange, \ie they have a changing adjacency. To describe the resulting phenomena requires a framework beyond the present one, which is limited by having displacements $\displacementVec{\nvec}{t}$ assigned to a fixed regular lattice $\nvec \in \siteSet$.

\begin{acknowledgments}
The authors thank 
Thibault Bertrand,
Callum Britton,
Jacob Knight 
and
Emir Sezik 
for useful discussions.
C.R.\ acknowledges support from the Engineering and Physical Sciences Research Council (Grant No. 2478322).
\end{acknowledgments}

\bibliography{references}
\onecolumngrid
\appendix

\section{Harmonic approximation of the triangular active crystal}
\seclabel{triangular_active_crystal}
\begin{figure}
    \centering
    \includegraphics[width=0.6\columnwidth, trim = 6.5cm 10cm 5cm 1.5cm, clip]{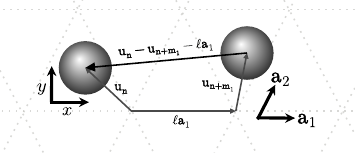}
    \caption{
    \justifying
    Illustration of the distance vector $\displacementVec{\nvec}{t}-\displacementVec{\nvec+\mvec_q}{t}-\latticeConstant\latticeVector_q$ between two particles that have been displaced away from their equilibrium lattice positions.}
    \flabel{VectorDisplacement}
\end{figure}

The notation in the following is essentially that of Ashcroft and Mermin's \textit{Solid State Physics} \cite[][Ch.~22, Eq.~(22.45)]{ashcroft1976solid}. All vector-valued variables will be bold, as are vector-valued indices on the lattice or in Fourier space. Matrices are uppercase bold sans-serif symbols. We aspire to be stricter about parameter dependencies as well as the use of lower-case and upper-case bold characters. For example, Ashcroft and Mermin use $\Rvec$ to indicate the position of a site on the regular lattice, when we write $\siteVec_\nvec$, and they write the dynamical matrix as a function $\gpvec{D}(\Rvec-\Rvec')$ whereas we write $\dynamicalMatrix{\nvec}{\nvec'}$. As an unfortunate compromise, we use superscripts to indicate components of vectors and matrices. We highlight when the superscripts are exponents whenever ambiguity arises.

We consider $V=\NOne\NTwo$ particles, densely packed into a triangular lattice and indexed by $\nvec=(n^1,n^2)\in\{1,2,\ldots,\NOne\}\otimes\{1,2,\ldots,\NTwo\}=\siteSet$, with periodic (Born-van Karman) boundary conditions \cite{ashcroft1976solid,vonKarmanBorn:1912,BornvonKarman:1913}, so that $(n_1,n_2)$ refers to the same particle as $(n^1+\NOne,n^2)$ and $(n^1,n^2+\NTwo)$. 
Using the two dimensionless primitive lattice vectors $\latticeVector_1 = (1,0)^\transpose$ and $\latticeVector_2 = (1/2,\sqrt{3}/2)^\transpose$ and the dimensionful lattice spacing $\latticeConstant$, the particle indices describe a physical lattice via $\siteVec_\nvec=(\latticeVector_1 n^1 + \latticeVector_2 n^2)\latticeConstant$. The position of every particle relative to its associated lattice site shall be given by the displacement $\displacementVec{\nvec}{t}$, \Fref{VectorDisplacement}.

Every particle $\nvec$ shall be surrounded by exactly $Q=6$ nearest neighbours, indexed $\nvec'=\nvec+\mvec_q$ for $q=1,2,\ldots,6$, where $\{\mvec_1,\ldots,\mvec_6\}=\{(1,0)^\transpose,(0,1)^\transpose,(-1,1)^\transpose,(-1,0)^\transpose,(0,-1)^\transpose,(1,-1)^\transpose\}$, chosen so that $\mvec_q=-\mvec_{q+3}$ for $q=1,2,3$. The indices translate to displacements via the lattice vectors 
\begin{equation}\elabel{latticeVector_from_m}
\latticeVector_q = m_q^1 \latticeVector_1 + m_q^2 \latticeVector_2 = \latticeMatrix \mvec_q\ ,
\end{equation}
where $\mvec_q=(m_q^1,m_q^2)^\transpose$ and $\latticeMatrix=(\latticeVector_1,\latticeVector_2)$ is the matrix composed of the column vectors $\latticeVector_1$ and $\latticeVector_2$. With these definitions $\siteVec_{\nvec+\mvec_q} = \siteVec_\nvec + \latticeConstant\latticeVector_q$.

The neighbourhood of every particle is thought to be constant, \ie particles are linked to each other permanently. This does not mean, however, that particle displacements remain finite, as the lattice as a whole moves by the net displacement
\begin{equation}\elabel{def_ubar}
\displacementVecBar{t} = \frac{1}{V}\sum_{\nvec} \displacementVec{\nvec}{t}
\ ,
\end{equation}
yet self-consistency breaks down if the \emph{relative} displacements of neighbouring particles were to grow without bound, and the \emph{nearest} neighbours change, spoiling the fixed adjacency.

Every nearest neighbouring pair $\nvec$ and $\nvec+\mvec_q$ of particles shall be subject to a pair potential $\pairPot(|\displacementVec{\nvec}{}-\displacementVec{\nvec+\mvec_q}{}-\latticeConstant\latticeVector_q|)$, which is a function of their absolute distance,
\begin{equation}
\Big|
\big(\displacementVec{\nvec}{t}+\siteVec_{\nvec}\big)
-
\big(\displacementVec{\nvec+\mvec_q}{t}+\siteVec_{\nvec+\mvec_q}\big)
\Big|
=
|\displacementVec{\nvec}{t}-\displacementVec{\nvec+\mvec_q}{t}-\latticeConstant\latticeVector_q|
\ .
\end{equation}
This modulus operation renders the lattice dynamics non-linear, even when the potential $\pairPot$ is harmonic. The pair potential is a function of a scalar and we will use dashes in the form $\pairPot'$ to indicate its derivatives such as $\pairPot'(z)=\plaind \pairPot(z)/\plaind z$.

The particles shall further be subject to thermal noise $\thermalNoiseVec_\nvec(t)$ with correlation matrix (outer product)
\begin{equation}\elabel{def_thermalNoise_App}
    \ave{\thermalNoiseVec_\nvec(t) \thermalNoiseVec^\transpose_{\nvec'}(t')}
    = \Gamma^2 \ident_2 \delta_{\nvec,\nvec'} \delta(t-t')\ ,
\end{equation}
where $\ident_2$ is the $2\times2$ identity matrix, and active noise $\activeNoiseVec_\nvec(t)$ with correlation matrix
\begin{equation}\elabel{activeNoise_correlator_App}
\ave{\activeNoiseVec_\nvec(t) \activeNoiseVec^\transpose_{\nvec'}(t')}
    = \activity^2 \ident_2 \delta_{\nvec,\nvec'} \exp{-\noiseRate|t-t'|}\ ,
\end{equation}
implementing the self-propulsion of speed $\activity$ and correlation time $1/\noiseRate$. Thermal and active noises are entirely independent, 
$\ave{\activeNoiseVec_{\nvec}(t)\thermalNoiseVec^\transpose_{\nvec'}(t')}=0$.
Both noise correlators are proportional to identities, so that different components are uncorrelated, which simplifies the calculations below. Initially located at $\siteVec_{\nvec}$, so that $\displacementVec{\nvec}{}=\nullvec$ for all $\nvec$, the particles are thus subject to the overdamped Langevin equation
\begin{equation}\elabel{initial_Langevin}
    \displacementVecDot{\nvec}{t} = 
	\thermalNoiseVec_\nvec(t) + \activeNoiseVec_\nvec(t)
	- \left.\nabla_{\displacementVec{\nvec}{}}\right|_{\displacementVec{\nvec}{}=\displacementVec{\nvec}{t}} 
	\sum_{q=1}^Q \pairPot(|\displacementVec{\nvec}{}-\displacementVec{\nvec+\mvec_q}{}-\latticeConstant\latticeVector_q|)
    \ ,
\end{equation}
where we highlight the formality that the pair potential is differentiated with respect to the vector $\displacementVec{\nvec}{}$ and evaluated at $\displacementVec{\nvec}{t}$, rather than being differentiated with respect to the function $\displacementVecDot{\nvec}{t}$.

In the following, we analyse \Eref{initial_Langevin} in the harmonic approximation, \ie expanding $\nabla\pairPot(|\ldots|)$ to leading order in $\displacementVec{\nvec}{t}$ for any $\nvec$. Firstly, we use that (higher orders in \Sref{higher_order_terms})
\begin{equation}\elabel{AbsExpansion}
    \begin{split}
        |\displacementVec{\nvec}{} - \displacementVec{\nvec + \mvec_q}{} - \latticeConstant \latticeVector_q| &= \sqrt{(\displacementVec{\nvec}{} - \displacementVec{\nvec + \mvec_q}{})^2 - 2(\displacementVec{\nvec}{} - \displacementVec{\nvec + \mvec_q}{})\cdot \latticeConstant \latticeVector_q + \latticeConstant^2 }\\
        &= \latticeConstant - (\displacementVec{\nvec}{} - \displacementVec{\nvec + \mvec_q}{})\cdot \latticeVector_q + \frac{1}{2\latticeConstant} \left(|\displacementVec{\nvec}{} - \displacementVec{\nvec + \mvec_q}{}|^2 - \big[(\displacementVec{\nvec}{} - \displacementVec{\nvec + \mvec_q}{})\cdot \latticeVector_q\big]^2\right)\\
        &~+ \mathcal{O}\left( (\displacementVec{\nvec}{} - \displacementVec{\nvec + \mvec_q}{})^3 \right)
    \end{split}
\end{equation}
to show the pair potential approximated to second order in the relative displacements is
\begin{equation}\elabel{PairPotentialApproximation}
    \begin{split}
        \pairPot(|\displacementVec{\nvec}{} - \displacementVec{\nvec + \mvec_q}{} - \latticeConstant \latticeVector_q|) &= \pairPot(\latticeConstant) \\
        &~- (\displacementVec{\nvec}{} - \displacementVec{\nvec + \mvec_q}{})\cdot \latticeVector_q \pairPot'(\latticeConstant)
        + \frac{1}{2\latticeConstant} \left(|\displacementVec{\nvec}{} - \displacementVec{\nvec + \mvec_q}{}|^2 - \big[(\displacementVec{\nvec}{} - \displacementVec{\nvec + \mvec_q}{})\cdot \latticeVector_q\big]^2\right) \pairPot'(\latticeConstant)\\
        &~+ \frac{1}{2} \left((\displacementVec{\nvec}{} - \displacementVec{\nvec + \mvec_q}{})\cdot \latticeVector_q\right)^2 \pairPot''(\latticeConstant) + \mathcal{O}\left( |\displacementVec{\nvec}{} - \displacementVec{\nvec + \mvec_q}{}|^3 \right),
    \end{split}
\end{equation}
where we have allowed $\pairPot'(\latticeConstant) \neq 0$ in order to model a particle lattice under tension or pressure so that the particles are unable to relax to their pairwise force-free separations. If the pairwise interactions are thought of as being mediated by springs, then this amounts to the springs being under constant pressure or tension. \emph{In the harmonic approximation used henceforth, we ignore the cubic and higher terms in \Eref{PairPotentialApproximation}.}

As a brief interlude, we discuss three aspects of the harmonic approximation. Firstly, the presence or absence of a force $\pairPot'(\latticeConstant)$ between a particle pair when their separation is the lattice spacing $\latticeConstant$, e.g. when both particles reside at their respective $\siteVec_\nvec$. Secondly, the potential $\pairPot$ being quadratic, \ie a ``harmonic \emph{potential}''. Thirdly, $\pairPot(|\displacementVec{\nvec}{} - \displacementVec{\nvec + \mvec_q}{} - \latticeConstant \latticeVector_q|)$ being analytic and at most quadratic in the $\displacementVec{\nvec}{}$, \ie the ``harmonic \emph{approximation}''.

Firstly, when the particles are not displaced from $\siteVec_\nvec$, so that $\displacementVec{\nvec}{}=\zerovec$ for all $\nvec\in\siteSet$, there are no forces between any particle pair \emph{only} if $\pairPot'(\latticeConstant)$ vanishes, \ie $\latticeConstant$ coincides with the relaxed length $\relaxedLength$ of the pair potential. Despite this, if $\latticeConstant$ deviates from $\relaxedLength$, there are no \emph{net} forces acting on any particle if all particles are located at their respective lattice sites, $\displacementVec{\nvec}{}=\zerovec$, since all forces acting on all particles cancel. If $\pairPot$ is harmonic, then being force-free at $\latticeConstant$ implies $\pairPot(z)=(\springConstant/2)(z-\latticeConstant)^2$ with spring constant $\springConstant$.

Secondly, the pair potential in \Eref{initial_Langevin} being harmonic, \ie Hookian, and therefore quadratic in its \emph{scalar} argument, say
\begin{equation}\elabel{harmonic_pot}
    \pairPot(z)=\half\springConstant(z-\relaxedLength)^2
\end{equation}
with relaxed length $\relaxedLength$, \emph{does not render exact} the harmonic approximation we are using here, because of the inherent non-linearity of the distance $z=|\displacementVec{\nvec}{} - \displacementVec{\nvec + \mvec_q}{} - \latticeConstant \latticeVector_q|$. Equation~\eref{harmonic_pot} becomes bilinear in $\displacementVec{\nvec}{} $ only when $\relaxedLength=0$. In other words, a harmonic (quadratic, Hookian) potential, \Eref{harmonic_pot}, like that borne by ideal springs, does not amount to a harmonic approximation. Anticipating the summation of the contributions to the potential across all nearest neighbours, \Fref{FullHarmonicU1_3} shows the potential landscape of a single particle, as a function of its displacement, surrounded by its six nearest neighbours at their non-displaced positions.

\begin{figure*}[t!]
    \centering
    \begin{subfigure}[t]{0.5\textwidth}
       \centering
       \begin{tikzpicture}
           \node at (0,0) {\includegraphics[width=0.98\linewidth]{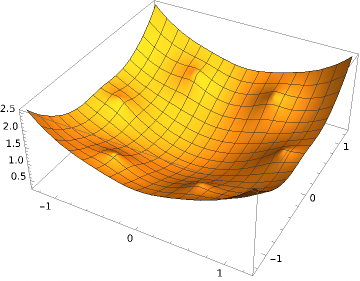}};
            \node at (-3.2,-0.5) {$\sum\pairPot$};
            \node[rotate=-18] at (0.2,-2.5) {$x$};
            \node[rotate=60] at (3.65,0) {$y$};
       \end{tikzpicture}
\caption{
\flabel{FullHarmonicU1_3} Pairwise Hookian \emph{potential} as a function of (anharmonic) distance.}
    \end{subfigure}%
    ~ 
    \begin{subfigure}[t]{0.5\textwidth}
        \centering
       \begin{tikzpicture}
           \node at (0,0) {\includegraphics[width=0.98\linewidth]{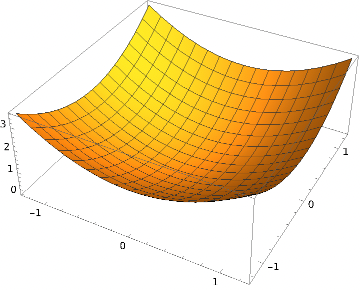}};
            \node at (-3.4,-0.5) {$\sum\pairPot$};
            \node[rotate=-18] at (0.2,-2.5) {$x$};
            \node[rotate=60] at (3.65,0) {$y$};
       \end{tikzpicture}
        \caption{\flabel{FullApproxU1_3}
        Harmonic \emph{approximation}.}
    \end{subfigure}
\caption{\flabel{U1_3}
\justifying
Potential landscape $\sum_{q=1}^6\pairPot(|\displacementVec{\nvec}{}-\latticeConstant\latticeVector_q|)$ of a single particle $\nvec$ as a function of its displacement, $\displacementVec{\nvec}{}=(x,y)^\transpose$, assuming all surrounding particles are motionless, $\displacementVec{\nvec+\mvec_q}{}=\nullvec$ for $q=1,\ldots,6$. 
Parameters: $\springConstant=1$, $\relaxedLength=1.3$ and $\latticeConstant=1$, \Eref{harmonic_pot}.
(a) Harmonic pair potentials, \Eref{harmonic_pot}, as a function of the distance $|\displacementVec{\nvec}{} - \displacementVec{\nvec + \mvec_q}{} - \latticeConstant \latticeVector_q|$. Since the potential is a function of the mutual \emph{distances}, some noticeable peaks appear at the position of neighbouring particles. (b)
Harmonic approximation, \Eref{harmonicApproximation_explicit}, using the same parameters as in subfigure (a).
}
\end{figure*}

Thirdly, the harmonic approximation that we use henceforth, \ie \Eref{PairPotentialApproximation} truncated before the cubic term, 
\begin{equation}\elabel{harmonicApproximation_explicit}
    \begin{split}
        \pairPot(|\displacementVec{\nvec}{} - \displacementVec{\nvec + \mvec_q}{} - \latticeConstant \latticeVector_q|) &\simeq \pairPot(\latticeConstant) \\
        &~- (\displacementVec{\nvec}{} - \displacementVec{\nvec + \mvec_q}{})\cdot \latticeVector_q \pairPot'(\latticeConstant)
        + \frac{1}{2\latticeConstant} \left((\displacementVec{\nvec}{} - \displacementVec{\nvec + \mvec_q}{})^2 - \big[(\displacementVec{\nvec}{} - \displacementVec{\nvec + \mvec_q}{})\cdot \latticeVector_q\big]^2\right) \pairPot'(\latticeConstant)\\
        &~+ \frac{1}{2} \left((\displacementVec{\nvec}{} - \displacementVec{\nvec + \mvec_q}{})\cdot \latticeVector_q\right)^2 \pairPot''(\latticeConstant)\ ,
    \end{split}
\end{equation}
means we assume the pair potential is analytic and at most quadratic in $\displacementVec{\nvec}{}$, but not just, say, $\springConstant(\displacementVec{\nvec}{} - \displacementVec{\nvec + \mvec_q}{})^2/2$.  As a result, there are no cusps or kinks in the potential, it is parabolic and smooth everywhere, \Fref{FullApproxU1_3}. 
The harmonic approximation captures well the potential as a function of the \emph{distance} between particles only for small $|\displacementVec{\nvec}{}|$. In this sense, it is a local harmonic-potential approximation.

Simplifying the harmonic \emph{approximation}, \Eref{harmonicApproximation_explicit}, further by assuming the harmonic \emph{potential}, \Eref{harmonic_pot}, results merely in 
\begin{equation}\elabel{harmonicApproximation_with_harmonic_pot}
    \begin{split}
        \pairPot(|\displacementVec{\nvec}{} - \displacementVec{\nvec + \mvec_q}{} - \latticeConstant \latticeVector_q|) =
        \half \springConstant\big(\latticeConstant-\relaxedLength- (\displacementVec{\nvec}{} - \displacementVec{\nvec + \mvec_q}{})\cdot\latticeVector_q \big)^2 +\half \springConstant\big(\latticeConstant-\relaxedLength\big)\frac{1}{\latticeConstant}
        \Big( 
        |\displacementVec{\nvec}{} - \displacementVec{\nvec + \mvec_q}{}|^2
        -
        \big((\displacementVec{\nvec}{} - \displacementVec{\nvec + \mvec_q}{})\cdot\latticeVector_q\big)^2
        \Big) \ ,
    \end{split}
\end{equation}
with the second term with prefactor $(\latticeConstant - \relaxedLength)$ dropping out whenever the relaxed length of the spring $\relaxedLength$ equals the lattice constant $\latticeConstant$, or when the relative displacement $\displacementVec{\nvec}{} - \displacementVec{\nvec + \mvec_q}{}$ is parallel to the lattice vector $\latticeVector_q$. As the first term in \Eref{harmonicApproximation_with_harmonic_pot} contains terms linear in $\displacementVec{\nvec}{}$, the potential is fully bilinear in $\displacementVec{\nvec}{}$ only when $\latticeConstant=\relaxedLength$ and even then, there are no $\latticeVector_q$ that render it diagonal, as $\latticeVector_q\latticeVector_q^\transpose$ cannot be proportional to $\ident_2$.

In the following, we will carefully distinguish the harmonic \emph{approximation}, \Eref{harmonicApproximation_explicit}, from the harmonic \emph{pair potential}, \Eref{harmonic_pot}, which are conceptually and mathematically independent notions.

In the harmonic approximation, \Eref{harmonicApproximation_explicit}, the gradient term in \Eref{initial_Langevin} can be written as
\begin{equation}\elabel{ForceApproximation}
    \begin{split}
        -\sum_{q=1}^Q \nabla_{\displacementVec{\nvec}{}}\pairPot(|\displacementVec{\nvec}{} - \displacementVec{\nvec + \mvec_q}{} - \latticeConstant \latticeVector_q|) &= -\frac{\pairPot'(\latticeConstant)}{\latticeConstant} \sum_{q=1}^6 \left(\displacementVec{\nvec}{} - \displacementVec{\nvec + \mvec_q}{} \right) - \left(\pairPot''(\latticeConstant) -\frac{\pairPot'(\latticeConstant)}{\latticeConstant} \right)\sum_{q=1}^6  \latticeVector_q\latticeVector_q \cdot \left(\displacementVec{\nvec}{} - \displacementVec{\nvec + \mvec_q}{} \right)  ,
    \end{split}
\end{equation}
where the linear term in \Eref{PairPotentialApproximation} vanishes in \Eref{ForceApproximation} as $\pairPot'(\latticeConstant)\sum_{q}\latticeVector_q = \boldsymbol{0}$ due to the mirror symmetry of the lattice vectors. Physically, this corresponds to the forces acting on every particle cancelling when the particles are not displaced.
In the harmonic approximation, the force on particle $\nvec$, \Eref{ForceApproximation}, can be expressed equivalently in terms of the dynamical (Hessian) matrix $\dynamicalMatrix{\nvec}{\nvec'}$ \cite{ashcroft1976solid},
\begin{equation}\elabel{ForceDmatrix}
-\sum_{q=1}^6 \nabla_{\displacementVec{\nvec}{}}
\pairPot(|\displacementVec{\nvec}{} - \displacementVec{\nvec + \mvec_q}{} - \latticeConstant \latticeVector_q|) = 
-\sum_{\nvec'}
\dynamicalMatrix{\nvec}{\nvec'} \displacementVec{\nvec'}{}\ ,
\end{equation}
where 
\begin{equation}\elabel{def_dynamicalMatrix_HarmonicAppendix}
    \dynamicalMatrix{\nvec}{\nvec'} = 
	\begin{cases}
	-\frac{\pairPot'(\latticeConstant)}{\latticeConstant} \ident_2 - 
	\left(
	\pairPot''(\latticeConstant) - \frac{\pairPot'(\latticeConstant)}{\latticeConstant}
	\right) \latticeVector_q \latticeVector^\transpose_q,
	&\quad 
    \text{for any $\nvec'$ such that there is a $q\in\{1,\ldots,Q\}$ 
    so that $\nvec'=\nvec+\mvec_q$},\\
	3 \left(
	\pairPot''(\latticeConstant) + \frac{\pairPot'(\latticeConstant)}{\latticeConstant}
	\right) \ident_2,		
	&\quad \text{for } \nvec'=\nvec,\\
	0, & \quad\text{otherwise},
    \end{cases}
\end{equation}
as deduced by comparing \Erefs{ForceApproximation} and \eref{ForceDmatrix}, and using that
\begin{equation}\elabel{sum_aqaq}
\sum_{q=1}^6\latticeVector_q \latticeVector^\transpose_q = 3\ident_2\ .
\end{equation}
Equation~\eref{ForceDmatrix} is linear in $\displacementVec{\nvec}{}$ even when the local potential is not fully bilinear, \Eref{harmonicApproximation_with_harmonic_pot}, but 
that does not render the remaining derivations trivial, as $\dynamicalMatrix{\nvec}{\nvec+\mvec_q}$ depends on $q$ and is not diagonal \cite{caprini2023inhomogeneous,caprini2020active,caprini2021spatial}. 
As a matter of convenience, we may introduce 
\begin{equation}
\curlyDynamicalMatrix_q = 
\frac{\pairPot'(\latticeConstant)}{\latticeConstant} \ident_2 + 
	\left(
	\pairPot''(\latticeConstant) - \frac{\pairPot'(\latticeConstant)}{\latticeConstant}
	\right) \latticeVector_q \latticeVector^\transpose_q\ ,
\end{equation}
which may be used in \Eref{def_dynamicalMatrix_HarmonicAppendix} to write it in terms of only three matrices, as $\curlyDynamicalMatrix_q=\curlyDynamicalMatrix_{q+3}$ for $q=1,2,3$, and $\curlyDynamicalMatrix_1+\curlyDynamicalMatrix_2+\curlyDynamicalMatrix_3=\dynamicalMatrix{\nvec}{\nvec}/2$. We further state explicitly the three relevant matrices 
\begin{align}
\latticeVector_1 \latticeVector^\transpose_1
=
\begin{pmatrix}
1 & 0\\
0 & 0
\end{pmatrix}
&\qquad&
\latticeVector_2 \latticeVector^\transpose_2
=
\begin{pmatrix}
1/4 & \sqrt{3}/4\\
\sqrt{3}/4 & 3/4
\end{pmatrix}
&\qquad&
\latticeVector_3 \latticeVector^\transpose_3
=
\begin{pmatrix}
1/4 & -\sqrt{3}/4\\
-\sqrt{3}/4 & 3/4
\end{pmatrix}~.
\end{align}

To solve the equation of motion \eref{initial_Langevin} in the harmonic approximation, we introduce a Fourier transform on the indices, to make use of the translational invariance of  $\dynamicalMatrix{\nvec}{\nvec'}$, \Eref{def_dynamicalMatrix_HarmonicAppendix},
\begin{subequations}
\begin{align}
\elabel{FourierFromDisplacement}
\displacementVecFourier{\pvec}{t} &= \sum_{\nvec} \exp{-\imag \BriVec{\pvec}\cdot\nvec} \displacementVec{\nvec}{t}\ ,\\
\elabel{displacementFromFourier}
\displacementVec{\nvec}{t} &= \frac{1}{V} \sum_{\pvec} \exp{\imag \BriVec{\pvec}\cdot\nvec} \displacementVecFourier{\pvec}{t}\ ,
\end{align}
\end{subequations}
and similarly 
\begin{subequations}\elabel{def_FourierD}
\begin{align}
\dynamicalMatrixFourier{\pvec}{\pvec'} &= \sum_{\nvec,\nvec'} \exp{-\imag (\BriVec{\pvec}\cdot\nvec+\BriVec{\pvec'}\cdot\nvec')} \dynamicalMatrix{\nvec}{\nvec'}\ ,\\
\elabel{dynMatrixFromFourier}
\dynamicalMatrix{\nvec}{\nvec'} &= \frac{1}{V^2} \sum_{\pvec,\pvec'} \exp{\imag (\BriVec{\pvec}\cdot\nvec+\BriVec{\pvec'}\cdot\nvec')} \dynamicalMatrixFourier{\pvec}{\pvec'}\ ,
\end{align}
\end{subequations}
and correspondingly for the thermal and active noise, $\thermalNoiseVec(t)$ and $\activeNoiseVec(t)$ respectively, with correlators
\begin{subequations}
\begin{align}
\ave{\thermalNoiseVecFourier_\pvec(t)\thermalNoiseVecFourier^\transpose_{\pvec'}(t')} &= \Gamma^2
V \delta_{\pvec+\pvec',\nullvec}  \delta(t-t')\ident_2 \elabel{ThermalNoiseCorrelationFourier}\\
\ave{\activeNoiseVecFourier_\pvec(t)\activeNoiseVecFourier^\transpose_{\pvec'}(t')} &= \activity^2
V \delta_{\pvec+\pvec',\nullvec}  
\exp{-\noiseRate|t-t'|} \ident_2\ . \elabel{ActiveNoiseCorrelationFourier}
\end{align}
\end{subequations}
Henceforth, $\nvec\in\siteSet$ is reserved to index a lattice site
and $\pvec\in\{0,1,\ldots,\NOne-1\}\otimes\{0,1,\ldots,\NTwo-1\}=\Pset$ to index wave vectors. With this indexing, real $\displacementVec{\nvec}{}$ result in $\displacementVecFourier{\pvec}{}^*=\displacementVecFourier{\pvec'}{}$ with $\pvec'=\Nvec-\pvec$. Beyond $\Pset$, Fourier coefficents for $\pvec$ equal those for $\pvec+(q^1 \NOne, q^2\NTwo)^\transpose$ for any $\qvec\in\Zset^2$. The wave vectors are determined from $\pvec=(p^1,p^2)^\transpose$ by $\BriVec{\pvec}=(2\pi p^1/\NOne, 2\pi p^2/\NTwo)^\transpose$. Shifting $\Pset$ to $\{-\NOne/2,\ldots,\NOne/2-1\}\otimes\{-\NTwo/2,\ldots,\NTwo-1\}$ for even $\NOne$ and $\NTwo$ might be a more convenient choice to accommodate the wave vectors in the first Brillouin zone (of indices).

Because of the translational invariance, the Fourier-transformed dynamical $2\times2$ matrix depends effectively only on a single wave-vector index $\pvec$, 
\begin{equation}\elabel{dynamicalMatrixFourier_p}
\begin{split}
\dynamicalMatrixFourier{\pvec}{\pvec'} &= V \delta_{\pvec+\pvec',\nullvec} \dynamicalMatrixFourier{\pvec}{} \\
&= V \delta_{\pvec+\pvec',\nullvec}
\sum_{q=1}^3
2[1-\cos(\BriVec{\pvec}\cdot\mvec_q)]
\left\{
\frac{\pairPot'(\latticeConstant)}{\latticeConstant} \ident_2
+
\left(
\pairPot''(\latticeConstant) - \frac{\pairPot'(\latticeConstant)}{\latticeConstant}
\right)\latticeVector_q\latticeVector^\transpose_q
\right\}\\
&= 2 V \delta_{\pvec+\pvec',\nullvec} \sum_{q=1}^3
\oneMcos{q}{\pvec} \curlyDynamicalMatrix_q \ ,
\end{split}
\end{equation}
where we have introduced, in the first equality, $\dynamicalMatrixFourier{\pvec}{}$, the Fourier transform of the dynamical matrix depending on only one wave vector index $\pvec$ and 
\begin{equation}\elabel{def_oneMcos_general}
\oneMcos{q}{\pvec}=1-\cos(\BriVec{\pvec}\cdot\mvec_q)
\end{equation}
in the last equality.
As a result, the Fourier transform of the Langevin \Eref{initial_Langevin} simplifies to
\begin{equation}\elabel{2d_SDE_Fourier}
	\displacementVecDotFourier{\pvec}{t} = \thermalNoiseVecFourier_{\pvec}(t) + \activeNoiseVecFourier_{\pvec}(t) - \dynamicalMatrixFourier{\pvec}{}\displacementVecFourier{\pvec}{t}\ .
\end{equation}
Its solution is now a matter of finding eigenvalues and eigenvectors of the $2\times2$ dynamical matrix $\dynamicalMatrixFourier{\pvec}{}$, which is straightforward in principle. Systematically characterising them, however, is more challenging. 

\subsection{Stability}\seclabel{stability}
By inspection of \Eref{2d_SDE_Fourier}, (linear) stability --- and equally the existence of a steady state --- requires the two eigenvalues $\eval_\pvec^+$ and $\eval_\pvec^-$ of $\dynamicalMatrixFourier{\pvec}{}$ to be non-negative for all $\pvec$ simultaneously. We further introduce the eigenvectors $\evec^+_\pvec$ and $\evec^-_\pvec$,
\begin{equation}
\dynamicalMatrixFourier{\pvec}{}
\evec^\pm_\pvec
=
\eval_\pvec^\pm
\evec^\pm_\pvec
    \ .
\end{equation}
The eigensystem is determined explicitly in \Sref{eigensystem} below. In the following, we discuss the preliminaries.

It is a matter of self-consistency that the eigenvalues of the dynamical matrix are non-negative, because any positive eigenvalue means there is no steady state, so that the lattice ultimately unravels into infinitely separated particles, as (some) separating forces increase with distance. Whether they do or do not increase with distance is a matter of parameters which come in two pairs. On the one hand, $\pairPot'(\latticeConstant)/\latticeConstant$ and $\pairPot''(\latticeConstant)$, and on the other the two components of $\pvec$, which determine $\BriVec{\pvec}$ and thus $\oneMcos{q}{\pvec}$. In particular 
\begin{equation}\elabel{def_oneMcos}
\oneMcos{1}{\pvec}=1-\cos(2\pi p^1/\NOne)\ ,
\qquad
\oneMcos{2}{\pvec}=1-\cos(2\pi p^2/\NTwo)\ ,
\qquad
\oneMcos{3}{\pvec}=1-\cos(2\pi p^2/\NTwo-2\pi p^1/\NOne)
\ ,
\end{equation}
which are pairwise independent.

For physical reasons, we demand $\pairPot''(\latticeConstant)>0$, because if this ``spring constant'' was negative, separating forces would ultimately increase with separation. However, it is \latin{a priori} not clear whether there are conditions on $\pairPot'(\latticeConstant)$.

A sufficient condition for non-negative eigenvalues can be derived by considering the trace and determinant of $\dynamicalMatrixFourier{\pvec}{}$. The trace produces
\begin{equation}\elabel{traceDynamicalMatrix}
\begin{split}
    \eval_\pvec^+ + \eval_\pvec^- = \Trace(\dynamicalMatrixFourier{\pvec}{}) = 2 \left(\pairPot''(\latticeConstant) + \frac{\pairPot'(\latticeConstant)}{\latticeConstant}\right)
\big(
\oneMcos{1}{\pvec}+
\oneMcos{2}{\pvec}+
\oneMcos{3}{\pvec}
\big)\ ,
\end{split}
\end{equation}
which is non-negative for all $p^1$ and $p^2$ provided $\pairPot''(\latticeConstant) + \pairPot'(\latticeConstant)/\latticeConstant \geq 0$, as $2\ge\oneMcos{q}{\pvec}\ge0$. To show non-negativity of the eigenvalues, it remains to show $\eval_\pvec^+ \eval_\pvec^- = \det{\dynamicalMatrixFourier{\pvec}{}} \geq 0$. To this end, we rewrite the dynamical matrix in the form 
\begin{equation}\elabel{dyn_in_S}
    \dynamicalMatrixFourier{\pvec}{} = \left(\pairPot''(\latticeConstant) - \frac{\pairPot'(\latticeConstant)}{\latticeConstant} \right) \SMatrixFourier{\pvec}{} + \frac{\pairPot'(\latticeConstant)}{\latticeConstant}\Trace(\SMatrixFourier{\pvec}{}) \ident_2 \ ,
\end{equation}
with 
\begin{equation}\elabel{def_Sp}
\SMatrixFourier{\pvec}{}=
\begin{pmatrix}
2\oneMcos{1}{\pvec} + (\oneMcos{2}{\pvec}+\oneMcos{3}{\pvec})/2 & \sqrt{3} (\oneMcos{2}{\pvec}-\oneMcos{3}{\pvec})/2\\
\sqrt{3} (\oneMcos{2}{\pvec}-\oneMcos{3}{\pvec})/2 & 3 (\oneMcos{2}{\pvec}+\oneMcos{3}{\pvec})/2
\end{pmatrix}
\ ,
\end{equation}
so that 
$\Trace(\SMatrixFourier{\pvec}{})=2(\oneMcos{1}{\pvec}+\oneMcos{2}{\pvec}+\oneMcos{3}{\pvec})$
and
$\det{\SMatrixFourier{\pvec}{}}=3(\oneMcos{1}{\pvec}\oneMcos{2}{\pvec}+\oneMcos{2}{\pvec}\oneMcos{3}{\pvec}+\oneMcos{3}{\pvec}\oneMcos{1}{\pvec})\ge0$ by direct evaluation and using $\oneMcos{1,2,3}{}\ge0$. Using the general formula 
$\det{\BMatrix + \CMatrix} = \det{\BMatrix} + \det{\CMatrix} + \Trace(\CMatrix^{-1}\BMatrix)\det{\CMatrix}$ for any $2\times2$ matrices $\BMatrix$ and $\CMatrix$ \cite{Marquis:2016}, and identifying $\BMatrix\propto\SMatrixFourier{\pvec}{}$ and $\CMatrix\propto\ident_2$, we obtain
\begin{equation}
\elabel{detDynamicalMatrix}
\det{\dynamicalMatrixFourier{\pvec}{}} =
\left(\pairPot''(\latticeConstant) - \frac{\pairPot'(\latticeConstant)}{\latticeConstant}\right)^2 \det{\SMatrixFourier{\pvec}{}}
+
\pairPot''(\latticeConstant)\frac{\pairPot'(\latticeConstant)}{\latticeConstant}[\Trace(\SMatrixFourier{\pvec}{})]^2 \ .
\end{equation}
Since $\det{\SMatrixFourier{\pvec}{}}\ge0$ for all $\pvec$ and given that $\pairPot''(\latticeConstant)>0$, the sign of $\det{\dynamicalMatrixFourier{\pvec}{}}$ is necessarily non-negative provided only that $\pairPot'(\latticeConstant)/\latticeConstant\ge0$. This is a stricter condition than $\pairPot''(\latticeConstant) + \pairPot'(\latticeConstant)/\latticeConstant \geq 0$ obtained for the non-negativity of the trace found above. In summary, given $\pairPot''(\latticeConstant)\ge0$, the condition $\pairPot'(\latticeConstant)/\latticeConstant\ge0$ is \emph{sufficient} to guarantee non-negativity of the eigenvalues $\eval_\pvec^\pm$ of $\dynamicalMatrix{\pvec}{}$ for all $\pvec$ and thus the stability of the triangular lattice. In the following, we analyse the eigensystem in further detail.

\subsubsection{Eigensystem}\seclabel{eigensystem}
From \Eref{def_Sp} the eigenvalues of $\SMatrixFourier{\pvec}{}$ are readily calculated via its trace and determinant, $\SMatrixFourier{\pvec}{}\evec^\pm_\pvec=\SMatrixEval^\pm_{\pvec}\evec^\pm_\pvec$,
resulting in 
\begin{equation}\elabel{def_sigma}
    \SMatrixEval^\pm_{\pvec} = (\oneMcos{1}{\pvec}+\oneMcos{2}{\pvec}+\oneMcos{3}{\pvec})
    \pm
    \sqrt{
(\oneMcos{1}{\pvec}^2+\oneMcos{2}{\pvec}^2+\oneMcos{3}{\pvec}^2)
-
(\oneMcos{1}{\pvec}\oneMcos{2}{\pvec}+\oneMcos{2}{\pvec}\oneMcos{3}{\pvec}+\oneMcos{3}{\pvec}\oneMcos{1}{\pvec})
    }
\end{equation}
and for $\pvec\ne\nullvec$
\begin{equation}\elabel{def_evecs}
    \evec^\pm_\pvec=\frac{1}{\NC_{\evec^\pm_\pvec}}
    \left(
    \frac{\sqrt{3}}{2}
    (\oneMcos{2}{\pvec}-\oneMcos{3}{\pvec}),
    \half(\oneMcos{2}{\pvec}+\oneMcos{3}{\pvec})-\oneMcos{1}{\pvec} 
    \pm 
    \sqrt{
(\oneMcos{1}{\pvec}^2+\oneMcos{2}{\pvec}^2+\oneMcos{3}{\pvec}^2)
-
(\oneMcos{1}{\pvec}\oneMcos{2}{\pvec}+\oneMcos{2}{\pvec}\oneMcos{3}{\pvec}+\oneMcos{3}{\pvec}\oneMcos{1}{\pvec})
    }
    \right)^\transpose
    \ ,
\end{equation}
which are normalised and orthogonal, given that both $\SMatrixFourier{\pvec}{}$
and $\dynamicalMatrixFourier{\pvec}{}$ are symmetric. The particular case of $\oneMcos{1}{\pvec}=0$ and $\oneMcos{2}{\pvec}=\oneMcos{3}{\pvec}$ discussed in \Sref{buckling_transition} results in $\evec^-_\pvec=\nullvec$, but is recovered from \Eref{def_evecs} by suitable limits, say $\oneMcos{2}{\pvec}-\oneMcos{3}{\pvec}=\epsilon$ and $\oneMcos{2}{\pvec}-\oneMcos{3}{\pvec}=\oneMcos{}{}$.

As $\oneMcos{q}{\pvec}=\oneMcos{q}{\pvec^*}$ for $\pvec,\pvec^*$ such that $\kvec_\pvec^*=(2\pi,2\pi)^\transpose-\kvec_\pvec$, it follows that $\SMatrixEval^\pm_{\pvec}=\SMatrixEval^\pm_{\pvec^*}$ and $\evec^\pm_\pvec=\evec^\pm_{\pvec^*}$. Similarly, from the $\oneMcos{q}{\pvec}$ being even in $\pvec$ it follows that 
$\SMatrixEval^\pm_{\pvec}=\SMatrixEval^\pm_{-\pvec}$ and $\evec^\pm_\pvec=\evec^\pm_{-\pvec}$. The normalisation $\NC_{\evec^\pm_\pvec}$ of $\evec^\pm_\pvec$ is somewhat cumbersome,
\begin{multline}\elabel{def_evec_norm}
    \NC^2_{\evec^\pm_\pvec} = 2 
    \Big(
(\oneMcos{1}{\pvec}^2+\oneMcos{2}{\pvec}^2+\oneMcos{3}{\pvec}^2)
-
(\oneMcos{1}{\pvec}\oneMcos{2}{\pvec}+\oneMcos{2}{\pvec}\oneMcos{3}{\pvec}+\oneMcos{3}{\pvec}\oneMcos{1}{\pvec})    
    \Big)\\
    \pm \big( 
    \oneMcos{3}{\pvec}+\oneMcos{2}{\pvec}
    -2\oneMcos{1}{\pvec}
    \big)
\sqrt{
(\oneMcos{1}{\pvec}^2+\oneMcos{2}{\pvec}^2+\oneMcos{3}{\pvec}^2)
-
(\oneMcos{1}{\pvec}\oneMcos{2}{\pvec}+\oneMcos{2}{\pvec}\oneMcos{3}{\pvec}+\oneMcos{3}{\pvec}\oneMcos{1}{\pvec})
    }    
\ .
\end{multline}
For $\pvec=\nullvec$, matrix $\SMatrixFourier{\pvec}{}$ vanishes, so that $\SMatrixEval_{\pvec=\nullvec}^\pm=0$, \Eref{def_sigma}, and we may choose $\evec^+_\nullvec=(1,0)^\transpose$ and $\evec^-_\nullvec=(0,1)^\transpose$.

Using the eigenvalues \Eref{def_sigma} of $\SMatrixFourier{\pvec}{}$
with \Eref{dyn_in_S}, the eigenvalues of $\dynamicalMatrixFourier{\pvec}{}$
are
\begin{equation}\elabel{eval_formula}
    \eval_\pvec^\pm =
    \left(\pairPot''(\latticeConstant) + \frac{\pairPot'(\latticeConstant)}{\latticeConstant} \right) (\oneMcos{1}{\pvec}+\oneMcos{2}{\pvec}+\oneMcos{3}{\pvec})
    \pm
    \left(\pairPot''(\latticeConstant) - \frac{\pairPot'(\latticeConstant)}{\latticeConstant} \right)
    \sqrt{
(\oneMcos{1}{\pvec}^2+\oneMcos{2}{\pvec}^2+\oneMcos{3}{\pvec}^2)
-
(\oneMcos{1}{\pvec}\oneMcos{2}{\pvec}+\oneMcos{2}{\pvec}\oneMcos{3}{\pvec}+\oneMcos{3}{\pvec}\oneMcos{1}{\pvec})
    } \ , 
\end{equation}
and its eigenvectors are identical to those of $\SMatrixFourier{\pvec}{}$, \Eref{def_evecs}. Again, from the even parity of the $\oneMcos{q}{\pvec}$, it follows that $\eval_\pvec^\pm=\eval_{-\pvec}^\pm$.

The argument of the square root may be written as 
\begin{equation}\elabel{sqrt_arg}
(\oneMcos{1}{}^2+\oneMcos{2}{}^2+\oneMcos{3}{}^2)
-
(\oneMcos{1}{}\oneMcos{2}{}+\oneMcos{2}{}\oneMcos{3}{}+\oneMcos{3}{}\oneMcos{1}{})
=
\left(
\begin{array}{l}
    \oneMcos{1}{}\\
    \oneMcos{2}{}\\
    \oneMcos{3}{}
\end{array}
\right)^\transpose
\begin{pmatrix}
1 & -1/2 & -1/2 \\
-1/2 & 1 & -1/2 \\
-1/2 & -1/2 & 1
\end{pmatrix}
\left(
\begin{array}{l}
    \oneMcos{1}{}\\
    \oneMcos{2}{}\\
    \oneMcos{3}{}
\end{array}
\right)
\end{equation}
where the symmetric matrix on the right-hand side of the equality has eigenvalues $0$, $3/2$ and $3/2$. Using the orthogonality of its eigenvectors, which span $\Rset^3$, one can then show any such product on the right-hand side of \Eref{sqrt_arg} is non-negative, provided only that $\oneMcos{q}{}\in\Rset$.

We shall now return to the question about the sign of $\eval_\pvec^\pm$ and thus that of the stability of the harmonic approximation. From above, it is clear $\pairPot''(\latticeConstant)>0$ and $\pairPot'(\latticeConstant)\ge0$ will result in $\eval_\pvec^\pm\ge0$, \ie the lattice is guaranteed to be stable for $\pairPot'(\latticeConstant)\ge0$. The physically interesting region to look for instability when $\pairPot''(\latticeConstant)>0$ is thus $\pairPot'(\latticeConstant)<0$. Given negative $\pairPot'(\latticeConstant)$, the pre-factor in front of the root in \Eref{eval_formula} is always positive, so that $\eval_\pvec^+\ge\eval_\pvec^-$. It is also clear that if $\pairPot'(\latticeConstant)$ is so negative that $\pairPot''(\latticeConstant)+\pairPot'(\latticeConstant)/\latticeConstant<0$, then $\eval_\pvec^-<0$ for all $\pvec\ne\nullvec$, \ie the lattice is guaranteed to be unstable for $\pairPot'(\latticeConstant)/\latticeConstant<-\pairPot''(\latticeConstant)$. Hence, we ask: is there a negative $\pairPot'(\latticeConstant)/\latticeConstant>-\pairPot''(\latticeConstant)$, for which at least one of $\eval_\pvec^+$ or $\eval_\pvec^-$ is negative for some $\pvec$?

If $0>\pairPot'(\latticeConstant)/\latticeConstant>-\pairPot''(\latticeConstant)$, the first term on the right-hand side of \Eref{eval_formula} is bound to be positive, as is the pre-factor in front of the square root. In which case, $\eval_\pvec^+\ge0$ and thus only $\eval_\pvec^-$ can be negative. The following discussion focuses on the question of when this occurs, \ie how negative $\pairPot'$ needs to be to produce $\eval_\pvec^-<0$ for some $\pvec$. First, we address this question by extremising $\eval_\pvec^\pm$ as expressed in \Eref{eval_formula}

\begin{table}[t!]
    \centering
{\renewcommand{\arraystretch}{1.5}%
\begin{tabular}{>{$}l<{$}|>{$}l<{$}>{$}l<{$}>{$}l<{$}|>{$}l<{$}>{$}l<{$}|>{$}l<{$}>{$}l<{$}|>{$}l<{$}>{$}l<{$}}
\kvec_\pvec^\transpose & 
\oneMcos{1}{\pvec} &
\oneMcos{2}{\pvec} &
\oneMcos{3}{\pvec} &
\sum &
\sqrt{\sdots} &
\eval_\pvec^+&
\eval_\pvec^-&
\evec_\pvec^+&
\evec_\pvec^-
\\
\hline
\hline
(0,0)              &0  &0  &0      & 0 & 0 & 0 & 0 & (0,1) \circmark & (1,0) \circmark \\
\hline
(0,\pi)            &0  &2  &2      & 4 & 2 & 6\pairPot''+2\pairPot'/\latticeConstant & 2\pairPot''+6\pairPot'/\latticeConstant & (0,1) & (1,0) \downarrow \\
(\pi,0)              &2  &0  &2      & 4 & 2 & 6\pairPot''+2\pairPot'/\latticeConstant & 2\pairPot''+6\pairPot'/\latticeConstant & -(\sin \frac{2\pi}{3},\cos \frac{2\pi}{3}) & (\sin (\frac{2\pi}{3}\!+\!\frac{\pi}{2}),\cos (\frac{2\pi}{3}\!+\!\frac{\pi}{2})) \\
(\pi,\pi)            &2  &2  &0      & 4 & 2 & 6\pairPot''+2\pairPot'/\latticeConstant & 2\pairPot''+6\pairPot'/\latticeConstant & -(\sin \frac{4\pi}{3},\cos \frac{4\pi}{3}) & -(\sin (\frac{4\pi}{3}\!+\!\frac{\pi}{2}),\cos (\frac{4\pi}{3}\!+\!\frac{\pi}{2})) \\
\hline
\big(\frac{2\pi}{3},\frac{4\pi}{3}\big)    &3/2&3/2&3/2    &9/2& 0 & \frac{9}{2}(\pairPot''+\pairPot'/\latticeConstant) & \frac{9}{2} (\pairPot''+\pairPot'/\latticeConstant) & (0,1) \circmark & (1,0) \circmark\\
\big(\frac{4\pi}{3},\frac{2\pi}{3}\big)    &3/2&3/2&3/2    &9/2& 0 & \frac{9}{2}(\pairPot''+\pairPot'/\latticeConstant) & \frac{9}{2} (\pairPot''+\pairPot'/\latticeConstant)
& (0,1) \circmark & (1,0) \circmark
\end{tabular}}
    \caption{
    \justifying
    Values of $\kvec_\pvec$ for which the two terms that make up $\eval^\pm_\pvec$ in \Eref{eval_formula} are extremal. The three columns labelled by $\oneMcos{q}{\pvec}$ show the coefficients of \Eref{def_oneMcos_general}, the column labelled $\Sigma$ shows $\oneMcos{1}{\pvec}+\oneMcos{2}{\pvec}+\oneMcos{3}{\pvec}$, the column labelled $\sqrt{\sdots}$ shows the value of the square root in \Eref{eval_formula}  (the square root of \Eref{sqrt_arg}). Apart from the eigenvalues as of \Eref{eval_formula}, the corresponding eigenvectors are shown in the right-most columns. Vectors labelled by a downward arrow $\downarrow$ are obtained by taking a suitable limit (or by direct evaluation). Those labelled by $\circmark$ are obtained directly from \Erefs{dyn_in_S} and \eref{def_Sp}, which produces $\dynamicalMatrixFourier{\pvec}{}\propto\ident_2$, so that the eigenvectors are a matter of choice. The first row shows the $0$-mode that is always present, the following three rows show the buckling instability to be discussed in \Sref{buckling_transition}, and the final two rows show an excitation that may be mistaken for being chiral with phases as shown in \Fref{not-chiral}.
    }
    \label{tab:extremal_kp}
\end{table}

Differentiating \Eref{eval_formula} with respect to $p^1$ and $p^2$, as if they were continuous, it is easy to find \emph{sufficient} conditions for the eigenvalues to be extremal by demanding that both terms in \Eref{eval_formula}, the sum over $\oneMcos{q}{\pvec}$ and the square root, vanish individually. It turns out that the square root is extremal for $\oneMcos{1}{\pvec}=\oneMcos{2}{\pvec}=\oneMcos{3}{\pvec}$ or for $\sin(2\pi p^1/\NOne) = - \sin (2\pi p^2/\NTwo)= \sin (2\pi p^2/\NTwo-2\pi p^1/\NOne)$, but that those two conditions coincide. It further turns out that sum and square root are extremal simultaneously. The extremal $\kvec_\pvec$ are listed in \Tref{extremal_kp}.

The search for negative $\eval_\pvec^-$ simplifies to the search for negative $\eval_\pvec^-\eval_\pvec^+$, when $\eval_\pvec^+>0$. In other words, we may as well search for negative 
\begin{multline}
    \eval_\pvec^-\eval_\pvec^+
    =
\left(\pairPot''(\latticeConstant)+\frac{\pairPot'(\latticeConstant)}{\latticeConstant}\right)^2
\left(
\begin{array}{l}
    \oneMcos{1}{\pvec}\\
    \oneMcos{2}{\pvec}\\
    \oneMcos{3}{\pvec}
\end{array}
\right)^\transpose
\begin{pmatrix}
1 & 1 & 1 \\
1 & 1 & 1 \\
1 & 1 & 1
\end{pmatrix}
\left(
\begin{array}{l}
    \oneMcos{1}{\pvec}\\
    \oneMcos{2}{\pvec}\\
    \oneMcos{3}{\pvec}
\end{array}
\right)\\
-
    \left(\pairPot''(\latticeConstant)-\frac{\pairPot'(\latticeConstant)}{\latticeConstant}\right)^2
\left(
\begin{array}{l}
    \oneMcos{1}{\pvec}\\
    \oneMcos{2}{\pvec}\\
    \oneMcos{3}{\pvec}
\end{array}
\right)^\transpose
\begin{pmatrix}
1 & -1/2 & -1/2 \\
-1/2 & 1 & -1/2 \\
-1/2 & -1/2 & 1
\end{pmatrix}
\left(
\begin{array}{l}
    \oneMcos{1}{\pvec}\\
    \oneMcos{2}{\pvec}\\
    \oneMcos{3}{\pvec}
\end{array}
\right)
\ .
\end{multline}
To find the \emph{onset} of negative $\eval_\pvec^-$ we look for the root of the expression above, 
$\eval_\pvec^-\eval_\pvec^+=0$, which produces
\begin{equation}\elabel{pairPot_to_eps}
    \frac
    {\left(\pairPot''(\latticeConstant)+\pairPot'(\latticeConstant)/\latticeConstant\right)^2}
    {\left(\pairPot''(\latticeConstant)-\pairPot'(\latticeConstant)/\latticeConstant\right)^2}
    =
    \frac
    {
    \left(
\begin{array}{l}
    \oneMcos{1}{\pvec}\\
    \oneMcos{2}{\pvec}\\
    \oneMcos{3}{\pvec}
\end{array}
\right)^\transpose
\begin{pmatrix}
1 & -1/2 & -1/2 \\
-1/2 & 1 & -1/2 \\
-1/2 & -1/2 & 1
\end{pmatrix}
\left(
\begin{array}{l}
    \oneMcos{1}{\pvec}\\
    \oneMcos{2}{\pvec}\\
    \oneMcos{3}{\pvec}
\end{array}
\right)
    }
    {
    \left(
\begin{array}{l}
    \oneMcos{1}{\pvec}\\
    \oneMcos{2}{\pvec}\\
    \oneMcos{3}{\pvec}
\end{array}
\right)^\transpose
\begin{pmatrix}
1 & 1 & 1 \\
1 & 1 & 1 \\
1 & 1 & 1
\end{pmatrix}
\left(
\begin{array}{l}
    \oneMcos{1}{\pvec}\\
    \oneMcos{2}{\pvec}\\
    \oneMcos{3}{\pvec}
\end{array}
\right)
    }
=-\half + \frac{3}{2}
\frac
{1}
{1+2\oneMcosFunc(\pvec)}\ ,
\end{equation}
where 
\begin{equation}\elabel{def_oneMcosFunc}
    \oneMcosFunc(\pvec) = \frac
    {\oneMcos{1}{\pvec}\oneMcos{2}{\pvec}+\oneMcos{2}{\pvec}\oneMcos{3}{\pvec}+\oneMcos{3}{\pvec}\oneMcos{1}{\pvec}}
    {\oneMcos{1}{\pvec}^2+\oneMcos{2}{\pvec}^2+\oneMcos{3}{\pvec}^2}
    \ ,
\end{equation}
which depends solely on $\pvec$ and not on any properties of $\pairPot$. As it is periodic in $\pvec=(p^1,p^2)^\transpose$, the global minimum of $\oneMcosFunc(\pvec)$ can be found by (numerical) inspection of all non-trivial $\pvec$. This produces $\oneMcosFunc(\pvec)\ge1/2$, with the equality realised for the three cases $(2\pi p^1/\NOne,2\pi p^2/\NTwo)\in\{ (0,\pi), (\pi,0), (\pi,\pi)\}$, assuming even $\NOne$ and $\NTwo$, so that, for example, $p^1=0$ and $p^2=\NTwo/2$.
Using $\oneMcosFunc(\pvec)\ge1/2$ in \Eref{pairPot_to_eps} gives, as a necessary condition for $\eval_\pvec^-\eval_\pvec^+ \leq 0$ when $\pairPot''\ge0\ge\pairPot'/\latticeConstant\ge -\pairPot''$,
\begin{equation}\elabel{instability_condition}
\pairPot'(\latticeConstant)/\latticeConstant
\le 
-\frac{1}{3}
\pairPot''(\latticeConstant) \ .
\end{equation}
To summarise the above, assuming $\pairPot''(\latticeConstant)>0$ throughout, if $\pairPot'(\latticeConstant)/\latticeConstant>0$, both $\eval_\pvec^+$ and $\eval_\pvec^-$ are bound to be strictly positive for all $\pvec\ne\nullvec$. If $\pairPot'(\latticeConstant)/\latticeConstant<-\pairPot''(\latticeConstant)$, then $\eval_\pvec^-<0$ for all $\pvec\ne\nullvec$. If $0>\pairPot'(\latticeConstant)/\latticeConstant> -\pairPot''(\latticeConstant)$, then $\eval_\pvec^+>0$ for all $\pvec\ne\nullvec$. Equation~\eref{instability_condition} is the condition for the existence of a $\pvec$ (strictly, for a $\kvec_\pvec$) with $\eval_\pvec^+\eval_\pvec^-=0$ and thus $\eval_\pvec^-=0$. By linearity of \Eref{eval_formula} in $\pairPot'(\latticeConstant)/\latticeConstant$ and by continuity, for $\pairPot'(\latticeConstant)/\latticeConstant<-\pairPot''(\latticeConstant)/3$ there are $\pvec$ (strictly $\kvec_\pvec$) such that $\eval_\pvec^-<0$, while $\pairPot'(\latticeConstant)/\latticeConstant>-\pairPot''(\latticeConstant)/3$ results in $\eval_\pvec^\pm>0$ for all $\pvec\ne\nullvec$. Condition \Eref{instability_condition} is thus the \emph{onset} of the instability. Using $\pairPot'(\latticeConstant)/\latticeConstant< -\pairPot''(\latticeConstant)/3$ in \Tref{extremal_kp}, for $\kvec_\pvec^\transpose\in\{(0,\pi),(\pi,0),(\pi,\pi)\}$, $\eval_\pvec^-$ indeed becomes negative. These give rise to the buckling transition discussed in the next section.

If the pair potential is harmonic, \Eref{harmonic_pot}, then $\pairPot''(z)=\springConstant>0$ independent of $z$, and $\pairPot'(\latticeConstant)/\latticeConstant=\springConstant(1-\relaxedLength/\latticeConstant)$. In this case, the instability is triggered when the lattice is compressed too much, \ie when the distance between particles becomes too small, $\latticeConstant < (3/4) \relaxedLength$.

\begin{figure}[t!]
    \centering
    \begin{subfigure}[t]{0.5\textwidth}
        \centering
        \includegraphics[height=1.2in]{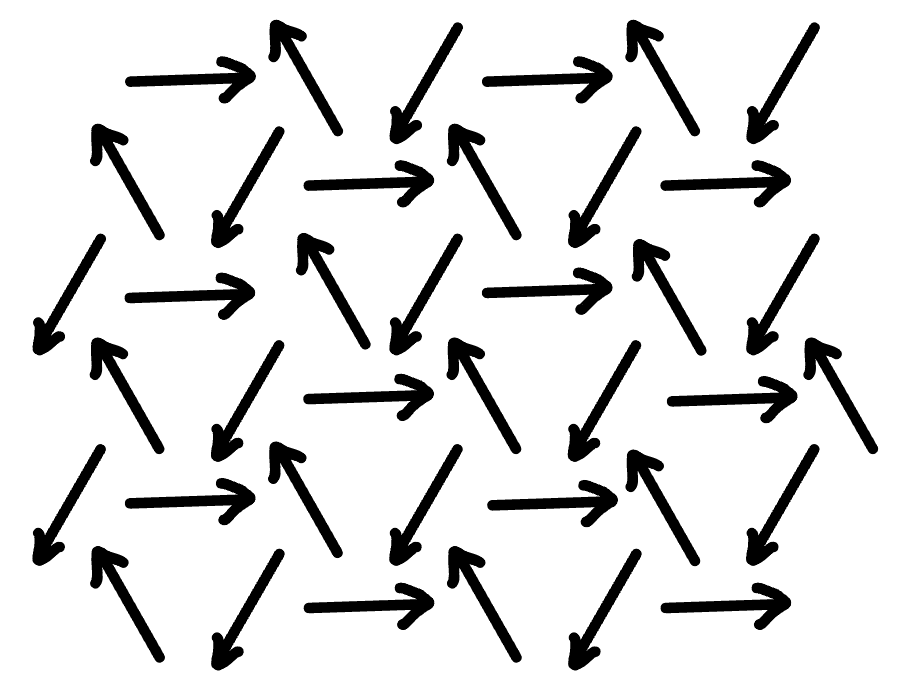}
        \caption{$\kvec_\pvec=(2\pi/3,4\pi/3)^\transpose$}
    \end{subfigure}%
    ~ 
    \begin{subfigure}[t]{0.5\textwidth}
        \centering
        \includegraphics[height=1.2in]{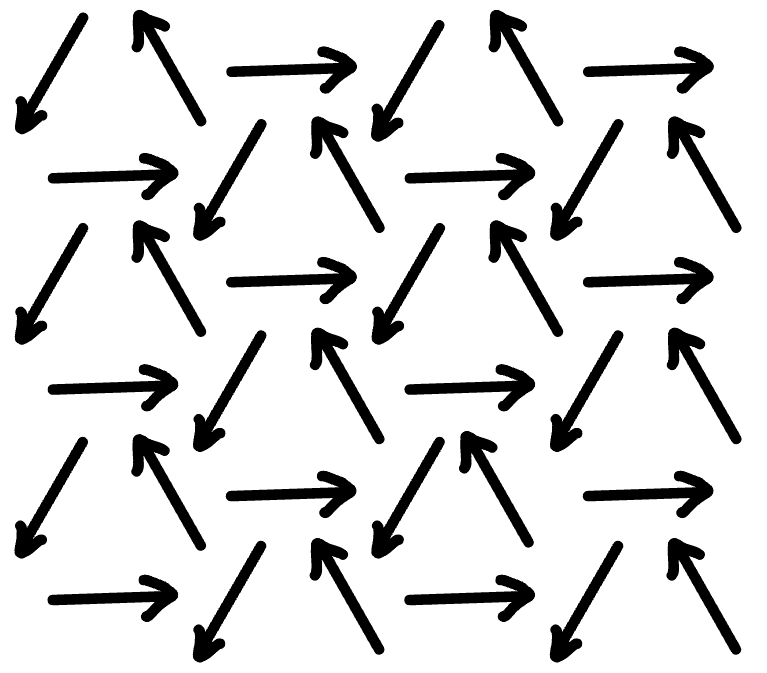}
        \caption{$\kvec_\pvec=(4\pi/3,2\pi/3)^\transpose$}
    \end{subfigure}
    \caption{
    \justifying
    Phases (sets of $\displacementVec{\nvec}{t}$, shown by the arrows) as obtained from the last two rows of \Tref{extremal_kp}, $\kvec_\pvec=(2\pi/3,4\pi/3)^\transpose$ and $\kvec_\pvec=(4\pi/3,2\pi/3)^\transpose$ respectively, using $\displacementVec{\nvec}{t}=\exp{\imag\kvec_\pvec\cdot\nvec}(1,\imag)^\transpose+\exp{\imag\kvec_{\pvec^*}\cdot\nvec}(1,-\imag)^\transpose$ with $\kvec_{\pvec^*}=(2\pi,2\pi)^\transpose-\kvec_\pvec$, shown by the arrows. The two phases shown in each subfigure look chiral, but can be transformed into each other by translation and rotation of the lattice.\flabel{not-chiral}}
\end{figure}

\subsubsection{Buckling transition}\seclabel{buckling_transition}
To understand the effect of negative eigenvalues, 
it is instructive to consider first a square lattice, which indeed produces the necessary stability condition $\pairPot'(\latticeConstant)>0$, different from the condition obtained on the triangular lattice above. The condition $\pairPot'(\latticeConstant)=0$, when the \emph{buckling transition} takes place on a square lattice, corresponds to the configuration when all pairs of particles are relaxed if they are not displaced, \ie to leading order, there is no force $\pairPot'$ between any pair of particles placed on a lattice with spacing $\latticeConstant$.

On the square lattice, as soon as $\pairPot'(\latticeConstant)<0$, \ie the particles are crammed together such that they exert forces on each other, the lattice buckles with the most-negative eigenvalue corresponding to an excitation where, say, $1-\cos(2\pi p^1/\NOne)$ vanishes and $1-\cos(2\pi p^2/\NTwo)$ is maximal, $p^1=0$ and $p^2=\NTwo/2$. The corresponding eigenvector $\evec=(1,0)^\transpose$ is orthogonal to $\BriVec{\pvec}$, resulting in a zig-zag configuration as illustrated in \Fref{SquareLatticeBuckling}. Right \emph{at} the transition, $\pairPot'(\latticeConstant)=0$, all modes in the $x$- and $y$-direction are ``massless'' at once, as illustrated for the triangular lattice in \Fref{TriangularLatticeBuckling}.

\begin{figure}
\includegraphics[height=5.5cm]{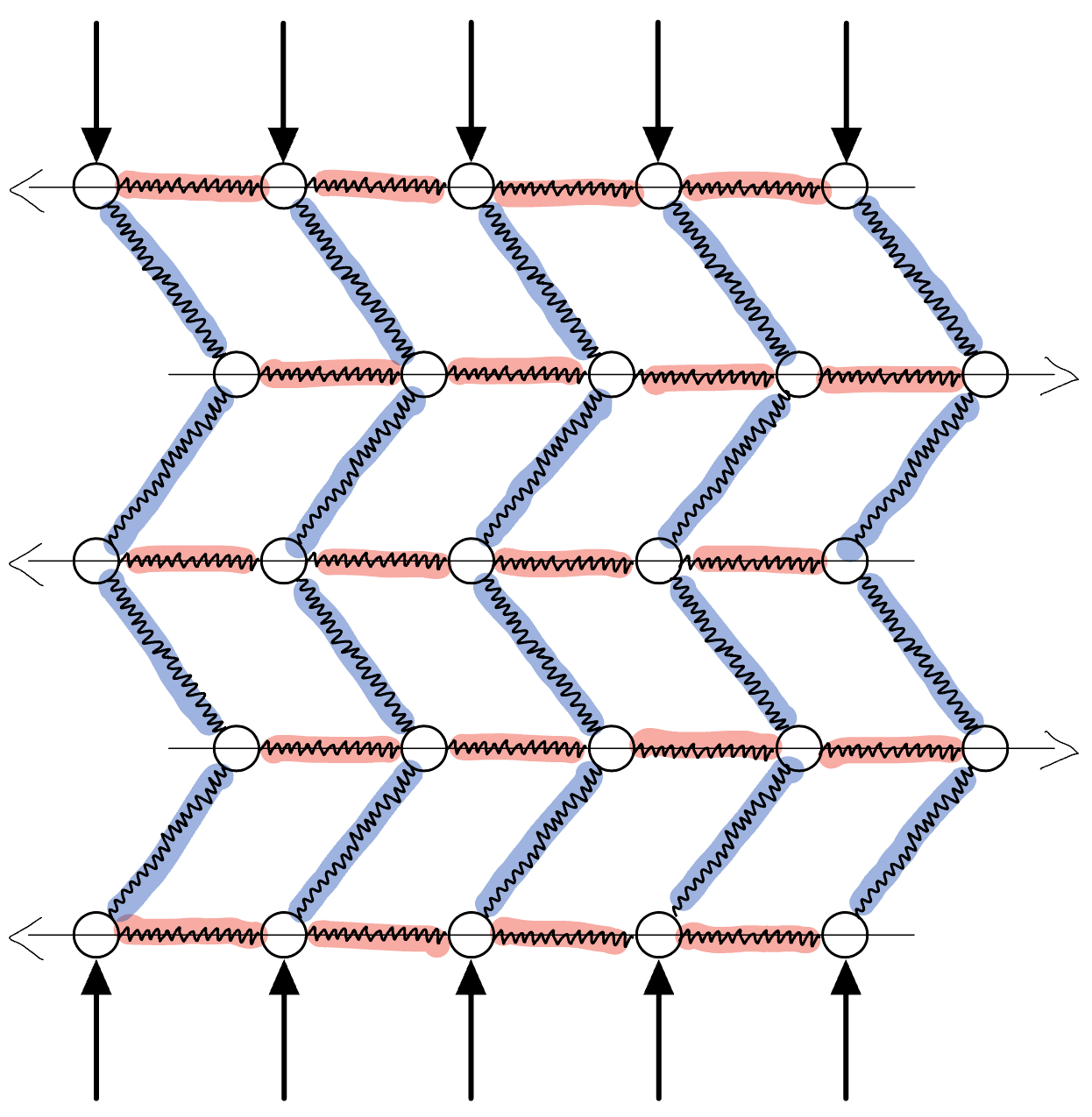}
\caption{\flabel{SquareLatticeBuckling}
\justifying
Zig-zag configuration of a square lattice under external pressure. The distance between rows of particles is identical to the distance between particles in a row. The bold vertical arrows suggest the external forces squeezing the lattice throughout (not shown on the sides), the thin horizontal arrows indicate the resulting displacements of entire rows of connected particles, which in the harmonic approximation (unphysically) keep moving. The resulting configuration gives way to the pressure applied in the vertical direction. Springs highlighted in blue are more relaxed than those highlighted in red.}
\end{figure}

A similar buckling can be observed on the triangular lattice where the eigenvalues $\eval_\pvec^-$ for $\kvec_\pvec^\transpose\in\{(0,\pi),(\pi,0),(\pi,\pi)\}$ become negative as
$\pairPot'(\latticeConstant)/\latticeConstant$ drops below $-\pairPot''(\latticeConstant)/3$, \Tref{extremal_kp}. All three wave vectors have in common that one of the $\oneMcos{q}{\pvec}$ vanishes, while two are equal and maximal. In the following, we focus on $\kvec_\pvec^\transpose=(0,\pi)$ which produces $\oneMcos{1}{\pvec}=0$ and $\oneMcos{2}{\pvec}=\oneMcos{3}{\pvec}=2$. The eigenvalue $\eval_\pvec^-=2\pairPot''(\latticeConstant)+6\pairPot'(\latticeConstant)/\latticeConstant$ is indeed negative for $\pairPot'(\latticeConstant)/\latticeConstant<-\pairPot''(\latticeConstant)/3$, with eigenvector $\evec_\pvec^-=(1,0)^\transpose$. \emph{At} the transition, $\pairPot'(\latticeConstant)/\latticeConstant=-\pairPot''(\latticeConstant)/3$, this eigenvalue vanishes. To determine whether $\eval_\pvec^-$ vanishes for a whole range of $\pvec$ when $\pairPot'(\latticeConstant)/\latticeConstant=-\pairPot''(\latticeConstant)/3$, we inspect the dynamical matrix at $\pairPot'(\latticeConstant)/\latticeConstant=-\pairPot''(\latticeConstant)/3$ and $\pvec=p(0,1)^\transpose$, where $p\in\Zset$, so that $\oneMcos{1}{\pvec}=0$ and $\oneMcos{2}{\pvec}=\oneMcos{3}{\pvec}$, 
\begin{equation}
    \dynamicalMatrixFourier{\pvec}{}=4\pairPot''(\latticeConstant)\oneMcos{2}{\pvec}
    \begin{pmatrix}
        0 & 0 \\
        0 & 1
    \end{pmatrix}\ ,
\end{equation}
which has eigenvalues $\eval_\pvec^+=4\pairPot''(\latticeConstant)\oneMcos{2}{\pvec}$ and indeed $\eval_\pvec^-=0$ for all $\pvec=p(0,1)^\transpose$ with eigenvectors $(0,1)^\transpose$ and $(1,0)^\transpose$ respectively.
At the transition, $\pairPot'(\latticeConstant)/\latticeConstant=-\pairPot''(\latticeConstant)/3$, the eigenvalues $\eval_\pvec^-$ of $\pvec=p(0,1)$ thus vanish for all $p\in\Zset$ simultaneously, allowing arbitrary excitations along $\evec_\pvec^-$ which happens to be transverse to $\pvec$. The same considerations apply to the eigenvalues along the other main axes of the lattice, $\pvec=p(1,0)$ with $\oneMcos{2}{\pvec}=0$, and $\pvec=p(1,1)$ with $\oneMcos{3}{\pvec}=0$. The harmonic lattice thus ``melts'' due to a multitude of modes simultaneously.

The buckling transition is illustrated in \Fref{TriangularLatticeBuckling}. That eigenvalues $\eval_\pvec^-$ for all modes $\pvec=(0,p^2\in\Zset)$ vanish means that any mode $\kvec_\pvec$ travelling purely in the $y$-direction with displacements purely in the $x$-direction is permissible without energetic costs. In other words, these massless excited modes produce displacements $\evec_\pvec^-$ along the $x$-direction but not the $y$-direction, so that, given periodic boundary conditions, entire rows of particles move in unison. They do so independently, as all modes $\kvec_\pvec$ in the $y$-direction equally vanish. In \Fref{TriangularLatticeBuckling} this is illustrated by rows of particles moving freely relative to each other in the $x$-direction.

\begin{figure}
\includegraphics[height=5.5cm]{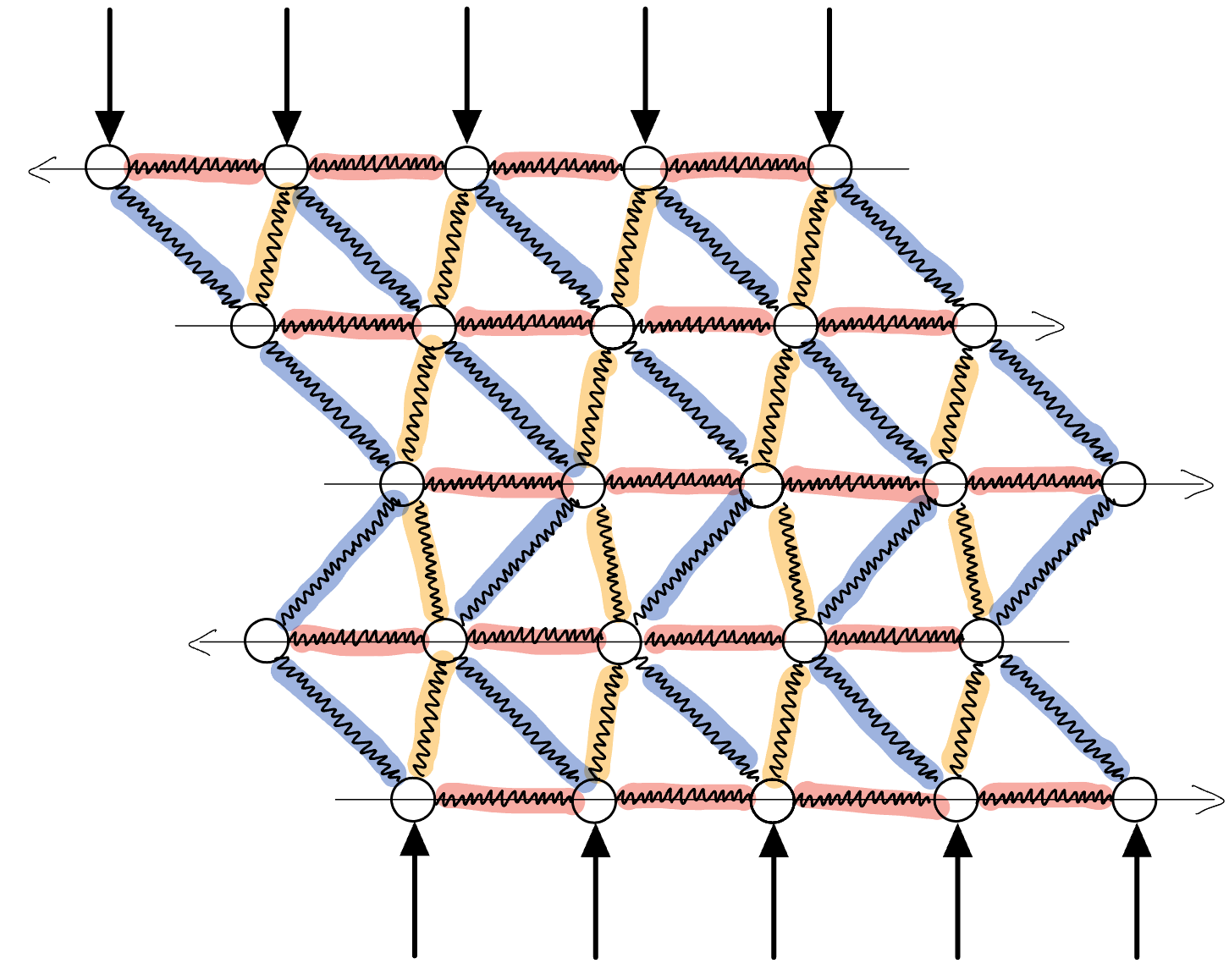}
\caption{\flabel{TriangularLatticeBuckling}
\justifying
Example configuration of a triangular lattice at the buckling transition. \emph{At} the transition, all modes modulating in the $y$-direction --- resulting in displacement in the $x$-direction, namely those with $\oneMcos{1}{\pvec}=0$ --- become massless simultaneously as $\eval_\pvec^-=0$ and $\evec_\pvec^-=(1,0)^\transpose$ for all $\pvec=p(0,1)^\transpose$, where $p\in\Zset$ (\Fref{SquareLatticeBuckling} shows only the most-negative eigenvalues on the square lattice \emph{beyond} the buckling transition). 
The same happens for modes with $\oneMcos{2}{\pvec}=0$ and $\oneMcos{3}{\pvec}=0$ (not shown).
As periodic boundary conditions apply, entire rows are displaced along the thin horizontal arrows away from their regular positions in response to the surrounding external pressure (thick arrows, not shown on the sides). The displacement results in more relaxed springs (blue) as well as more compressed springs (orange), but also in different projections along the direction of movement, \Fref{TriangularLatticeBucklingMechanics}.
}
\end{figure}

The indefinite slipping of an entire row of particles at the buckling transition, \Fref{TriangularLatticeBuckling}, is of course unphysical. In the case of pair-interacting particles, once the lattice rearranges, the changing neighbourhood is not captured in the harmonic approximation with fixed adjacency. Even if particles are permanently interacting by Hookian springs, the lateral instability that arises from equal and opposite repulsion cannot lead to an indefinite displacement. However, the \emph{onset} of the instability \emph{is} physical. Self-consistency of the harmonic approximation requires $\pairPot'(\latticeConstant)/\latticeConstant>-\pairPot''(\latticeConstant)/3$ and the rearrangement of the lattice that occurs for $\pairPot'(\latticeConstant)/\latticeConstant<-\pairPot''(\latticeConstant)/3$ is curbed by higher-order terms, \Sref{higher_order_terms}.

\begin{figure*}[t!]
    \centering
    \begin{subfigure}[t]{0.5\textwidth}
       \centering
       \begin{tikzpicture}
           \node at (0,0)        
       {\includegraphics[width=0.95\linewidth]{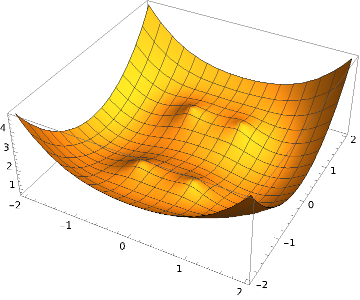}};
            \node at (-3.4,-0.4) {$\sum\pairPot$};
            \node[rotate=-18] at (0.4,-2.5) {$x$};
            \node[rotate=60] at (3.6,0.2) {$y$};
       \end{tikzpicture}       
\caption{
\flabel{FullHarmonicU1_6} Pairwise Hookian potential as a function of distance.}
    \end{subfigure}%
    ~ 
    \begin{subfigure}[t]{0.5\textwidth}
        \centering
       \begin{tikzpicture}
           \node at (0,0)        
       {\includegraphics[width=0.95\linewidth]{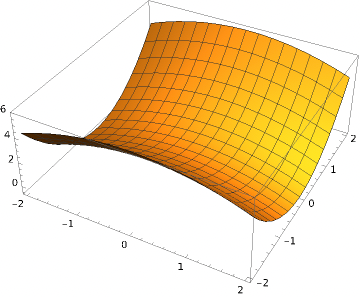}};
            \node at (-3.4,-0.3) {$\sum\pairPot$};
            \node[rotate=-18] at (0.4,-2.5) {$x$};
            \node[rotate=60] at (3.7,0.3) {$y$};
       \end{tikzpicture}              
        \caption{
\flabel{FullApproxU1_6} Harmonic approximation.}
    \end{subfigure}
\caption{\flabel{U1_6}
\justifying
As in \Fref{U1_3}, but on a slightly larger scale, potential landscape $\sum_{q\in\{2,3,5,6\}}\pairPot(|\displacementVec{\nvec}{}-\latticeConstant\latticeVector_q|)$ of a particle as a function of its displacement assuming, however, that particles along the $x$-axis maintain their distance at $\latticeConstant$, whereas all other surrounding particles are standing still. 
Parameters: $\springConstant=1$, $\relaxedLength=1.6$ and $\latticeConstant=1$, \Eref{harmonic_pot}, so that $\pairPot'(\latticeConstant)/\latticeConstant=-0.6\pairPot'(\latticeConstant)$, well after the buckling transition at $\pairPot'(\latticeConstant)/\latticeConstant=-0.33\ldots\pairPot'(\latticeConstant)$, \Eref{instability_condition}.
(a) Hookian pair potentials, \Eref{harmonic_pot}, as a function of the distance $|\displacementVec{\nvec}{} - \displacementVec{\nvec + \mvec_q}{} - \latticeConstant \latticeVector_q|$. This potential landscape shows correctly that large displacements in the $x$-direction will eventually result in large restoring forces. (b)
Harmonic approximation, \Eref{harmonicApproximation_explicit}, using the same parameters as in subfigure (a). The negative curvature along the $x$-direction indicates an unphysical unstable direction.
}
\end{figure*}

The instability discussed above is also visible in the potential landscape, \Fref{U1_6}. As in \Fref{U1_3}, this shows the potential of a particle as a function of its displacement, but now assuming its neighbours in the $x$-direction move in unison with the particle, which happens as a whole row of particles if displaced by the same amount. Figure~\fref{FullHarmonicU1_6} shows the resulting potential assuming Hookian pair-wise potentials, \Eref{harmonic_pot}, as a function of the distance between particles. For sufficiently large distances, the potential picks up again, providing a potential trough. The harmonic approximation, \Eref{harmonicApproximation_explicit}, shown in \Fref{FullApproxU1_6}, loses stability for sufficiently negative $\pairPot'(\latticeConstant)/\latticeConstant$, suggesting an unphysical indefinite slip, as it captures the potential landscape only for small-enough displacements.

\subsection{Correlators}\seclabel{correlators}

We return to solving \Erefs{initial_Langevin} and \eref{2d_SDE_Fourier} in the harmonic approximation, assuming non-negative eigenvalues $\eval_\pvec^\pm$ for all $\pvec$. Having determined the normalised orthogonal eigenvectors $\evec_\pvec^\pm$, 
\Eref{def_evecs}, of $\dynamicalMatrixFourier{\pvec}{}$, we can write $\displacementVecFourier{\pvec}{t}$ in terms of the two components 
\begin{equation}
\elabel{displacementP_by_compo}
\displacementVecFourier{\pvec}{t} = 
\displacementFourier[+]{\pvec}{t} \evec^+_\pvec
+
\displacementFourier[-]{\pvec}{t} \evec^-_\pvec\ ,
\end{equation}
and similarly for the noises,
\begin{equation}
\thermalNoiseVecFourier_\pvec(t) =
\thermalNoiseFourierPM[+]_\pvec(t) \evec^+_\pvec
+
\thermalNoiseFourierPM[-]_\pvec(t) \evec^-_\pvec\ ,
\end{equation}
where
\begin{equation}\elabel{thermalNoiseFourier_correlator}
\ave{
\left(
\begin{array}{ll}
\thermalNoiseFourierPM[+]_\pvec(t) \thermalNoiseFourierPM[+]_{\pvec'}(t') &
\thermalNoiseFourierPM[+]_\pvec(t) \thermalNoiseFourierPM[-]_{\pvec'}(t') \\
\thermalNoiseFourierPM[-]_\pvec(t) \thermalNoiseFourierPM[+]_{\pvec'}(t') &
\thermalNoiseFourierPM[-]_\pvec(t) \thermalNoiseFourierPM[-]_{\pvec'}(t') 
\end{array}
\right)
}
=\Gamma^2V\delta_{\pvec+\pvec',\nullvec} \ident_2 \delta(t-t')\ ,
\end{equation}
and
\begin{equation}
\activeNoiseVecFourier_\pvec(t) =
\activeNoiseFourierPM[+]_\pvec(t) \evec^+_\pvec
+
\activeNoiseFourierPM[-]_\pvec(t) \evec^-_\pvec
\end{equation}
where
\begin{equation}\elabel{activeNoiseFourier_correlator}
\ave{
\left(
\begin{array}{ll}
\activeNoiseFourierPM[+]_\pvec(t) \activeNoiseFourierPM[+]_{\pvec'}(t') &
\activeNoiseFourierPM[+]_\pvec(t) \activeNoiseFourierPM[-]_{\pvec'}(t') \\
\activeNoiseFourierPM[-]_\pvec(t) \activeNoiseFourierPM[+]_{\pvec'}(t') &
\activeNoiseFourierPM[-]_\pvec(t) \activeNoiseFourierPM[-]_{\pvec'}(t') 
\end{array}
\right)
}
=\activity^2 V\delta_{\pvec+\pvec',\nullvec}  \ident_2 \exp{-\noiseRate |t-t'|}\ .
\end{equation}

Hence, \Eref{2d_SDE_Fourier} reduces to two decoupled scalar differential equations
\begin{equation}\elabel{SDE_on_pm}
\displacementDotFourier[\pm]{\pvec}{t} = - \eval_\pvec^\pm \displacementFourier[\pm]{\pvec}{t} 
+ \thermalNoiseFourier_\pvec^\pm(t)
+ \activeNoiseFourier_\pvec^\pm(t)\ ,
\end{equation}
with solution 
\begin{equation}
\elabel{displacement_by_integration}
\displacementFourier[\pm]{\pvec}{t}  = \int_0^t \dint{t'}
\exp{-\eval_\pvec^\pm (t-t')}
\Big\{
 \thermalNoiseFourier_\pvec^\pm(t')
 + \activeNoiseFourier_\pvec^\pm(t')
\Big\},
\end{equation}
as the initial condition is $\displacementFourier[\pm]{\pvec}{t=0}=0$. 
Given the independence of different modes, indicated by $\delta_{\pvec+\pvec',\nullvec}$ in \Erefs{thermalNoiseFourier_correlator} and \eref{activeNoiseFourier_correlator}, the correlator of $\displacementFourier[\pm]{\pvec}{t}$ shows similarly 
\begin{equation}\elabel{DisplacementComponentCorrelator_all}
    \ave{
\left(
\begin{array}{ll}
\displacementFourier[+]{\pvec}{t_1}
\displacementFourier[+]{\pvec'}{t_2} &
\displacementFourier[+]{\pvec}{t_1}
\displacementFourier[-]{\pvec'}{t_2}  \\
\displacementFourier[-]{\pvec}{t_1}
\displacementFourier[+]{\pvec'}{t_2}  &
\displacementFourier[-]{\pvec}{t_1}
\displacementFourier[-]{\pvec'}{t_2} 
\end{array}
\right)
}
=\delta_{\pvec+\pvec',\nullvec}
    \ave{
\left(
\begin{array}{ll}
\displacementFourier[+]{\pvec}{t_1}
\displacementFourier[+]{-\pvec}{t_2} &
0 \\
0 &
\displacementFourier[-]{\pvec}{t_1}
\displacementFourier[-]{-\pvec}{t_2} 
\end{array}
\right)
}.
\end{equation}
As
\begin{equation}
    \ave{\displacementFourier[+]{\pvec}{t_1}\displacementFourier[-]{-\pvec}{t_2}}
    =
    \ave{\displacementFourier[-]{\pvec}{t_1}\displacementFourier[+]{-\pvec}{t_2}}
    = 0
\end{equation}
the only non-trivial correlators are
\begin{subequations}\elabel{DisplacementComponentCorrelator}
    \begin{align}
        \ave{\displacementFourier[+]{\pvec}{t_1}\displacementFourier[+]{-\pvec}{t_2}} &= V \activityCorrInt_\pvec^+(t_1,t_2)
+
V \thermalCorrInt_\pvec^+(t_1,t_2)\ ,\\
        \ave{\displacementFourier[-]{\pvec}{t_1}\displacementFourier[-]{-\pvec}{t_2}} &= V \activityCorrInt_\pvec^-(t_1,t_2)
+
V \thermalCorrInt_\pvec^-(t_1,t_2)\ ,
    \end{align}
\end{subequations}
which separate into independent active $\activityCorrInt_\pvec^\pm(t_1,t_2)$ and thermal $\thermalCorrInt_\pvec^\pm(t_1,t_2)$ components. These components are the fundamental ingredients in all calculations to come and will be specified below. In the harmonic approximation, if all noises are Gaussian, then the correlators are also Gaussian and are therefore sufficient to characterise the system fully. 

The correlator of the active noise being proportional to the identity indicating isotropy, \Eref{activeNoise_correlator_App}, equally renders the correlator \Eref{activeNoiseFourier_correlator} isotropic. This is an important simplification. If the noise were originally anisotropic in \Eref{activeNoise_correlator_App}, say
\begin{equation}\elabel{activeNoise_correlator_alternative}
\ave{\activeNoiseVec_\nvec(t) \activeNoiseVec^\transpose_{\nvec'}(t')}
    = \Wmatrix  \delta_{\nvec,\nvec'} \exp{-\noiseRate|t-t'|}\ ,
\end{equation}
possibly with off-diagonal terms, such that different components were correlated,
this would result in correlators for the Fourier transform, \Eref{ActiveNoiseCorrelationFourier}, that are not proportional to the identity. This would produce correlations between 
$\activeNoiseFourierPM[+]_\pvec(t)$
and
$\activeNoiseFourierPM[-]_{-\pvec}(t)$
in the correlator \eref{activeNoiseFourier_correlator}, as the off-diagonal (in $\pm$) elements $\evec_\pvec^{+} \cdot \Wmatrix \evec_\pvec^{-}$ would generally not vanish, $\activeNoiseFourierPM[+]_\pvec(t)=\evec_\pvec^{+} \cdot \activeNoiseVecFourier_\pvec(t)$.

The displacement correlators in \Erefs{DisplacementComponentCorrelator} are rather cumbersome because they are composed of the double integrals
\begin{subequations}\elabel{def_WXi}
\begin{equation}\elabel{def_activityCorrInt}
\begin{split}
V \activityCorrInt_\pvec^\pm(t_1,t_2) &:=
\int_0^{t_1}\dint{t'}
\int_0^{t_2}\dint{t''}
\exp{-\eval_\pvec^\pm (t_1-t')}
\exp{-\eval_{-\pvec}^\pm (t_2-t'')}
\ave{\hat{\activeNoise}_{\pvec}^{\pm}(t')\hat{\activeNoise}_{-\pvec}^{\pm}(t'')}\\
&=
\int_0^{t_1}\dint{t'}
\int_0^{t_2}\dint{t''}
V\exp{-\eval_\pvec^\pm (t_1-t'+t_2-t'')}
\activity^2 
\exp{-\noiseRate|t'-t''|}\ ,
\end{split}
\end{equation}
\begin{equation}
\begin{split}
V \thermalCorrInt_\pvec^\pm(t_1,t_2) &:=
\int_0^{t_1}\dint{t'}
\int_0^{t_2}\dint{t''}
\exp{-\eval_\pvec^\pm (t_1-t')}
\exp{-\eval_{-\pvec}^\pm (t_2-t'')}\ave{\hat{\thermalNoise}_{\pvec}^{\pm}(t')\hat{\thermalNoise}_{-\pvec}^{\pm}(t'')}\\
&=
\int_0^{t_1}\dint{t'}
\int_0^{t_2}\dint{t''}
V\exp{-\eval_\pvec^\pm (t_1-t'+t_2-t'')}
\Gamma^2 
\delta(t'-t'')\ , \elabel{def_thermalCorrInt}
\end{split}
\end{equation}
\end{subequations}
which, of course, can be evaluated in closed form, as in \Sref{ExactIntegrals}, but are too unwieldy for the analysis that follows. To simplify the expressions, we focus on 
large times $t$. 

If any of the eigenvalues $\eval_\pvec^\pm$ are negative, then, by inspection, both $\activityCorrInt_\pvec^\pm(t,t)$ and $\thermalCorrInt_\pvec^\pm(t,t)$ will diverge for large $t$. As discussed in \Sref{stability}, the harmonic approximation becomes inconsistent for negative eigenvalues. In the following, we therefore focus on non-negative eigenvalues $\eval_\pvec^\pm$.

A steady state would require $t_{1,2}\gg 1/\noiseRate$ and $t_{1,2}\gg 1/\eval_\pvec^\pm\ge0$. However, the latter can never be achieved for $\pvec=\nullvec$, as $\eval_\nullvec^\pm$ vanishes, resulting in the equal-time correlators behaving 
for large $t$ like
\begin{subnumcases}{\elabel{def_activityCorrInt_Larget}
\activityCorrInt_\pvec^\pm(t,t)=}
\frac{2 \activity^2}{\noiseRate}
t,
& for  $\pvec=\nullvec$ and $t\gg 1/\noiseRate$, 
\elabel{def_activityCorrInt_p0}
\\
\frac{\activity^2}{\eval_\pvec^\pm (\eval_\pvec^\pm+\noiseRate)}, 
&  for $\pvec\ne\nullvec,~t\gg 1/\noiseRate$ and $t\gg 1/\eval_\pvec^\pm$,
\end{subnumcases}
and similarly
\begin{subnumcases}{\elabel{def_thermalCorrInt_Larget}
\thermalCorrInt_\pvec^\pm(t,t) =}
\Gamma^2 t,
& for $\pvec=\nullvec$, 
\elabel{def_thermalCorrInt_p0}
\\
\frac{\Gamma^2}{2\eval_\pvec^\pm}, 
&  for $\pvec\ne\nullvec \text{ and }t\gg 1/\eval_\pvec^\pm$.
\end{subnumcases}
The exceptional nature of the $\nullvec$-mode has its physical origin in the diffusion of the lattice as a whole. As the lattice structure ties the displaced sites together, the resulting net displacement decreases with the number of sites $V$. The resulting correlator is, 
using \Erefs{displacementFromFourier},
\eref{displacementP_by_compo}, and \eref{DisplacementComponentCorrelator},
given by the outer product,
\begin{multline}\elabel{displacement_corr}
\ave{\displacementVec{\nvec}{t} \displacementVec[\transpose]{\nvec'}{t}} = \frac{1}{V} 
\left(
\Gamma^2 + \frac{2\activity^2}{\noiseRate}
\right) t 
\ident_2
+ 
\frac{1}{V}\starredsum_{\pvec}
\exp{\imag \BriVec{\pvec} \cdot (\nvec-\nvec')}
\left(
\Big[
\activityCorrInt_\pvec^+(t,t)
+
\thermalCorrInt_\pvec^+(t,t)
\Big]
\Ematrix_\pvec^+
+
\Big[
\activityCorrInt_\pvec^-(t,t)
+
\thermalCorrInt_\pvec^-(t,t)
\Big]
\Ematrix_\pvec^-
\right),
\end{multline}
with $2 \times 2$ matrices 
\begin{equation}\elabel{def_Ematrix}
\Ematrix_\pvec^\pm = \evec_\pvec^\pm\left(\evec_\pvec^\pm\right)^\transpose
\end{equation}
which are the outer products of the eigenvectors in \Eref{def_evecs}. Above, we assume $t\gg1/\noiseRate$ in order to use the result for $\activityCorrInt_\pvec^\pm(t,t)$ at $\pvec=\nullvec$,
as stated in \Eref{def_activityCorrInt_p0}.
The starred sum in \Eref{displacement_corr} denotes the summation over all $\pvec$ except $\pvec=\nullvec$. 

In order to keep the stationary limit $t\to\infty$ well-defined, we introduce the displacement of the centre of mass,
\begin{equation}\elabel{def_ubar_again}
	\displacementVecBar{t} = \frac{1}{V} \sum_{\nvec} \displacementVec{\nvec}{t} = \frac{1}{V} \displacementVecFourier{\pvec=\nullvec}{t} \ ,
\end{equation}
which, for $t\gg 1/\noiseRate$, has correlator 
\begin{equation}\elabel{var_rigid_mode}
\ave{\displacementVecBar{t} \displacementVecBar[\transpose]{t}}\\
= \frac{1}{V} 
\left(
\Gamma^2 + \frac{2\activity^2}{\noiseRate}
\right) t 
\ident_2
\end{equation}
according to \Eref{displacement_corr} and equally from the correlator of $\displacementVecFourier{\pvec=\nullvec}{t}$, via \Erefs{displacementP_by_compo}, \eref{DisplacementComponentCorrelator_all}, \eref{DisplacementComponentCorrelator} and \eref{def_ubar_again}.
Equation~\eref{var_rigid_mode} is the first term on the right-hand side of \Eref{displacement_corr}, linear in $t$. It produces the expected effective diffusion constant of the lattice, $(\Gamma^2/2 + \activity^2/\noiseRate)/V$, which equals that of an active Brownian particle with translational diffusion constant $\mathsf{D}_t=\Gamma^2/(2V)$, squared self-propulsion speed $w_0^2=2\activity^2/V$ and rotational diffusion constant $\mathsf{D}_r=\noiseRate$, \cite[Eq.~(29)]{zhang2024field}.

Deducting the centre-of-mass displacement from the overall displacement $\displacementVec{\nvec}{t}$ gives the relative displacement
\begin{equation}
	\displacementVecRel{\nvec}{t} := \displacementVec{\nvec}{t} - \displacementVecBar{t} 
	=
	\elabel{displacementRelFromFourier}
\frac{1}{V} \starredsum_{\pvec} \exp{\imag \BriVec{\pvec}\cdot\nvec} \displacementVecFourier{\pvec}{t}\ ,
\end{equation}
with the sum replaced by the starred sum compared to \Eref{displacementFromFourier}. With this adjustment, 
the first term on the right-hand side of \Eref{displacement_corr} disappears from the equal-time two-point correlator of the relative displacements so that, in the steady-state
limit 
\begin{equation}\elabel{displacementRel_corr}
\lim_{t\to\infty}
\ave{\displacementVecRel{\nvec}{t}\displacementVecRel[\transpose]{\nvec'}{t}}
=
\\
\frac{1}{V}\starredsum_{\pvec}
\exp{\imag \BriVec{\pvec} \cdot (\nvec-\nvec')}
\left(
\left[
\frac{\activity^2}{\eval_\pvec^+ (\eval_\pvec^++\noiseRate)}
+
\frac{\Gamma^2}{2\eval_\pvec^+}
\right]
\Ematrix_\pvec^+
+
\left[
\frac{\activity^2}{\eval_\pvec^- (\eval_\pvec^-+\noiseRate)}
+
\frac{\Gamma^2}{2\eval_\pvec^-}
\right]
\Ematrix_\pvec^-
\right)\ .
\end{equation}
While \Eref{displacementRel_corr} is the full correlation matrix, its trace is the more standard two-point correlator.
In the particular case of $\nvec=\nvec'$, \Eref{displacementRel_corr} simplifies in the obvious way that the exponential from the Fourier transform disappears from the expression,
\begin{equation}\elabel{divergent_var}
\lim_{t\to\infty}
\ave{\displacementVecRel{\nvec}{t}\cdot\displacementVecRel{\nvec}{t}}
=
\lim_{t\to\infty}
\Trace 
\ave{\displacementVecRel{\nvec}{t}\displacementVecRel{\nvec'}{t}}
=
\frac{1}{V}\starredsum_{\pvec}
\left(
\left[
\frac{\activity^2}{\eval_\pvec^- (\eval_\pvec^-+\noiseRate)}
+
\frac{\Gamma^2}{2\eval_\pvec^-}
\right]
+
\left[
\frac{\activity^2}{\eval_\pvec^+ (\eval_\pvec^++\noiseRate)}
+
\frac{\Gamma^2}{2\eval_\pvec^+}
\right]
\right),
\end{equation}
using that 
$\Trace \Ematrix^\pm=\evec^\pm\cdot\evec^\pm=1$.
By inspection, \Eref{divergent_var} diverges logarithmically in large $V$, as $\eval_\pvec^\pm$ vanishes like $|\pvec|^2$ for fixed 
$\pvec$
as $V\to\infty$, \Eref{eval_formula}. At large $V$, this observable of the active crystal differs thus from the passive one
by the thermal noise parameterised by $\Gamma^2$ getting boosted by the activity $\activity^2/\noiseRate$. 
As discussed in the main text, 
in the harmonic approximation, the divergence of the variance of the relative displacement indicates the absence of crystalline order according to Mermin's original criterion \cite{mermin1968crystalline}.

However, the crystalline \emph{integrity} is much better probed via the steady-state mean squared particle separation,
\begin{equation}
    \ave{|\displacementVec{\nvec}{} - \displacementVec{\nvec + \mvec_q}{} - \latticeConstant \latticeVector_q|^2} =
    \ave{|\displacementVec{\nvec}{} - \displacementVec{\nvec + \mvec_q}{}|^2} + \latticeConstant^2 =
    \ave{|\displacementVecRel{\nvec}{} - \displacementVecRel{\nvec + \mvec_q}{}|^2} + \latticeConstant^2
\end{equation}
using that $\ave{\displacementVec{\nvec}{}}=0$ and $|\latticeVector|^2=1$ as well as \Eref{displacementRelFromFourier}.
From \Erefs{displacement_corr} and \eref{displacementRel_corr}, this is, up to a shift by 
\begin{equation}
\latticeConstant^2=\left|\ave{\displacementVec{\nvec}{} - \displacementVec{\nvec + \mvec_q}{} - \latticeConstant \latticeVector_q}\right|^2
\end{equation}
the variance
\begin{equation}\elabel{distance_variance_2D}
  \lim_{t\to\infty}\ave{|\displacementVec{\nvec}{t} - \displacementVec{\nvec+\mvec_q}{t}|^2}
  =
\frac{1}{V}\starredsum_{\pvec}
2
\oneMcos{q}{\pvec}
\left(
\left[
\frac{\activity^2}{\eval_\pvec^+ (\eval_\pvec^++\noiseRate)}
+
\frac{\Gamma^2}{2\eval_\pvec^+}
\right]
+
\left[
\frac{\activity^2}{\eval_\pvec^- (\eval_\pvec^-+\noiseRate)}
+
\frac{\Gamma^2}{2\eval_\pvec^-}
\right]
\right),
\end{equation}
which is generally not independent of $q$ because of the lattice having different extent in different directions, even when $\NOne=\NTwo$. In other words, the lattice geometry leads to anisotropy of the local fluctuations. Mirror symmetry results in $q=1$ and $q=4$ having the same mean squared separation, as do all of $q=2,3,5,6$, see \Fref{ActiveCrystalSchematic}.

There are two scenarios that result in the crystal disintegrating by a divergent inter-particle distance. Firstly, when some eigenvalues $\eval_\pvec^\pm$ are negative, the lattice disintegrates, as the harmonic approximation is no longer self-consistent, \Sref{stability}, and equivalently the correlators $\activityCorrInt_\pvec^\pm(t,t)$ and $\thermalCorrInt_\pvec^\pm(t,t)$ diverge, \Erefs{def_WXi}. To arrive at \Eref{distance_variance_2D} we had to assume $\eval_\pvec^\pm\ge0$, in fact all eigenvalues in the summands are strictly positive, $\eval_\pvec^\pm>0$ for $\pvec\ne\nullvec$. However, the sum in \Eref{distance_variance_2D} might still diverge as the number of terms diverges, depending on whether and how $\eval_\pvec^\pm$ approach $0$ as $V\to\infty$. As discussed in detail in \Sref{convergence} below, it turns out that $\oneMcos{q}{\pvec}/\eval_\pvec^\pm$ remains finite in the limit of large $V$, so that the summands in \Eref{distance_variance_2D} remain bounded and the variance converges.

If all eigenvalues are non-negative, \ie if $\pairPot'(\latticeConstant)/\latticeConstant>-\pairPot''(\latticeConstant)/3$, then in the limit of $\NOne=\NTwo\to\infty$ the (Riemann) sum, \Eref{distance_variance_2D}, converges to the integral 
\begin{multline}\elabel{distance_variance_2D_integral}
  \lim_{\NOne=\NTwo\to\infty}\lim_{t\to\infty}\ave{|\displacementVec{\nvec}{t} - \displacementVec{\nvec+\mvec_q}{t}|^2}
  \\
  =
2 \int_0^{2\pi} \dbar k^1 \int_0^{2\pi} \dbar k^2
\oneMcosCont{q}(\kvec)
\left(
\left[
\frac{\activity^2}{\eval^+(\kvec) (\eval^+(\kvec)+\noiseRate)}
+
\frac{\Gamma^2}{2\eval^+(\kvec)}
\right]
+
\left[
\frac{\activity^2}{\eval^-(\kvec) (\eval^-(\kvec)+\noiseRate)}
+
\frac{\Gamma^2}{2\eval^-(\kvec)}
\right]
\right),
\end{multline}
independent of $q$, as briefly explained below.
In \Eref{distance_variance_2D_integral} we have used the notation of 
\begin{equation}\elabel{def_oneMcosFunc_again}
\oneMcosCont{q}(\kvec)=1-\cos(\mvec_q\cdot\kvec)
\end{equation}
as a function of $\kvec=(k^1,k^2)^\transpose$,
closely related to $\oneMcos{q}{\pvec}$ that is a function of $\pvec$ rather than $\kvec$, \Eref{def_oneMcos_general}. The eigenvalues $\eval^\pm(\kvec)$ in \Eref{distance_variance_2D_integral} are correspondingly given by \Eref{eval_formula} with $\oneMcos{q}{\pvec}$ replaced by $\oneMcosCont{q}(\kvec)$ in \Eref{def_oneMcosFunc_again}. The right-hand side of \Eref{distance_variance_2D_integral} simplifies further by non-dimensionalising --- in particular, introducing $\pairPotCoeff=\pairPot'(\latticeConstant)/(\latticeConstant\pairPot''(\latticeConstant))$. However, even $\activity=0$ still requires numerical, rather than analytical, evaluation. The melting of the lattice --- as signalled by the divergence of the variance of the inter-particle distance, \Eref{distance_variance_2D_integral} --- occurs at $\pairPotCoeff = -1/3$.

The independence of the variance $\ave{|\displacementVec{\nvec}{t} - \displacementVec{\nvec+\mvec_q}{t}|^2}$ from $q$ in \Eref{distance_variance_2D_integral}, physically, is due to the thermodynamic limit restoring isotropy, which is absent in \Eref{distance_variance_2D}. Mathematically, the independence is demonstrated by a substitution of the integration variables $k^1$ and $k^2$ and symmetries of $\oneMcosCont{q}(\kvec)=1-\cos(\mvec_q\cdot\kvec)$, \Eref{def_oneMcosFunc_again}. As $\oneMcosCont{q}(\kvec)$ and therefore the eigenvalues $\eval^\pm(\kvec)$, \Eref{eval_formula}, are even in $\kvec$, \Eref{distance_variance_2D_integral} is invariant under changing $q\in\{1,2,3\}$ to $q+3$, using that $\mvec_q=-\mvec_{q+3}$
as introduced before \Eref{latticeVector_from_m}. Similarly, \Eref{distance_variance_2D_integral} for $q=1$ and $q=2$ produces the same result, as changing the integration variables from $k^1$ and $k^2$ to $k'^1=k^2$ and $k'^2=k^1$ maps $\oneMcosCont{1}(\kvec)$ to $\oneMcosCont{2}(\kvec')$ and \latin{vice versa}, leaving the eigenvalues unchanged, because they are invariant under permutations of the labels $q$ of $\oneMcosCont{q}(\kvec)$, \Eref{eval_formula}. Demonstrating that \Eref{distance_variance_2D_integral} for $q=3$ is identical to the result $q=1$ uses the mapping $k'^1=k^2-k^1$ and $k'^2=-k^1$, so that $k'^2-k'^1=-k^2$ and therefore 
$\oneMcosCont{1}(\kvec')=\oneMcosCont{3}(\kvec)$,
$\oneMcosCont{2}(\kvec')=\oneMcosCont{1}(\kvec)$ and
$\oneMcosCont{3}(\kvec')=\oneMcosCont{2}(\kvec)$. This substitution draws on the periodicity of $\oneMcosCont{q}(\kvec)$, as it requires a translation of the integration limits.

\subsubsection{Convergence of \Erefs{2D_crystalline_integrity} and \eref{distance_variance_2D}}\seclabel{convergence}
Even when all eigenvalues are non-negative, for the correlator in \Erefs{2D_crystalline_integrity} and \eref{distance_variance_2D} it is relevant to know \emph{how} eigenvalues vanish (if they do), in order to show that $(1-\cos(\BriVec{\pvec}\cdot\mvec_q))/\eval_\pvec^\pm=\oneMcos{q}{\pvec}/\eval_\pvec^\pm$ is bounded, such that the sum \Eref{2D_crystalline_integrity} can be determined in the limit of large $\NOne$ and $\NTwo$. By inspection of \Eref{eval_formula}, $\eval_\pvec^\pm$ vanish trivially whenever $\pvec=(0,0)^\transpose$, resulting in $\BriVec{\pvec}=(0,0)^\transpose$ and thus $\oneMcos{q}{\pvec}=0$ for all $q$. 

In the following, we assume all eigenvalues $\eval_\pvec^\pm$ for $\pvec\ne\nullvec$ are strictly positive, so that $\pvec = \nullvec$ produces the unique global minimum of $\eval_{\pvec=\nullvec}^\pm=0$. In other words, we assume $\pairPot'(\latticeConstant)/\latticeConstant>-\pairPot''(\latticeConstant)/3$. By periodicity, there are some $\pvec\in\{0,1,\ldots,\NOne-1\}\otimes\{0,1,\ldots,\NTwo-1\}=\Pset$ that produce $\BriVec{\pvec}$ which for large $\NOneTwo$ come arbitrarily close to, say, $(2\pi,2\pi)^\transpose$, but a suitable shift of $\Pset$ will result in $\BriVec{\pvec}\in[-\pi,\pi)^2$, so that $\oneMcos{q}{\pvec}$ attain their maximum at the boundaries of the ``first Brillouin zone of indices''. Despite this rearrangement, it remains a (very reasonable) assumption in the following that, even for large $\NOne$ and $\NTwo$, the eigenvalues $\eval_{\pvec}$ are bounded away from $0$ everywhere, except in the vicinity of $\pvec=\nullvec$. In large $V$, the smallest, yet positive, $\eval_{\pvec}^\pm$ are thus to be found around $\pvec = \nullvec$. 

In a perturbation theory about $\BriVec{\pvec}=(0,0)^\transpose$, for the convergence of \Erefs{2D_crystalline_integrity} and \eref{distance_variance_2D} it suffices to show 
\begin{equation}\elabel{lim_of_ratio}
\lim_{\NOne,\NTwo\to\infty} \left|\frac{\oneMcos{q}{\pvec}}{\eval_\pvec^\pm}\right|\le\Upsilon<\infty
\ ,
\end{equation}
with some $\Upsilon\in\Rset$, so that every summand in \Erefs{2D_crystalline_integrity} and \eref{distance_variance_2D} is bounded, and therefore the sum divided by $V=\NOne\NTwo$ is as well.
\latin{A priori}, the limit in \Eref{lim_of_ratio} depends on $\pvec$ and the details of how $\NOne\to\infty$ and $\NTwo\to\infty$. If this results in, say $\oneMcos{1}{\pvec}\ll\oneMcos{2}{\pvec}$, then $\oneMcos{2}{\pvec}$ converges to $\oneMcos{3}{\pvec}$, as can be seen by expanding $\oneMcos{q}{\pvec}$ for large $\NOne,\NTwo$, \ie small arguments of the cosine. The dynamical matrix \Eref{dynamicalMatrixFourier_p} can then be written in the form 
\begin{equation}\elabel{dynamicalMatrixFourier_with_one_gamma0}
\dynamicalMatrixFourier{\pvec}{} 
= 2 \oneMcos{2}{\pvec} 
\left(
2 \frac{\pairPot'(\latticeConstant)}{\latticeConstant} \ident_2 + 
	\left(
	\pairPot''(\latticeConstant) - \frac{\pairPot'(\latticeConstant)}{\latticeConstant}
	\right)
\begin{pmatrix}
1/2 & 0 \\
0 & 3/2
\end{pmatrix}
\right)
\end{equation}
with the matrix corresponding to $\latticeVector_1 \latticeVector^\transpose_1 + \latticeVector_2 \latticeVector^\transpose_2$. This dynamical matrix has eigenvalues 
\begin{subequations}
\elabel{evals_with_one_gamma0}
\begin{align}
\eval_\pvec^-&= 2 \oneMcos{2}{\pvec} 
\left(
\frac{3}{2} \frac{\pairPot'(\latticeConstant)}{\latticeConstant} 
+
\frac{1}{2} \pairPot''(\latticeConstant)
\right)\ ,\\
\eval_\pvec^+&= 2 \oneMcos{2}{\pvec} 
\left(
\frac{1}{2} \frac{\pairPot'(\latticeConstant)}{\latticeConstant} 
+
\frac{3}{2} \pairPot''(\latticeConstant)
\right)\ ,
\end{align}
\end{subequations}
so that 
\begin{equation}
\lim_{\NOne,\NTwo\to\infty} \frac{\oneMcos{1}{\pvec}}{\eval_\pvec^\pm} = 0
\end{equation}
and
\begin{equation}
\lim_{\NOne,\NTwo\to\infty} \frac{\oneMcos{2,3}{\pvec}}{\eval_\pvec^\pm} = 
\begin{cases}
\left(
\frac{\pairPot'(\latticeConstant)}{\latticeConstant} 
+
3\pairPot''(\latticeConstant)
\right)^{-1}\\
\left(
3 \frac{\pairPot'(\latticeConstant)}{\latticeConstant} 
+
\pairPot''(\latticeConstant)
\right)^{-1}
\end{cases} \ .
\end{equation}
Just like $\oneMcos{1}{\pvec}\to0$ implies $\oneMcos{2}{\pvec}\to\oneMcos{3}{\pvec}$, the same argument applies to $\oneMcos{2}{\pvec}\ll\oneMcos{3}{\pvec}$ or $\oneMcos{3}{\pvec}\ll\oneMcos{1}{\pvec}$, 
as $\oneMcos{2}{\pvec}=0$ implies $\oneMcos{3}{\pvec}=\oneMcos{1}{\pvec}$,
and $\oneMcos{3}{\pvec}=0$ implies $\oneMcos{1}{\pvec}=\oneMcos{2}{\pvec}$. 

If the limit is taken such that $\NOne/\NTwo$ converges to a finite ratio, then, in general, to leading order all three $\oneMcos{q}{\pvec}$ can be written as multiples of, say $(1/\NOne)^2$, as can be seen by expanding $\oneMcos{q}{\pvec}$. As a result, the eigenvalues $\eval_\pvec^\pm$ are to leading order linear in $(1/\NOne)^2$, which cancels with the leading $(1/\NOne)^2$ of $\oneMcos{q}{\pvec}$ in $\oneMcos{1}{\pvec}/\eval_\pvec^\pm$. 

In summary, independent of the details on how the limit $\NOne,\NTwo\to\infty$ is taken, the ratio $\oneMcos{1}{\pvec}/\eval_\pvec^\pm$ converges, so that the summands in \Eref{2D_crystalline_integrity} and \Eref{distance_variance_2D} remain finite and the sum converges.

\subsection{Internal energy}\seclabel{energy}
Given the force term in the overdamped Langevin \Eref{initial_Langevin}, the total internal energy of the active crystal is, up to a constant shift,
\begin{equation}\elabel{initial_energy}
    \internalEnergy(t) = \half \sum_{\nvec}\sum_{q=1}^Q
    \pairPot\big(|\displacementVec{\nvec}{t}-\displacementVec{\nvec+\mvec_q}{t}-\latticeConstant\latticeVector_q|\big) \ ,
\end{equation}
where the factor of $1/2$ corrects for the double counting of each pair potential in the double sum.
Correspondingly, in the harmonic approximation, \Erefs{harmonicApproximation_explicit} and \eref{def_dynamicalMatrix_HarmonicAppendix}, the energy reduces to the essentially bilinear (yet not diagonal)
\begin{equation}\elabel{def_internalEnergyHarmonic}
\internalEnergyHarmonic(t) = 3V \pairPot(\latticeConstant) +
\half \sum_{\nvec}\sum_{\nvec'}
\displacementVec{\nvec}{t}\cdot
\dynamicalMatrix{\nvec}{\nvec'} \displacementVec{\nvec'}{t}\ ,
\end{equation}
using 
$\sum_q \latticeVector_q = \nullvec$ as well as 
$\sum_q \latticeVector_q\latticeVector_q^\transpose=3\ident_2$, \Eref{sum_aqaq},
and 
$\sum_\nvec\sum_q \displacementVec{\nvec+\mvec_q}{}\!\!\cdot\latticeVector_q=\sum_\nvec\displacementVec{\nvec}{}\!\cdot\!\sum_q\latticeVector_q=0$, which follows from the periodicity of the labelling. The local force, \Eref{ForceDmatrix}, follows immediately by differentiating \Eref{initial_energy}.

After rewriting \Eref{def_internalEnergyHarmonic} in terms of the Fourier series, \Erefs{displacementFromFourier} and \eref{dynMatrixFromFourier}, and using the orthogonal eigensystem $\dynamicalMatrixFourier{\pvec}{}\evec_\pvec^\pm =\eval_\pvec^\pm\evec_\pvec^\pm$, \Sref{correlators}, the sum over $\nvec$ in \Eref{def_internalEnergyHarmonic} can be rewritten as
\begin{equation}\elabel{U_sum}
\sum_{\nvec}\sum_{\nvec'}
\displacementVec{\nvec}{t}\cdot
\dynamicalMatrix{\nvec}{\nvec'} \displacementVec{\nvec'}{t}
=
\frac{1}{V}
\sum_\pvec 
\displacementVecFourier{-\pvec}{t}
\cdot
\dynamicalMatrixFourier{\pvec}{}
\displacementVecFourier{\pvec}{t}
=
\frac{1}{V}
\sum_\pvec 
\displacementFourier[+]{-\pvec}{t}
\eval^+_\pvec
\displacementFourier[+]{\pvec}{t}
+
\displacementFourier[-]{-\pvec}{t}
\eval^+_\pvec
\displacementFourier[-]{\pvec}{t}
\ .
\end{equation}
The expectation is easily taken using \Erefs{DisplacementComponentCorrelator}. In the steady state, \Erefs{def_activityCorrInt_Larget} and \eref{def_thermalCorrInt_Larget}, the harmonic approximation gives 
\begin{equation}\elabel{stationaryinternalEnergyHarmonic}
\begin{split}
    \lim_{t\to\infty}
    \ave{\internalEnergyHarmonic(t)} 
    &= 3V \pairPot(\latticeConstant) +
    \half \sum_\pvec^* 
\left[\left(
\frac{\Gamma^2}{2} + 
\frac{\activity^2}{\eval_\pvec^++\noiseRate}
\right)
+
\left(
\frac{\Gamma^2}{2} + 
\frac{\activity^2}{\eval_\pvec^-+\noiseRate}
\right)\right]\\
&= 3V \pairPot(\latticeConstant) +  d (V-1) \frac{\Gamma^2}{4} +
    \half \sum_\pvec^* 
\left(\frac{\activity^2}{\eval_\pvec^++\noiseRate}
+
\frac{\activity^2}{\eval_\pvec^-+\noiseRate} \right) \ ,
\end{split}
\end{equation}
where $\eval^\pm_{\pvec=\nullvec}=0$ in the full sum of \Eref{U_sum} removes the $\pvec=\nullvec$ term. The factor $d=2$ in \Eref{stationaryinternalEnergyHarmonic} is the size of the matrix $\dynamicalMatrix{\nvec}{\nvec'}$, so that $d (V-1)$ is the total number of excitable degrees of freedom. In equilibrium, by equipartition, each such degree of freedom is to carry energy $k_BT/2$, corresponding to $\Gamma^2/4$ according to \Eref{def_thermalNoise_App}, as $\Gamma^2$ corresponds to twice the diffusion constant or twice $k_BT$. The contribution $d (V-1)\Gamma^2/4$ to the harmonic energy in \Eref{stationaryinternalEnergyHarmonic} is thus in line with equilibrium thermodynamics.

A further sanity check of \Eref{stationaryinternalEnergyHarmonic} is obtained by taking the limit $\noiseRate\to\infty$ in a way that renders the active noise thermal, \Eref{activeNoise_correlator_App}. This is most easily implemented by taking $\activity^2=\GammaTilde\noiseRate/2$, as $\int\dint{t}\exp{-\noiseRate|t|}=2/\noiseRate$, \Eref{activeNoise_correlator_App}. In the limit $\noiseRate\to\infty$, all $d=2$ terms in the summand of the starred sum of \Eref{stationaryinternalEnergyHarmonic} then become $\GammaTilde/2$, resulting in  $d(V-1)\GammaTilde/4$ overall, in line with the preceding term.

In general, the activity contributes more to the energy of a mode $\pvec$ when $\eval_\pvec^\pm\ll\noiseRate$, which is the case for $\pvec$ close to $\nullvec$ or close to $(\NOne-1,\NTwo-1)^\transpose$. For any $\pvec$ with  $\eval_\pvec^\pm\ll\noiseRate$, the activity contributes $2\activity^2/\noiseRate$. The number of modes that have this property grows linearly in $V$ for $\NOne\propto \NTwo$, 
so that generally in large $V$ the expected total energy scales like $\ave{\internalEnergyHarmonic} \propto V (\Gamma^2/2 + \CC \activity^2/\noiseRate)$ with some constant $\CC$.

\subsection{Steady-state entropy production}\seclabel{Entropy}
Before calculating the entropy production of the active crystal, one final ingredient is the cross correlator,
\begin{equation}\elabel{CrossCorrelationDefn}
\ave{\activeNoiseFourierPM[\pm]_\pvec(t_1)\displacementFourier[\pm]{-\pvec}{t_2}} = \int_0^{t_2}\dint{t''}
\exp{-\eval_{-\pvec}^\pm (t_2-t'')}
\ave{\hat{\activeNoise}_{\pvec}^{\pm}(t_1)\hat{\activeNoise}_{-\pvec}^{\pm}(t'')}
=
\int_0^{t_2}\dint{t''}
V\exp{-\eval_\pvec^\pm (t_2-t'')}
\activity^2 
\exp{-\noiseRate|t_1-t''|}\ ,
\end{equation}
where we have inserted the solution for $\displacementFourier[\pm]{-\pvec}{t_2}$ from \Eref{displacement_by_integration}, and have used that thermal and active noise are independent, $\ave{\hat{\activeNoise}_{\pvec}^{\pm}(t_1) \thermalNoiseFourier_{\pvec'}^\pm(t')}=0$ as well as $\eval_{-\pvec}^\pm = \eval_{\pvec}^\pm$, \Erefs{def_oneMcos_general} and \eref{eval_formula}. The cross-correlator
\begin{equation}\elabel{CrossCorrelationDefn_Off}
\ave{\activeNoiseFourierPM[\pm]_\pvec(t_1)\displacementFourier[\mp]{-\pvec}{t_2}} = 0
\end{equation}
vanishes as the solution $\displacementFourier[\mp]{-\pvec}{t_2}$ draws only on $\activeNoiseFourierPM[\mp]_\pvec(t)$, whose correlator with $\activeNoiseFourierPM[\pm]_\pvec(t_1)$ vanishes, \Eref{activeNoiseFourier_correlator},  and on $\thermalNoiseFourier_{\pvec'}^\mp(t')$, which is independent anyway.

Performing the integration in \Eref{CrossCorrelationDefn}, we obtain the exact expression
\begin{equation}\elabel{CrossCorrelationExplicit}
\ave{\activeNoiseFourierPM[\pm]_\pvec(t_1)\displacementFourier[\pm]{-\pvec}{t_2}} = \begin{cases}
            V\frac{\activity^2 }{\eval_\pvec^\pm + \noiseRate}\left(e^{-\noiseRate(t_1 - t_2)} - e^{-\eval_\pvec^\pm t_2} e^{-\noiseRate t_1} \right), &~\text{for } t_1 \geq t_2, \\
            V\frac{\activity^2}{\eval_\pvec^\pm + \noiseRate}\left(e^{-\eval_\pvec^\pm(t_1 - t_2)} - e^{-\eval_\pvec^\pm t_2} e^{-\noiseRate t_1} \right) + V\frac{\activity^2}{\eval_\pvec^\pm - \noiseRate}\left(e^{-\noiseRate(t_2 - t_1)} - e^{-\eval_\pvec^\pm (t_2-t_1)} \right) , &~\text{for } t_2 \geq t_1,
        \end{cases}
\end{equation}
which does not diverge in large $t_1$ and $t_2$ when $\pvec = \zerovec$, in contrast to the $\ave{\displacementFourier[\pm]{\pvec}{t_1}\displacementFourier[\pm]{-\pvec}{t_2}}$ correlator, \Erefs{def_activityCorrInt_p0} and similarly \eref{def_thermalCorrInt_p0}.
In the equal-time steady-state limit $t_1=t_2\to\infty$, the cross-correlation, \Eref{CrossCorrelationExplicit}, becomes
\begin{equation}\elabel{CrossCorrelationSteadyState}
\lim_{t\to\infty}\ave{\activeNoiseFourierPM[\pm]_\pvec(t)\displacementFourier[\pm]{-\pvec}{t}} = V\frac{\activity^2}{\eval_\pvec^\pm + \noiseRate}\ .
\end{equation}
By comparing to the explicit expression for $\activityCorrInt_\pvec^\pm(t,t)$ at large times $t \gg 1/\eval_\pvec^\pm$ and $t\gg1/\noiseRate$, \Eref{def_activityCorrInt_Larget}, we may write 
\begin{equation}\elabel{CrossCorrelationSteadyState_ActivityIntegral}
\lim_{t\to\infty}\eval_\pvec^\pm\ave{\activeNoiseFourierPM[\pm]_\pvec(t)\displacementFourier[\pm]{-\pvec}{t}} 
= \begin{cases}
0, & \text{for } \pvec=\nullvec, \\
(\eval_\pvec^\pm)^2 \lim_{t\to\infty}V\activityCorrInt_\pvec^\pm(t,t) = 
V\frac{\activity^2 \eval_\pvec^\pm}{\eval_\pvec^\pm + \noiseRate}. \\
\end{cases}\end{equation}
which is the form the correlator features in the calculation of the steady-state entropy production below.

The entropy production of an active crystal is most easily determined using the framework outlined in Ref.~\cite{pruessner2025field}, which introduces a notion of instantaneous internal \emph{local} entropy-production rate in the steady-state limit, 
\begin{equation}\elabel{EPR_instantaneous}
\entropyProductionDensity_\nvec = 
-
\sum_{q=1}^6
\nabla_{\displacementVec{\nvec}{}}^2 \pairPot\big(|\displacementVec{\nvec}{}-\displacementVec{\nvec+\mvec_q}{}-\latticeConstant\latticeVector_q | \big)
+
\frac{2}{\Gamma^2}
\left(
\activeNoiseVec_\nvec - \sum_{q=1}^6
\nabla_{\displacementVec{\nvec}{}}\pairPot\big(|\displacementVec{\nvec}{}-\displacementVec{\nvec+\mvec_q}{}-\latticeConstant\latticeVector_q | \big)
\right)^2    
\ .
\end{equation}
so that the total steady-state entropy production per site is just $\ave{\entropyProductionDensity_\nvec}$, independent of $\nvec$.
In the harmonic approximation, the force from the neighbouring particles can be written as 
$\sum_q \nabla_{\displacementVec{\nvec}{}}\pairPot(|\displacementVec{\nvec}{} - \displacementVec{\nvec + \mvec_q}{} - \latticeConstant\latticeVector_q |) = \sum_{\nvec'}\dynamicalMatrix{\nvec}{\nvec'}\displacementVec{\nvec'}{}$,
\Erefs{ForceDmatrix} and \eref{def_dynamicalMatrix_HarmonicAppendix}. From this, it follows that $\sum_q \nabla_{\displacementVec{\nvec}{}}^2
\pairPot(|\displacementVec{\nvec}{} - \displacementVec{\nvec + \mvec_q}{} - \latticeConstant\latticeVector_q |) = 
\Trace\dynamicalMatrix{\nvec}{\nvec}$.
Replacing the gradient and Laplacian of the potential in \Eref{EPR_instantaneous} with these expressions, followed by expanding the brackets and taking the expectation of \Eref{EPR_instantaneous}, we arrive at 
\begin{equation}\elabel{EPR_derivation_step1}
\ave{\entropyProductionDensity_\nvec} = 
-
\Trace\dynamicalMatrix{\nvec}{\nvec}
+
\frac{2}{\Gamma^2}
\left(
\ave{\activeNoiseVec_\nvec \cdot \activeNoiseVec_\nvec} - \sum_{\nvec'} 2\ave{\activeNoiseVec_\nvec \cdot \dynamicalMatrix{\nvec}{\nvec'}\displacementVec{\nvec'}{} } + \sum_{\nvec'}\sum_{\nvec''}\ave{\big(\dynamicalMatrix{\nvec}{\nvec'}\displacementVec{\nvec'}{}\big) \cdot \big(\dynamicalMatrix{\nvec}{\nvec''}\displacementVec{\nvec''}{}\big)} \right) \ ,
\end{equation}
with all terms to be evaluated at equal times in the steady state.
The expectations on the right-hand side of \Eref{EPR_derivation_step1} are most efficiently evaluated in their Fourier representation, \Erefs{FourierFromDisplacement} and \eref{def_FourierD}. The calculation is further simplified by taking the sum over all sites on the left using translational invariance, resulting in the total entropy production
\begin{equation}
    \ave{\entropyProduction} = \sum_\nvec \ave{\entropyProductionDensity_\nvec}
    = 
-
\sum_\nvec \Trace \dynamicalMatrix{\nvec}{\nvec}
+    \frac{2}{V\Gamma^2}
    \sum_\pvec
\left(
\ave{\activeNoiseVecFourier_\pvec \cdot \activeNoiseVecFourier_{-\pvec}} - 2\ave{\activeNoiseVecFourier_\pvec \cdot \dynamicalMatrixFourier{-\pvec}{}\displacementVecFourier{-\pvec}{} } + \ave{\dynamicalMatrixFourier{\pvec}{}\displacementVecFourier{\pvec}{} \cdot \dynamicalMatrixFourier{-\pvec}{}\displacementVecFourier{-\pvec}{}} \right) \ ,
\end{equation}
where we have replaced $\dynamicalMatrix{\nvec}{\nvec}$ in \Eref{EPR_derivation_step1} by $\dynamicalMatrix{\nullvec}{\nullvec}$. The trace of $\dynamicalMatrix{\nvec}{\nvec}$ is immediately evaluated, \Eref{def_dynamicalMatrix_HarmonicAppendix}, but it is more instructive to use
instead
\begin{equation} 
    \sum_\nvec \Trace \dynamicalMatrix{\nvec}{\nvec} = 
    \sum_\pvec \Trace \dynamicalMatrixFourier{\pvec}{} =
    \starredsum_\pvec \Trace \dynamicalMatrixFourier{\pvec}{} =
    \starredsum_\pvec \left( \eval_\pvec^+ + \eval_\pvec^-\right) \ ,
\end{equation}
using the Fourier representation, \Eref{def_FourierD}. 
The expectation $\ave{\activeNoiseVecFourier_\pvec(t) \cdot \activeNoiseVecFourier_{-\pvec}(t)}=d\activity^2
V$ in the steady state follows immediately from \Eref{ActiveNoiseCorrelationFourier} with dimension $d=2$. The remaining two terms simplify using the representation \Eref{displacementP_by_compo} of $\displacementVecFourier{\pvec}{}$ in terms of the eigenvectors \Eref{def_evecs} of $\dynamicalMatrixFourier{\pvec}{}$. In the steady state,
\begin{equation}\elabel{EPR_term1}
    \ave{\activeNoiseVecFourier_\pvec \cdot \dynamicalMatrixFourier{-\pvec}{}\displacementVecFourier{-\pvec}{} }
    =
    V\activity^2\left(
    \frac{\eval_\pvec^+}{\eval_\pvec^+ + \noiseRate}
    +
    \frac{\eval_\pvec^-}{\eval_\pvec^- + \noiseRate}
\right)
\end{equation}
using \Erefs{CrossCorrelationDefn_Off} and \eref{CrossCorrelationSteadyState}.
Equation~\eref{EPR_term1} vanishes for $\pvec=\nullvec$ provided only $\noiseRate\ne0$.
Similarly, for $\pvec\ne\nullvec$,
\begin{equation}
    \lim_{t\to\infty}
    \ave{\dynamicalMatrixFourier{\pvec}{}\displacementVecFourier{\pvec}{} \cdot \dynamicalMatrixFourier{-\pvec}{}\displacementVecFourier{-\pvec}{}}
= 
(\eval_\pvec^+)^2 V \left( 
\frac{\activity^2}{\eval_\pvec^+ (\eval_\pvec^++\noiseRate)}
+
\frac{\Gamma^2}{2\eval_\pvec^+}
\right)
+
(\eval_\pvec^-)^2 V \left( 
\frac{\activity^2}{\eval_\pvec^- (\eval_\pvec^-+\noiseRate)}
+
\frac{\Gamma^2}{2\eval_\pvec^-}
\right)
\end{equation}
using \Erefs{DisplacementComponentCorrelator_all}, \eref{DisplacementComponentCorrelator}, \eref{def_activityCorrInt_Larget} and \eref{def_thermalCorrInt_Larget}. As $\dynamicalMatrixFourier{\pvec}{}\displacementVecFourier{\pvec}{}$ vanishes for $\pvec=\nullvec$ at any finite $t$, so does the limit. In summary,
\begin{equation}
    \ave{\entropyProduction} = \frac{2d\activity^2}{\Gamma^2}
    +
    \starredsum_\pvec \left\{
    -\eval_\pvec^+ - \eval_\pvec^-
+ \frac{2d\activity^2}{\Gamma^2}
- \frac{4 \activity^2}{\Gamma^2}
\left(
    \frac{\eval_\pvec^+}{\eval_\pvec^+ + \noiseRate}
    +
    \frac{\eval_\pvec^-}{\eval_\pvec^- + \noiseRate}
\right)
+
\frac{2}{\Gamma^2}
\left(
\frac{\activity^2 \eval_\pvec^+}{\eval_\pvec^++\noiseRate}
+
\frac{\Gamma^2 \eval_\pvec^+}{2}
+
\frac{\activity^2 \eval_\pvec^-}{\eval_\pvec^+-\noiseRate}
+
\frac{\Gamma^2 \eval_\pvec^-}{2}
\right)
    \right\} \ ,
\end{equation}
where the first term on the right-hand side is due to $\pvec=\nullvec$ not being included in the starred sum, which suits all terms except $\ave{\activeNoiseVecFourier_\pvec(t) \cdot \activeNoiseVecFourier_{-\pvec}(t)}=d\activity^2
V$, which does not vanish for $\pvec=\nullvec$. Simplifying along the obvious lines finally gives
\begin{equation}\elabel{entropyProduction_final}
\ave{\entropyProduction} = \frac{2d\activity^2}{\Gamma^2} +
\frac{2\activity^2}{\Gamma^2} \sum_\pvec^* \left\{
\frac{\noiseRate}{\eval_\pvec^++\noiseRate}
+
\frac{\noiseRate}{\eval_\pvec^-+\noiseRate}
\right\}
=
\frac{2\activity^2}{\Gamma^2} \sum_\pvec \left\{
\frac{\noiseRate}{\eval_\pvec^++\noiseRate}
+
\frac{\noiseRate}{\eval_\pvec^-+\noiseRate}
\right\}
\end{equation}
where we have used $d=2$ explicitly, as well as $\eval_\nullvec=0$. In higher dimensions, the increased $d$ is matched by an increased number of eigenvalues, which can always be paired in the suggested form, resulting in a sum over $1-\eval/(\eval+\noiseRate)=\noiseRate/(\eval+\noiseRate)$. 

Even in the stable case, $\eval_\pvec^\pm\ge0$, the total entropy production, \Eref{entropyProduction_final}, generally diverges as $V\to\infty$ on account of the summands being bounded from below, as $\oneMcos{q}{\pvec}$ in \Eref{eval_formula} are bounded from above, \Eref{def_oneMcos}. On the other hand, the entropy production per site $\ave{\entropyProduction}/V$ converges in large $V$. 

On the right-hand side of the first equality of \Eref{entropyProduction_final}, the first term, $2d\activity^2/\Gamma^2$ is the entropy production of a self-propelled particle in $d$ dimensions, with squared velocity amplitude $\activity^2/V$ in each dimension and subject to thermal diffusion $\Gamma^2/(2V)$, \Eref{var_rigid_mode} and discussion thereafter. This is the entropy production due to the $\nullvec$-mode, which displaces the lattice as a whole, as if it was a self-propelled Brownian particle  \cite{cocconi2020entropy}. The terms inside the starred sum of \Eref{entropyProduction_final}, on the other hand, can be seen as the entropy production of an active particle, here an active mode, in a harmonic potential with stiffness $\eval_\pvec^\pm$ subject to diffusion with constant $\Gamma^2/(2V)$ and squared self-propulsion speed $\activity^2/V$ that has correlation time $1/\noiseRate$. In the notation of Ref.~\cite{garcia2021run}, this means $\alpha=\noiseRate$, as the self-propulsion correlator decays like $\exp{-\alpha t}$ there, $w^2$ there corresponds to $\activity^2/V$ here, $D=\Gamma^2/(2V)$ and $k=\eval_\pvec^\pm$. This may not come as a surprise as \Eref{SDE_on_pm} is in fact the equation of motion of a single ``particle'' (a mode) in a harmonic potential with stiffness $\eval_\pvec^\pm$.

The derivation of \Eref{entropyProduction_final} concludes the present section. It demonstrates that the entropy production of interacting spatial degrees of freedom can be written as the sum over the entropy production rates of \emph{collective} spatial degrees of freedom provided they do not interact, in line with the findings of Ref.~\cite{pruessner2025field}. The similarity of the terms in the entropy production, \Eref{entropyProduction_final}, and the expected (harmonic) energy, \Eref{stationaryinternalEnergyHarmonic}, is striking and suggestive of a deeper relationship between the two.

\section{Higher-order terms in the expansion of \Eref{PairPotentialApproximation}}\seclabel{higher_order_terms}

The Taylor expansion of the pair potential introduced in \Eref{initial_Langevin} can be carried on from \Eref{PairPotentialApproximation}
\begin{align}\elabel{Taylor1}
    \pairPot\big(|\zvec+\latticeConstant\latticeVector_q|\big) 
    &= \pairPot(\latticeConstant) \\
    \nonumber & + \left.\big(\zvec\cdot\nabla_\rvec\big)\right|_{\rvec=\nullvec} \pairPot\big(|\rvec+\latticeConstant\latticeVector_q|\big)\\
    \nonumber & + \half \left.\big(\zvec\cdot\nabla_\rvec\big)^2\right|_{\rvec=\nullvec} \pairPot\big(|\rvec+\latticeConstant\latticeVector_q|\big)\\
    \nonumber & + \frac{1}{3!}\left.\big(\zvec\cdot\nabla_\rvec\big)^3\right|_{\rvec=\nullvec} \pairPot\big(|\rvec+\latticeConstant\latticeVector_q|\big)\\
    \nonumber & + \frac{1}{4!}\left.\big(\zvec\cdot\nabla_\rvec\big)^4\right|_{\rvec=\nullvec} \pairPot\big(|\rvec+\latticeConstant\latticeVector_q|\big)
    + \text{higher-order terms}
\end{align}
with $\zvec=\displacementVec{\nvec+\mvec_q}{t}-\displacementVec{\nvec}{t}$ in \Eref{initial_Langevin}.
The above may more succinctly be written as
\begin{align}\elabel{Taylor2}
    \pairPot\big(|\zvec+\latticeConstant\latticeVector_q|\big) 
    & = \pairPot(\latticeConstant) \\
    \nonumber & + \sum_i \left. z^i \partial^i \right|_{\rvec=\nullvec} \pairPot\big(|\rvec+\latticeConstant\latticeVector_q|\big)\\
    \nonumber & + \half \sum_{ij} \left. z^iz^j \partial^i \partial^j \right|_{\rvec=\nullvec} \pairPot\big(|\rvec+\latticeConstant\latticeVector_q|\big)\\
    \nonumber & + \frac{1}{3!}\sum_{ijk} \left. z^iz^jz^k \partial^i \partial^j \partial^k \right|_{\rvec=\nullvec} \pairPot\big(|\rvec+\latticeConstant\latticeVector_q|\big)\\
    \nonumber & + \frac{1}{4!}\sum_{ijkm} \left. z^iz^jz^kz^m \partial^i \partial^j \partial^k \partial^m\right|_{\rvec=\nullvec} \pairPot\big(|\rvec+\latticeConstant\latticeVector_q|\big)
    + \text{higher-order terms}
\end{align}
where $\partial^i$ denotes the partial derivative with respect to $r^i$, which is the $i$th of $d$ components of $\rvec$. The first-order term is
\begin{equation}\elabel{first_derivative}
    \partial^i \pairPot\big(|\rvec+\latticeConstant\latticeVector_q|\big) 
    = (r^i+\latticeConstant\latticeVectorCompo^i_q)
    \frac{\pairPot'\big(|\rvec+\latticeConstant\latticeVector_q|\big)}{|\rvec+\latticeConstant\latticeVector_q|} \ .
\end{equation}
Similarly to the notation above, $\latticeVectorCompo^i_q$ is the $i$th component of $\latticeVector_q$. Dashes on the pair potential $\pairPot$ denotes its derivatives with respect to its scalar argument, \ie
\begin{equation}
    \pairPot'(z) = \frac{\plaind }{\plaind z} \pairPot(z) \ .
\end{equation}
The higher-order terms beyond first order in \Eref{Taylor2} are 
\begin{align}\elabel{second_order_full}
    \partial^i \partial^j \pairPot\big(|\rvec+\latticeConstant\latticeVector_q|\big) 
    &= 
    \delta^{ij}\ 
     \pairPot'\big(|\rvec+\latticeConstant\latticeVector_q|\big)\\
     \nonumber
     &+
     (r^i+\latticeConstant\latticeVectorCompo_q^i)
     (r^j+\latticeConstant\latticeVectorCompo_q^j)
     \left(
     - \frac{\pairPot'\big(|\rvec+\latticeConstant\latticeVector_q|\big)}{|\rvec+\latticeConstant\latticeVector_q|^3}
     + \frac{\pairPot''\big(|\rvec+\latticeConstant\latticeVector_q|\big)}{|\rvec+\latticeConstant\latticeVector_q|^2}
     \right)
\end{align}
and
\newcommand{\rla}[1]{(r^#1+\latticeConstant\latticeVectorCompo_q^#1)}
\begin{align}\elabel{third_order_full}
    \partial^i \partial^j \partial^k \pairPot\big(|\rvec+\latticeConstant\latticeVector_q|\big) 
    &= 
    \Big( \delta^{ij} \rla{k} + \delta^{jk} \rla{i} + \delta^{ki} \rla{j} \Big)\ 
     \left(
     - \frac{\pairPot'\big(|\rvec+\latticeConstant\latticeVector_q|\big)}{|\rvec+\latticeConstant\latticeVector_q|^3}
     + \frac{\pairPot''\big(|\rvec+\latticeConstant\latticeVector_q|\big)}{|\rvec+\latticeConstant\latticeVector_q|^2}
     \right)\\
     \nonumber
     &+\rla{i}\rla{j}\rla{k}
     \left(
     3 \frac{\pairPot'\big(|\rvec+\latticeConstant\latticeVector_q|\big)}{|\rvec+\latticeConstant\latticeVector_q|^5}
     -3 \frac{\pairPot''\big(|\rvec+\latticeConstant\latticeVector_q|\big)}{|\rvec+\latticeConstant\latticeVector_q|^4}
     + \frac{\pairPot'''\big(|\rvec+\latticeConstant\latticeVector_q|\big)}{|\rvec+\latticeConstant\latticeVector_q|^3}
     \right)
\end{align}
and
\begin{align}\elabel{fourth_order_full}
    \partial^i \partial^j \partial^k \partial^m \pairPot\big(|\rvec+\latticeConstant\latticeVector_q|\big) 
    &= 
    \Big( \delta^{ij} \delta^{km} + \delta^{jk} \delta^{im} + \delta^{ki} \delta^{jm} \Big)\ 
     \left(
     - \frac{\pairPot'\big(|\rvec+\latticeConstant\latticeVector_q|\big)}{|\rvec+\latticeConstant\latticeVector_q|^3}
     + \frac{\pairPot''\big(|\rvec+\latticeConstant\latticeVector_q|\big)}{|\rvec+\latticeConstant\latticeVector_q|^2}
     \right)\\
     \nonumber
    &+ \bigg(\  
      \delta^{ij} \rla{k}\rla{m} 
    + \delta^{km} \rla{i}\rla{j} \\
    \nonumber \quad
    &\ \ \;+ \delta^{jk} \rla{i}\rla{m} 
    + \delta^{im} \rla{j}\rla{k} \\
    \nonumber \quad
    &\ \ \;+ \delta^{ki} \rla{j}\rla{m} 
    + \delta^{jm} \rla{k}\rla{i}  \ \bigg) \\
    \nonumber & \qquad \times
     \left(
     3 \frac{\pairPot'\big(|\rvec+\latticeConstant\latticeVector_q|\big)}{|\rvec+\latticeConstant\latticeVector_q|^5}
     -3 \frac{\pairPot''\big(|\rvec+\latticeConstant\latticeVector_q|\big)}{|\rvec+\latticeConstant\latticeVector_q|^4}
     + \frac{\pairPot'''\big(|\rvec+\latticeConstant\latticeVector_q|\big)}{|\rvec+\latticeConstant\latticeVector_q|^3}
     \right)\\
     \nonumber
     &+\rla{i}\rla{j}\rla{k}\rla{m}\\
     \nonumber & \qquad \times
     \left(
     -15 \frac{\pairPot'\big(|\rvec+\latticeConstant\latticeVector_q|\big)}{|\rvec+\latticeConstant\latticeVector_q|^7}
     +15 \frac{\pairPot''\big(|\rvec+\latticeConstant\latticeVector_q|\big)}{|\rvec+\latticeConstant\latticeVector_q|^6}
     -6 \frac{\pairPot'''\big(|\rvec+\latticeConstant\latticeVector_q|\big)}{|\rvec+\latticeConstant\latticeVector_q|^5}
     + \frac{\pairPot^{(4)}\big(|\rvec+\latticeConstant\latticeVector_q|\big)}{|\rvec+\latticeConstant\latticeVector_q|^4}
     \right)\ .
\end{align}
Of course, in \Erefs{Taylor1} and \eref{Taylor2} these expressions are evaluated at $\rvec=\nullvec$, which renders them much more compact. 

To proceed, we consider the total energy, \Eref{initial_energy}, of the system given a set of displacements $\displacementVec{\nvec}{t}$. The expansion in \Eref{Taylor2} will therefore be summed over $q$ and $\nvec$.

Summing over $q$ simplifies some of the expressions above, for example
\begin{equation}\elabel{simple1}
    \sum_q^Q \latticeConstant\latticeVectorCompo_q^i = 0
\end{equation}
for any component $i$, as 
\begin{equation}
    \latticeVectorCompo_q^i
    =
    -\latticeVectorCompo_{q+Q/2}^i
\end{equation}
for $q\in\{1,\ldots,Q/2\}$. For the same reason,
\begin{equation}\elabel{simple2}
    \sum_q^Q 
    \latticeVectorCompo_q^i 
    \latticeVectorCompo_q^j
    \latticeVectorCompo_q^k 
    = 0 \ ,
\end{equation}
as shifting $q$ by $Q/2$ changes the sign of the summand for any $i,j,k$.

To see the effect of \Eref{simple1}, we  consider the total energy \Eref{initial_energy} to first order,
\begin{equation}
    \internalEnergy(t) = \half \sum_{\nvec}\sum_{q=1}^Q
    \left(
    \pairPot(\latticeConstant)
    +
    \sum_i 
    (\displacement[i]{\nvec+\mvec_q}{}-\displacement[i]{\nvec}{})
\latticeVectorCompo^i_q\pairPot'(\latticeConstant)
    \right)
\end{equation}
where we have dropped the time-dependence of the vector components $\displacement[i]{\nvec}{}$. At first sight, \Eref{simple1} seems to be of limited use, as it seems to merely entail vanishing $\sum_{\nvec}\sum_{q=1}^Q \displacement[i]{\nvec}{} \latticeVectorCompo^i_q$, but in general not vanishing $\sum_{\nvec}\sum_{q=1}^Q \displacement[i]{\nvec+\mvec_q}{} \latticeVectorCompo^i_q$. However, by periodicity of $\nvec$, the summation index can be shifted in a $q$-dependent manner
\begin{equation}\elabel{first_order_vanishes}
    \sum_{\nvec}\sum_{q=1}^Q
\displacement[i]{\nvec+\mvec_q}{}
\latticeVectorCompo^i_q
=
    \sum_{q=1}^Q\sum_{\nvec}
\displacement[i]{\nvec}{}
\latticeVectorCompo^i_q
=
    \sum_{\nvec}
    \sum_{q=1}^Q
\displacement[i]{\nvec}{}
\latticeVectorCompo^i_q
=0  
\end{equation}
so that the linear order drops out entirely, producing $\internalEnergy(t) = (VQ/2)\pairPot(\latticeConstant)$ to this order.

In the following, we express the internal energy, \Eref{initial_energy}, in a perturbation theory to fourth order using \Eref{Taylor2}.
To ease notation, we drop the time-dependence and write $\pairPot$ for $\pairPot(\latticeConstant)$ and similarly $\pairPot'$ for $\pairPot'(\latticeConstant)$ \etc We now list the contributions to the total energy as stated in \Eref{Taylor2}. For future reference, we further write
\begin{equation}\elabel{internalEnergy_with_1234}
    \internalEnergy = \internalEnergy_0 + \internalEnergy_1 + \internalEnergy_2 + \internalEnergy_3 + \internalEnergy_4 + \ldots
\end{equation}
with the following contributions to the energy at each order:
\newcommand{\membraneTensor}[2]{\mathcal{T}^{#2}_{#1}}

\noindent \underline{Contribution to the energy from $0$th order at $\rvec=\nullvec$}
\begin{equation}\elabel{iE_0}
    \internalEnergy_0=\half\sum_\nvec\sum_q \pairPot(\latticeConstant) =
    \frac{VQ}{2}\pairPot(\latticeConstant)\ .
\end{equation}

\noindent \underline{Contribution to the energy from $1$st order at $\rvec = \nullvec$}\newline
Equation~\eref{first_derivative} evaluated at $\rvec=\nullvec$ produces $\latticeVectorCompo^i_q \pairPot'$, which vanishes under summation, \ie
\begin{equation}\elabel{iE_1}
\internalEnergy_1=0
\end{equation}
as seen above, \Eref{first_order_vanishes}.

\noindent \underline{Contribution to the energy from $2$nd order at $\rvec = \nullvec$}\newline
Equation~\eref{second_order_full} evaluated at $\rvec=\nullvec$ gives
\begin{equation}\elabel{mT_2}
    \membraneTensor{2}{ij}:=\left.\partial^i \partial^j\right|_{\rvec=\nullvec}\pairPot\big(|\rvec+\latticeConstant\latticeVector_q|\big) 
=
\delta^{ij}\frac{\pairPot'}{\latticeConstant} + \latticeVectorCompo^i_q\latticeVectorCompo^j_q
\left(-\frac{\pairPot'}{\latticeConstant}+\pairPot''\right) \ ,
\end{equation}
where we have introduced $\membraneTensor{2}{ij}$ to simplify notation below.
The contribution to the energy $\internalEnergy$ to second order thus becomes
\begin{equation}
    \internalEnergy_2=\half\sum_\nvec\sum_q
    \half\sum_{ij}
    (\displacement[i]{\nvec+\mvec_q}{}-\displacement[i]{\nvec}{})
    (\displacement[j]{\nvec+\mvec_q}{}-\displacement[j]{\nvec}{})
    \left(
    \delta^{ij}\frac{\pairPot'}{\latticeConstant} + \latticeVectorCompo^i_q\latticeVectorCompo^j_q
\left(-\frac{\pairPot'}{\latticeConstant}+\pairPot''\right)
\right) \ ,
\end{equation}
which produces
\begin{equation}
    \internalEnergy_2=
    \half\sum_\nvec\sum_q\Bigg\{
    (\displacementVec{\nvec}{}-\displacementVec{\nvec+\mvec_q}{})\cdot\displacementVec{\nvec}{}\frac{\pairPot'}{\latticeConstant}
    +
    (\displacementVec{\nvec}{}-\displacementVec{\nvec+\mvec_q}{})\cdot\latticeVector_q \displacementVec{\nvec}{}\cdot\latticeVector_q \left(\pairPot''-\frac{\pairPot'}{\latticeConstant}\right)
    \Bigg\} \ ,
\end{equation}
and with the help of \Eref{sum_aqaq}, can be further simplified to \Erefs{def_internalEnergyHarmonic} and \eref{def_dynamicalMatrix_HarmonicAppendix},
\begin{equation}\elabel{iE_2}
    \internalEnergy_2=
    \frac{3}{2}\left(\pairPot''+\frac{\pairPot'}{\latticeConstant}\right)\sum_\nvec\displacementVec{\nvec}{}\cdot\displacementVec{\nvec}{}
    -\half\frac{\pairPot'}{\latticeConstant}\sum_\nvec\sum_q\displacementVec{\nvec}{}\cdot\displacementVec{\nvec+\mvec_q}{}
    -\half\left(\pairPot''-\frac{\pairPot'}{\latticeConstant}\right)\sum_\nvec\sum_q(\displacementVec{\nvec}{}\cdot\latticeVector_q)(\displacementVec{\nvec+\mvec_q}{}\cdot\latticeVector_q) \ .
\end{equation}

\noindent \underline{Contribution to the energy from $3$rd order at $\rvec = \nullvec$}\\
Equation~\eref{third_order_full} evaluated at $\rvec=\nullvec$ gives
\begin{equation}\elabel{mT_3}
    \membraneTensor{3}{ijk}:=\left.\partial^i \partial^j \partial^k\right|_{\rvec=\nullvec}\pairPot\big(|\rvec+\latticeConstant\latticeVector_q|\big) 
= 
\big(
\delta^{ij}\latticeVectorCompo^k_q
+
\delta^{jk}\latticeVectorCompo^i_q
+
\delta^{ki}\latticeVectorCompo^j_q
\big)
\left(
-\frac{\pairPot'}{\latticeConstant^2}
+\frac{\pairPot''}{\latticeConstant}
\right)
+\latticeVectorCompo^i_q\latticeVectorCompo^j_q\latticeVectorCompo^k_q
\left(
3\frac{\pairPot'}{\latticeConstant^2}
-3\frac{\pairPot''}{\latticeConstant}
+\pairPot'''
\right) \ ,
\end{equation}
which needs to be used in 
\begin{equation}
    \internalEnergy_3=\half\sum_\nvec\sum_q
    \frac{1}{6}\sum_{ijk}
    (\displacement[i]{\nvec+\mvec_q}{}-\displacement[i]{\nvec}{})
    (\displacement[j]{\nvec+\mvec_q}{}-\displacement[j]{\nvec}{})
    (\displacement[k]{\nvec+\mvec_q}{}-\displacement[k]{\nvec}{})
    \left.\partial^i \partial^j \partial^k\right|_{\rvec=\nullvec}\pairPot\big(|\rvec+\latticeConstant\latticeVector_q|\big)
\ .
\end{equation}
This can be simplified using identities such as
\begin{equation}
    \sum_\nvec\sum_q
    (\displacementVec{\nvec+\mvec_q}{}\cdot\displacementVec{\nvec+\mvec_q}{})(\displacementVec{\nvec+\mvec_q}{}\cdot\latticeVector_q)=
    \sum_\nvec\sum_q
    (\displacementVec{\nvec}{}\cdot\displacementVec{\nvec}{})(\displacementVec{\nvec}{}\cdot\latticeVector_q)
\end{equation}
and
\begin{equation}
    \sum_\nvec\sum_q
    (\displacementVec{\nvec}{}\cdot\displacementVec{\nvec}{})(\displacementVec{\nvec+\mvec_q}{}\cdot\latticeVector_q)=
    \sum_{\nvec'}\sum_q
    (\displacementVec{\nvec'-\mvec_q}{}\cdot\displacementVec{\nvec'-\mvec_q}{})(\displacementVec{\nvec'}{}\cdot\latticeVector_q)=
    -\sum_{\nvec'}\sum_{q'}
    (\displacementVec{\nvec'+\mvec_{q'}}{}\cdot\displacementVec{\nvec'+\mvec_{q'}}{})(\displacementVec{\nvec'}{}\cdot\latticeVector_{q'})
\end{equation}
where $\nvec'=\nvec+\mvec_q$ and $q'=q+Q/2$ with suitable periodic conditions applied, say $q'=7$ corresponding to $q'=1$, so that $\latticeVector_{q'}=-\latticeVector_q$ and $\mvec_{q'}=-\mvec_q$. It follows that
\begin{equation}\elabel{iE_3}
    \internalEnergy_3=
    \half\sum_\nvec\sum_q
    (\displacementVec{\nvec}{}\cdot\latticeVector_q)
    \left\{
    (2\displacementVec{\nvec}{}-\displacementVec{\nvec+\mvec_q}{})\cdot\displacementVec{\nvec+\mvec_q}{}
    \left(-\frac{\pairPot'}{\latticeConstant^2}+\frac{\pairPot''}{\latticeConstant} \right)
    -(\displacementVec{\nvec+\mvec_q}{}\cdot\latticeVector_q)^2\left(3\frac{\pairPot'}{\latticeConstant^2}-3\frac{\pairPot''}{\latticeConstant}+\pairPot'''\right)
    \right\}\ .
\end{equation}
It might surprise that this term, which is odd in $\displacementVec{\nvec}{}$, does not vanish. The change in sign of $\internalEnergy_3$ under $\{\displacementVec{\nvec}{}\}\to\{-\displacementVec{\nvec}{}\}$ seems to contradict the expected invariance of physics under mirror or inversion symmetry operations. However, $\displacementVec{\nvec}{}$ describes the \emph{displacement} of a particle away from its regular lattice site. Even in one dimension, changing the sign of $\displacement{n}{}$ results in a very different physical setup, say for just three sites, $\displacement{1}{}=0$, $\displacement{2}{}=(1/3)\latticeConstant$ and $\displacement{3}{}=\latticeConstant$ means that the distances between particles $1$ and $2$ is $(4/3)\latticeConstant$, between $2$ and $3$ is $(5/3)\latticeConstant$ and between $3$ and $1$ is $0$, whereas inverting the sign results in distances $(2/3)\latticeConstant$, $(1/3)\latticeConstant$ and $2\latticeConstant$. While the sum of the distances is indeed the same, consistent with $\internalEnergy_1$ vanishing, the sum of their cubes is not, consistent with $\internalEnergy_3\ne0$.

\newcommand{\delaa}[3]{\delta^{#1}\latticeVectorCompo^{#2}_q\latticeVectorCompo^{#3}_q}

\bigskip
\noindent \underline{Contribution to the energy from $4$th order at $\rvec = \nullvec$}\newline
Equation~\eref{fourth_order_full} evaluated at $\rvec=\nullvec$ gives
\begin{multline}\elabel{mT_4}
    \membraneTensor{4}{ijkm}:=\left.\partial^i\partial^j\partial^k\partial^m\right|_{\rvec=\nullvec}\pairPot\big(|\rvec+\latticeConstant\latticeVector_q|\big) 
=\big(
 \delta^{ij}\delta^{km}
+\delta^{jk}\delta^{im}
+\delta^{ki}\delta^{jm}
\big)
\left(
-\frac{\pairPot'}{\latticeConstant^3}
+\frac{\pairPot''}{\latticeConstant^2}
\right)\\
+
\big(
 \delaa{ij}{k}{m}+\delaa{km}{i}{j}
+\delaa{jk}{i}{m}+\delaa{im}{j}{k}
+\delaa{ki}{j}{m}+\delaa{jm}{k}{i}
\big)
\left(
3\frac{\pairPot'}{\latticeConstant^3}
-3\frac{\pairPot''}{\latticeConstant^2}
+\frac{\pairPot'''}{\latticeConstant}
\right)\\
+
\latticeVectorCompo^i_q\latticeVectorCompo^j_q\latticeVectorCompo^k_q\latticeVectorCompo^m_q
\left(
-15 \frac{\pairPot'}{\latticeConstant^3}
+15 \frac{\pairPot''}{\latticeConstant^2}
-6  \frac{\pairPot'''}{\latticeConstant}
+\pairPot^{(4)}
\right) \ .
\end{multline}
Performing the summation then produces the contribution
\begin{multline}\elabel{iE_4}
    \internalEnergy_4=
    \half\sum_\nvec\sum_q\Bigg\{
    \half \bigg(
    (\displacementVec{\nvec}{}-\displacementVec{\nvec+\mvec_q}{})\cdot\displacementVec{\nvec}{}
    \bigg)^2
    \left(
    -\frac{\pairPot'}{\latticeConstant^3}
    +\frac{\pairPot''}{\latticeConstant^2}
    \right)
    +\quarter 
    |\displacementVec{\nvec}{}-\displacementVec{\nvec+\mvec_q}|^2
    |(\displacementVec{\nvec}{}-\displacementVec{\nvec+\mvec_q}{})\cdot\latticeVector_q|^2
    \left(
    3\frac{\pairPot'}{\latticeConstant^3}
    -3\frac{\pairPot''}{\latticeConstant^2}
    +\frac{\pairPot'''}{\latticeConstant}
    \right)
    \\
    +\frac{1}{12}
    \left[
    (\displacementVec{\nvec}{}\cdot\latticeVector_q)^4 - 4 (\displacementVec{\nvec}{}\cdot\latticeVector_q)^3 (\displacementVec{\nvec+\mvec_q}{}\cdot\latticeVector_q) + 3 (\displacementVec{\nvec}{}\cdot\latticeVector_q)^2(\displacementVec{\nvec+\mvec_q}{}\cdot\latticeVector_q)^2
    \right]
    \left(
    -15 \frac{\pairPot'}{\latticeConstant^3}
    +15 \frac{\pairPot''}{\latticeConstant^2}
    -6  \frac{\pairPot'''}{\latticeConstant}
    +\pairPot^{(4)}
    \right)
    \Bigg\} \ .
\end{multline}

The internal energy $\internalEnergy$ to fourth order in $\displacementVec{\nvec}{}$ is thus given by \Eref{internalEnergy_with_1234} using \Erefs{iE_0}, 
\eref{iE_1}, 
\eref{iE_2}, 
\eref{iE_3}
and
\eref{iE_4}. To make the connection to the existing literature, \Sref{higher_order_modulus}, writing the internal energy \Eref{initial_energy} in the form \Eref{Taylor2}, but with the help of \Erefs{mT_2}, \eref{mT_3} and \eref{mT_4},
\begin{equation}
\internalEnergy = 
    \frac{VQ}{2}\pairPot(\latticeConstant)
+ \half\sum_\nvec\sum_q \left\{
 \half          \sum_{ij}   z_{\nvec,q}^i z_{\nvec,q}^j                             \membraneTensor{2}{ij}
+\frac{1}{6}    \sum_{ijk}  z_{\nvec,q}^i z_{\nvec,q}^j z_{\nvec,q}^k               \membraneTensor{3}{ijk}
+\frac{1}{24}   \sum_{ijkm} z_{\nvec,q}^i z_{\nvec,q}^j z_{\nvec,q}^k z_{\nvec,q}^m \membraneTensor{4}{ijkm}
+\ldots
\right\}
\end{equation}
with $\zvec_{\nvec,q}=\displacementVec{\nvec+\mvec_q}{}-\displacementVec{\nvec}{}$
is more telling, because in this form, $\zvec_{\nvec,q}$ is the lattice gradient of $\displacementVec{\nvec}{}$ in the $\latticeVector_q$-direction. In the continuum, $\zvec_{\nvec,q}$ becomes a gradient field and, depending on the order the expansion is written to, produces a Gaussian or a quartic field theory.

\subsection{Limitations of the harmonic approximation and connection to \texorpdfstring{$\varphi^4$}{phi4} theory}
\seclabel{higher_order_modulus}
The shift of entire rows of particles by an infinite amount at the buckling instability as of \Sref{buckling_transition} is clearly unphysical. As illustrated in \Frefs{TriangularLatticeBuckling} and \fref{U1_6}, in the harmonic approximation, particles under external pressure may seem to comply by slipping down an upside-down infinitely extended harmonic potential, while even just a harmonic pair potential $\pairPot(|\ldots|)$ would prevent this. In principle, the correct behaviour is captured by the theory if $\pairPot(|\ldots|)$ is expanded to high-enough order. However, this is only ``in principle'' because it is not guaranteed. Specifically, the expansion is about small displacements compared to $\latticeConstant$, so the behaviour in large displacements, in particular around any ``remote singularity'' due to a modulus, is not captured correctly, and how well the expansion is doing depends on the magnitude of the terms included and omitted. In fact, exploring this with computer algebra
indicates the in-file mode that is unstable in the harmonic approximation becomes stable at sixth order, but still leaves some directions unstable, say $12$ o'clock, as well as $2$, $4$ \etc

Assuming the validity of the expansion, it is natural to expect some universality at the transition from the stable lattice --- whose constituent particles fluctuate about the sites of the regular lattice structure --- to the buckled lattice. The relevant model appears to be that of the crumpling membrane \cite{PaczuskiKardarNelson:1988}, possibly with disorder \cite{DelzescauxMouhannaTissier:2024}. In this field theory, the field is a gradient --- indeed like it is here: $\displacementVec{\nvec+\mvec_q}{t}-\displacementVec{\nvec}{t} \ \corresponding \ \nabla\phivec(\xvec_\nvec,t)$. 

To render the present phenomenon a $\varphi^4$-theory, we require displacement $\displacementVec{\nvec}{t}$ and not just the difference in displacement to enter into the action. In other words, particles would need to be anchored at lattice sites. This is clearly not happening in clusters in motility-induced phase separation \cite{RednerHaganBaskaran:2013, cates2015motility}, which float freely like rafts. 

In simulations of active particles undergoing motility-induced phase separation, however, some features of the instability can be observed, as the constituents of clusters seem to chew and churn, with several particles in single file moving in unison at once, before coming to rest \cite{pruessner2025field}. 
This behaviour is incompatible with any potential that simply curtails large excursions and eventually pulls particles back to the origin. It challenges another approximation, namely that of fixed adjacency. Specifically, we have assumed throughout that particles maintain their neighbourhood. However, in (computer) experiments, particles interact with whoever they are in contact with or, more generally, with all particles with the interaction strength depending on the pair distance. In the present work, the set of \emph{nearest} neighbours is fixed, as if springs were attached between interaction partners. Handling changing interaction matrices is a matter of characterising Euclidean random matrices and well beyond the scope of the present work.

\section{Harmonic approximation of the one-dimensional active crystal}\seclabel{one_dimension}

\begin{figure}
    \centering
   \includegraphics[width=0.8\columnwidth, trim = 3.8cm 5cm 2.1cm 6.2cm, clip]{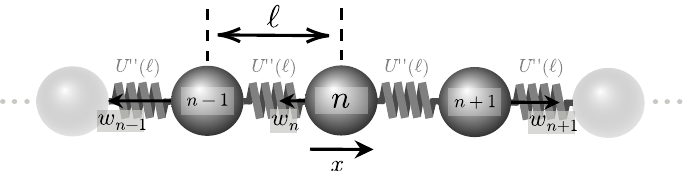}
    \caption{Schematic of the one-dimensional harmonic active chain.}
    \label{fig:ActiveChain}
\end{figure}

Analogous to \Sref{triangular_active_crystal}, in the following we will perform the derivation for the case of a one-dimensional harmonic active crystal, \ie a harmonic active chain, as illustrated in Fig.~\ref{fig:ActiveChain}. To limit repetition, we will omit much of the detail in the following since most of the mathematical machinery was already introduced for the two-dimensional case.

Under the harmonic approximation, \Sref{triangular_active_crystal}, the stochastic equation of motion is
\begin{equation}\elabel{Langevin1D}
    \displacementDot{n}{t} = \thermalNoise_n(t) + \activeNoise_n(t) - \sum_{n'=1}^{N} \dynamicalScalar{n}{n'} \displacement{n'}{t}
    \ ,
\end{equation}
where $\displacement{n}{t} \in \Rset$ represents the local displacement of the particle at a site indexed by $n \in\{1,2,\ldots,N \}$. As for the two-dimensional crystal, the zero-mean thermal and active noise have autocorrelation functions given by, respectively,
\begin{subequations}
\begin{align}
    \ave{\thermalNoise_\nvec(t) \thermalNoise_{\nvec'}(t')} &= \Gamma^2 \delta_{n,n'} \delta(t-t') \elabel{thermalNoise1D} \ ,\\
    \ave{\activeNoise_n(t) \activeNoise_{n'}(t')} &= \activity^2 \delta_{n,n'} \exp{-\noiseRate|t-t'|} \elabel{activeNoise1D} \ .
\end{align}
\end{subequations}
The sum $\sum_{n'}$ in \Eref{Langevin1D} is over the set $\{1,2,\ldots,N \}$ of all sites. However, the dynamical interaction kernel (dynamical matrix) selects only those $n'$ that are nearest neighbours of $n$, explicitly,
\begin{equation}\elabel{dynamicalScalar1D}
    \dynamicalScalar{n}{n'} = 
	\begin{cases}
	-
	\pairPot''(\latticeConstant),
	&\quad \text{for any $n'$ such that there is a $q\in\{1,2\}$ 
    so that $n'=n+m_q$,}\\
	2 \pairPot''(\latticeConstant),		
	&\quad \text{for $n'=n$,}
    \\
	0, & \quad\text{otherwise},
    \end{cases}
\end{equation}
\cf in two dimensions \Eref{def_dynamicalMatrix_HarmonicAppendix},
based on the harmonic approximation 
\begin{equation}\elabel{1DU_harmonic}
    \pairPot(|z-\latticeConstant|) \simeq \pairPot(\latticeConstant) - z \pairPot'(\latticeConstant)+\half z^2 \pairPot''(\latticeConstant)
\end{equation}
in \Sref{triangular_active_crystal}, where the set of allowed differences in the indices is now simply $\{m_1,m_2\} = \{1,-1\}$, with an associated set of displacements $\{\latticeConstant, -\latticeConstant\}$. 

Equation~\eref{dynamicalScalar1D} is exact if the pair potential is bilinear in the displacements $\displacement{n}{}$, say $\half\springConstant(\displacement{n}{}-\displacement{n+1}{}-\latticeConstant-\relaxedLength)^2$, but a \emph{harmonic approximation} if it is a function of the \emph{distance}, $\half\springConstant(|\displacement{n}{}-\displacement{n+1}{}-\latticeConstant|-\relaxedLength)^2$, which is fully captured by the harmonic approximation only when all differences of displacement are bounded by the lattice spacing, $|\displacement{n}{}-\displacement{n+1}{}|\le\latticeConstant$. Strikingly, the first derivative of the pair potential does not feature at all in \Eref{dynamicalScalar1D}. In the present approximation, the relaxed length $\relaxedLength$ of the potential is essentially irrelevant to the dynamics and any external pressure does not affect it.

As in the two-dimensional case, we solve the system by first introducing the Fourier transforms for the displacements,
\begin{subequations}
\begin{align}
\elabel{FourierDisplacement1D}
\displacementFourier{p}{t} &= \sum_{n=1}^{N} \exp{-\imag \wavenumber{p} n} \displacement{n}{t} \ ,\\
\elabel{InverseFourierDisplacement1D}
\displacement{n}{t} &= \frac{1}{N} \sum_{p=0}^{N-1} \exp{\imag \wavenumber{p}n} \displacementFourier{p}{t} \ ,
\end{align}
\end{subequations}
the dynamical interaction kernel,
\begin{subequations}
\begin{align}
\dynamicalScalarFourier{p}{p'} &= \sum_{n=1}^{N} \sum_{n'=1}^{N} \exp{-\imag (\wavenumber{p}n+\wavenumber{p'}n')} \dynamicalScalar{n}{n'} \ ,\\
\dynamicalScalar{n}{n'} &= \frac{1}{N} \sum_{p=0}^{N-1} \frac{1}{N} \sum_{p'=0}^{N-1} \exp{\imag (\wavenumber{p}n+\wavenumber{p'}n')} \dynamicalScalarFourier{p}{p'} \ ,
\end{align}
\end{subequations}
and the autocorrelation functions of the noises,
\begin{subequations}
\begin{align}
\ave{\thermalNoiseFourier_p(t)\thermalNoiseFourier_{p'}(t')} &= \Gamma^2
N \delta_{p+p',0}\delta(t-t') \ , \elabel{ThermalNoiseCorrelationFourier1D}\\
\ave{\activeNoiseFourier_p(t)\activeNoiseFourier_{p'}(t')} &= \activity^2
N \delta_{p+p',0} 
\exp{-\noiseRate|t-t'|} \  ,\elabel{ActiveNoiseCorrelationFourier1D}
\end{align}
\end{subequations}
where $k_p = 2\pi p/N$ is the wavenumber indexed by $p \in\{0,1,\ldots,N-1 \}$. After taking the above Fourier transforms, the Langevin \Eref{Langevin1D} simplifies to
\begin{equation}\elabel{LangevinFourier1D}
	\displacementDotFourier{p}{t} = - \dynamicalScalarFourier{p}{}\displacementFourier{p}{t} + \thermalNoiseFourier_{p}(t) + \activeNoiseFourier_{p}(t) \ ,
\end{equation}
where we introduced
\begin{equation}\elabel{DynamicalFourierEigenvalues1D}
\dynamicalScalarFourier{p}{p'} = N\delta_{p+p',0}\dynamicalScalarFourier{p}{} = N\delta_{p+p',0}2\pairPot''(\latticeConstant)[1-\cos(\wavenumber{p})] \ .
\end{equation}
The solution to \Eref{LangevinFourier1D} is thus
\begin{equation}\elabel{LangevinSolution1D}
\displacementFourier{p}{t} = \int_0^t \dint{t'}
\exp{- \dynamicalScalarFourier{p}{} (t-t')}
\Big\{
 \thermalNoiseFourier_p(t')
 + \activeNoiseFourier_p(t')
\Big\} \ ,
\end{equation}
with initial condition $\displacementFourier{p}{t=0}=0$.
By analogy to \Erefs{displacement_by_integration} and \eref{DisplacementComponentCorrelator}, it is clear the correlator is given by
\begin{equation}
\elabel{DisplacementCorrelatorFourier1D}
\ave{\displacementFourier{p}{t_1}\displacementFourier{-p}{t_2}} = N\activityCorrInt_p^{\mathrm{1D}}(t_1,t_2)
+
N\thermalCorrInt_p^{\mathrm{1D}}(t_1,t_2) \ ,
\end{equation}
where $\activityCorrInt_p^{\mathrm{1D}}(t_1,t_2)$ and $\thermalCorrInt_p^{\mathrm{1D}}(t_1,t_2)$ are given by the corresponding expressions for $\activityCorrInt_\pvec^\pm(t_1,t_2)$, \Eref{def_activityCorrInt}, and $\thermalCorrInt_\pvec^\pm(t_1,t_2)$, \Eref{def_thermalCorrInt}, after replacing all instances of $\eval^\pm_\pvec$, \Eref{eval_formula}, with $\dynamicalScalarFourier{p}{}$, \Eref{DynamicalFourierEigenvalues1D}.
Furthermore, the correlator in real space is given by
\begin{equation}\elabel{DisplacementCorrelatorReal1D}
\ave{\displacement{n}{t_1} \displacement{n'}{t_2}}
= 
\frac{1}{N}\sum_{p=0}^{N-1}
\exp{\imag \wavenumber{p} (n-n')}
\left(
\activityCorrInt_p^{\mathrm{1D}}(t_1,t_2)
+
\thermalCorrInt_p^{\mathrm{1D}}(t_1,t_2)
\right).
\end{equation}
We can also define the relative displacement of the particles by subtracting the average displacement $\displacementBar{t} = \displacementFourier{p=0}{t}/N$, \ie
\begin{equation}\elabel{DisplacementRelFourier1D}
	\displacementRel{n}{t} = \displacement{n}{t} - \displacementBar{t} =
\frac{1}{N} \sum_{p=1}^{N-1} \exp{\imag \wavenumber{p}n} \displacementFourier{p}{t} \ .
\end{equation}
This now permits an exact expression for the steady-state variance that does not \emph{trivially} diverge in $t$,
\begin{equation}\elabel{VarianceDisplacementRel1D}
\lim_{t\to\infty}
\ave{\displacementRel{n}{t}\displacementRel{n}{t}}
=
\frac{1}{N}\sum_{p=1}^{N-1}
\left(
\frac{\activity^2}{\dynamicalScalarFourier{p}{} (\dynamicalScalarFourier{p}{}+\noiseRate)}
+
\frac{\Gamma^2}{2\dynamicalScalarFourier{p}{}}
\right).
\end{equation}

To determine the behaviour of $\ave{\displacementRel{n}{t}\displacementRel{n}{t}}$ in large $N$, one may be tempted to write $N^{-1}\sum_{p=1}^{N-1}1/(1-\cos(2\pi p/N))$ as a Riemann sum. However, such a sum does not converge, because the integrand diverges as the integration limits written in $k=2\pi p/N$ converge to $0$ and $2\pi$ with decreasing integration mesh $1/N$. The sum is entirely dominated by the contributions of $p$ close to $1$ and close to $N-1$.

Hence, for large $N$ and fixed $p$, we may approximate $\dynamicalScalarFourier{p}{}$ by $\pairPot''(\latticeConstant)\wavenumber{p}^2$.  At sufficiently large $N$, we can further assume $\pairPot''(\latticeConstant)\wavenumber{p}^2\ll \noiseRate$, whence we obtain
\begin{equation}\elabel{VarianceDisplacementRel1D_LargeN_Derivation}
\lim_{t\to\infty}
\ave{\displacementRel{n}{t}\displacementRel{n}{t}}
=
\frac{1}{\pairPot''(\latticeConstant)}\left(
\frac{\activity^2}{\noiseRate}
+
\frac{\Gamma^2}{2}
\right)
\frac{2}{N}\sum_{p=1}^{(N-1)/2}
\frac{1}{\wavenumber{p}^2} + \OC(N^0) \ .
\end{equation}
The summands $1/\wavenumber{p}^2$ decay sufficiently fast with increasing $p$ that the upper limit of the sum in \Eref{VarianceDisplacementRel1D_LargeN_Derivation} can be safely extended to $\infty$. Then, using that $\sum_{p=1}^{\infty}1/\wavenumber{p}^2 = N^2/24$, we finally obtain
\begin{equation}\elabel{VarianceDisplacementRel1D_LargeN_Final}
\lim_{t\to\infty}
\ave{\displacementRel{n}{t}\displacementRel{n}{t}}
=
\frac{N}{12\pairPot''(\latticeConstant)}\left(
\frac{\activity^2}{\noiseRate}
+
\frac{\Gamma^2}{2}
\right) + \OC(N^0) \ .
\end{equation}
Hence, the variance of the relative displacement diverges linearly with the number of particles for the one-dimensional crystal, in contrast to the logarithmic divergence found for that of the two-dimensional crystal, \Eref{divergent_var}.

The crystalline integrity for the one-dimensional crystal is better probed by the steady-state mean squared particle separation, which is, up to a shift by $\latticeConstant^2$,
\begin{equation}\elabel{MeanSquaredSeparation1D_defn}
  \lim_{t\to\infty}\ave{|\displacement{n}{t} - \displacement{n+m_q}{t}|^2}
  =
\frac{1}{N}\sum_{p=1}^{N-1}
2\big(1- \cos(\wavenumber{p}) \big)
\left(
\frac{\activity^2}{\dynamicalScalarFourier{p}{} (\dynamicalScalarFourier{p}{}+\noiseRate)}
+
\frac{\Gamma^2}{2\dynamicalScalarFourier{p}{}}
\right),
\end{equation}
which, in contrast to the two-dimensional case, \Erefs{2D_crystalline_integrity} and \eref{distance_variance_2D}, we observe to be independent of $q$. After inserting the explicit expression for $\dynamicalScalarFourier{p}{}$, \Eref{DynamicalFourierEigenvalues1D}, the resulting sum
\begin{equation}\elabel{MeanSquaredSeparation1D_carriedOn}
  \lim_{t\to\infty}\ave{|\displacement{n}{t} - \displacement{n+m_q}{t}|^2}
  =
\frac{1}{N \pairPot''(\latticeConstant)}\left\{
\frac{N-1}{2}\Gamma^2 +  \sum_{p=1}^{N-1}
\frac{\activity^2}{\dynamicalScalarFourier{p}{}+\noiseRate}
\right\}
\end{equation}
has a closed form in the limit $N\to\infty$, because the summand is bounded from above and from below. In particular, using that 
\begin{equation}
\lim_{N\to\infty}\frac{1}{N} \sum_{p=1}^{N} \frac{1}{c-\cos(\wavenumber{p})} 
=\int_0^1 \dint{p}\frac{1}{c-\cos(\wavenumber{p})}
= \frac{1}{\sqrt{c^2 - 1}}
\end{equation}
we find
\begin{equation}\elabel{distance_variance_1D}
  \lim_{N\to\infty}\lim_{t\to\infty}\ave{|\displacement{n}{t} - \displacement{n+m_q}{t}|^2}
  =
\frac{1}{\pairPot''(\latticeConstant)}
\left(
\frac{\activity^2}{\noiseRate\sqrt{1+4\frac{\pairPot''(\latticeConstant)}{\mu}}}
+
\frac{\Gamma^2}{2}
\right).
\end{equation}

Next, we have the total instantaneous energy of the one-dimensional crystal, 
\begin{equation}\elabel{def_internalEnergy1D}
\internalEnergy(t) = 
\half \sum_{n=1}^{N}\sum_{q=1}^{2}
\pairPot(|\displacement{n}{t}-\displacement{n+m_q}{t} - \latticeConstant m_q|) \ ,
\end{equation}
with the factor $1/2$ accounting for the double counting of the pair potentials in the double sum, since $|\displacement{n}{t}-\displacement{n+m_q}{t} - \latticeConstant m_q|$ is the same for either $n$ with $q=1$, or $n+1$ with $q=2$.
Unlike the two-dimensional crystal, the average energy can be expressed in terms of the mean squared separation, \Eref{MeanSquaredSeparation1D_defn}, using the harmonic approximation, \Eref{1DU_harmonic}, \ie
\begin{equation}\elabel{EnergyHarmonic1D}
\ave{\internalEnergyHarmonic(t)} = 
N \pairPot(\latticeConstant)+
\half \sum_{n=1}^{N}\sum_{q=1}^{2}
\half \pairPot''(\latticeConstant)\ave{|\displacement{n}{t} - \displacement{n+m_q}{t}|^2},
\end{equation}
similar to \Eref{def_internalEnergyHarmonic}. 
Using \Erefs{DynamicalFourierEigenvalues1D} and \eref{MeanSquaredSeparation1D_carriedOn} we thus arrive at
\begin{equation}\elabel{expected_energy1D}
    \lim_{t\to\infty}
\ave{\internalEnergyHarmonic(t)} = 
N \pairPot(\latticeConstant)+
\quarter \Gamma^2 (N-1) + \half \activity^2
\sum_{p=1}^{N-1}
\frac{1}{\dynamicalScalarFourier{p}{}+\noiseRate} \ ,
\end{equation}
structurally identical to \Erefs{expected_energy_main} and \eref{stationaryinternalEnergyHarmonic}.
Using \Eref{distance_variance_1D}, and translational invariance in \Eref{EnergyHarmonic1D}, we can read off the steady-state average energy per particle in the thermodynamic limit, yielding
\begin{equation}\elabel{EnergyHarmonic1D_LargeN}
\lim_{N\to\infty}\lim_{t\to\infty}\frac{\ave{\internalEnergyHarmonic(t)}}{N} = 
\pairPot(\latticeConstant)+
\half \left(
\frac{\activity^2}{\noiseRate\sqrt{1+4\frac{\pairPot''(\latticeConstant)}{\mu}}}
+
\frac{\Gamma^2}{2}
\right).
\end{equation}

Lastly, we have the entropy production of the one-dimensional crystal. Starting from the general expression for the steady-state entropy production of an active crystal, \Eref{EPR_instantaneous}, we have for the one-dimensional case,
\begin{equation}\elabel{EPR1D_defn}
\entropyProductionDensity_n = 
\frac{2}{\Gamma^2}
\left(
\activeNoise_n - \sum_{n'=1}^{N} \dynamicalScalar{n}{n'} \displacement{n'}{}
\right)^2    
-
2\pairPot''(\latticeConstant) \ .
\end{equation}
The derivation of a closed-form expression for the average entropy production of the one-dimensional crystal proceeds analogously to the two-dimensional case in \Sref{Entropy}. Skipping over this repetitive detail, we ultimately obtain
\begin{equation}\elabel{Entropy1Dfinal}
\ave{\entropyProductionDensity_n} = \frac{2 \activity^2}{N \Gamma^2}\sum_{p=0}^{N-1}
\frac{\noiseRate}{\noiseRate+\dynamicalScalarFourier{p}{}},
\end{equation}
\cf \Erefs{EPR_derivation_step5_main}, \eref{Entropy1Dfinal_main}
and \eref{entropyProduction_final}. As expected, \Erefs{Entropy1Dfinal} and \eref{EnergyHarmonic1D_LargeN} conform to \Eref{generic_eps_sigma}.

\section{Exact expressions for components of fully time-dependent displacement correlator, \Eref{DisplacementComponentCorrelator}}\seclabel{ExactIntegrals}

Here, we state exact expressions for the active $\activityCorrInt_\pvec^\pm(t_1,t_2)$ and thermal $\thermalCorrInt_\pvec^\pm(t_1,t_2)$ components of the displacement correlator, \Eref{def_WXi}.

First, we consider the integral arising from the thermal-noise correlator, \Eref{def_thermalCorrInt} with \Eref{thermalNoiseFourier_correlator} assuming $\eval_\pvec^\pm\ne0$, 
\begin{equation}\elabel{ThermalNoiseIntegral}
        \thermalCorrInt_\pvec^\pm(t_1,t_2) 
        = \frac{\Gamma^2}{2\eval_\pvec^\pm}\left( e^{-\eval_\pvec^\pm|t_{2} - t_{1}|} - e^{-\eval_\pvec^\pm (t_1 + t_2)} \right) \ .
\end{equation}
In the case of $\eval_\pvec^\pm=0$, in particular for
the $\nullvec$-mode, we obtain instead 
\begin{equation}\elabel{ThermalNoiseIntegral_0mode}
    \thermalCorrInt_\nullvec^\pm(t_1,t_2) = \Gamma^2 \mathrm{min}(t_1 , t_2) \ .
\end{equation}
By considering the long-time $t_1=t_2 \gg 1/\eval_\pvec^\pm$ behaviour 
of \Erefs{ThermalNoiseIntegral} and \eref{ThermalNoiseIntegral_0mode}, we recover \Eref{def_thermalCorrInt_Larget} in \Sref{correlators}.

Next, we calculate the integral \Eref{def_activityCorrInt} arising from the active noise,
\begin{equation}\elabel{activityCorrInt_Derivation}
V\activityCorrInt_\pvec^\pm(t_1,t_2) :=
\int_0^{t_2}\dint{t''}
\int_0^{t_1}\dint{t'}
\exp{-\eval_\pvec^\pm (t_1-t')}
\exp{-\eval_{-\pvec}^\pm (t_2-t'')}
\ave{\hat{\activeNoise}_{\pvec}^{\pm}(t')\hat{\activeNoise}_{-\pvec}^{\pm}(t'')} \ ,
\end{equation}
which gives with \Eref{activeNoiseFourier_correlator} for $\eval_\pvec^\pm\ne0$
\begin{equation}\elabel{ActiveNoiseIntegral}
        \activityCorrInt_\pvec^\pm(t_1,t_2) 
= \frac{\activity^2}{(\eval_\pvec^\pm)^{2}-\noiseRate^{2}}\left(
\exp{-\noiseRate |t_2-t_1|}
- \exp{-\noiseRate t_1 - \eval_\pvec^\pm t_2}
- \exp{-\noiseRate t_2 - \eval_\pvec^\pm t_1}
- \frac{\noiseRate}{\eval_\pvec^\pm} \exp{-\eval_\pvec^\pm |t_2-t_1|}
\right) +
\frac{\activity^2}{\eval_\pvec^\pm (\eval_\pvec^\pm-\noiseRate)} \exp{- \eval_\pvec^\pm (t_1+t_2)}
\end{equation}
and similarly for $\eval_\pvec^\pm = 0$,
\begin{equation}\elabel{ActiveNoiseIntegral_0mode}
        \activityCorrInt_\nullvec^\pm(t_1,t_2) =
        \frac{2 \activity^2}{\noiseRate}\min(t_1,t_2)
        - 
        \frac{\activity^2}{\noiseRate^2}\left(
        1+\exp{-\noiseRate|t_2-t_1|}-\exp{-\noiseRate t_1}-\exp{-\noiseRate t_2}
        \right) \ .
\end{equation}
As above, considering $t_1=t_2 \gg 1/\eval_\pvec^\pm$ and $t_1=t_2 \gg 1/\noiseRate$ in \Erefs{ActiveNoiseIntegral} and \eref{ActiveNoiseIntegral_0mode} recovers \Eref{def_activityCorrInt_Larget} in \Sref{correlators}.

\end{document}